\documentclass[5p]{elsarticle}
\usepackage{CJKutf8}
\usepackage{float}
\usepackage{amsmath}
\usepackage{amssymb}
\usepackage{bm}
\usepackage{booktabs}
\usepackage{color}
\usepackage{dashrule}
\usepackage{epsfig}
\usepackage{epstopdf}
\usepackage{graphicx}
\usepackage{listings}
\usepackage{lipsum}
\usepackage{makecell}
\usepackage{multicol}
\usepackage{multirow}
\usepackage{newtxtext}
\usepackage{newtxmath}
\usepackage{natbib}
\usepackage{hyperref}
\usepackage{overpic}
\usepackage{pifont}
\usepackage{soul}
\usepackage{subfigure}
\usepackage{tikz}
\usepackage{tikz-3dplot}
\usepackage{verbatim} 
\hypersetup{
	colorlinks = true,
	urlcolor   = green,
	citecolor  = blue}
\usetikzlibrary{calc}
\usetikzlibrary{arrows.meta}
\newcommand\Larrow[1]{\mathrel{\raisebox{0.4mm}{\!\!
			\begin{tikzpicture}[>=stealth]
				\node[inner sep=1ex] (a) {$\scriptstyle #1$};
				\draw[<-, line width = 0.8pt] ($(a.south west)+(0,0.05045)$) --($(a.south east)+(0,0.05045)$);
\end{tikzpicture}}}}

\newcommand\Rarrow[1]{\mathrel{\raisebox{0.4mm}{\!\!
			\begin{tikzpicture}[>=stealth]
				\node[inner sep=1ex] (a) {$\scriptstyle #1$};
				\draw[->, line width = 0.8pt] ($(a.south west)+(0,0.05045)$) --($(a.south east)+(0,0.05045)$);
\end{tikzpicture}}}}

\makeatletter
\newsavebox{\@brx}
\newcommand{\llangle}[1][]{\savebox{\@brx}{\(\m@th{#1\langle}\)}%
	\mathopen{\copy\@brx\kern-0.5\wd\@brx\usebox{\@brx}}}
\newcommand{\rrangle}[1][]{\savebox{\@brx}{\(\m@th{#1\rangle}\)}%
	\mathclose{\copy\@brx\kern-0.5\wd\@brx\usebox{\@brx}}}
\makeatother

\newcommand\Rey{\mbox{\textit{Re}}}     % Reynolds number
\newcommand\Mach{\mbox{\textit{Ma}}}    % Mach number
\newcommand\Pran{\mbox{\textit{Pr}}}    % Prandtl number
\newcommand\Knud{\mbox{\textit{Kn}}}    % Knudsen number
\newcommand\Nuss{\mbox{\textit{Nu}}}    % Nusselt number
\newcommand\Stokes{\mbox{\textit{St}}}  % Stokes number

\usetikzlibrary{shapes.geometric,arrows}
\tikzstyle{startstop} = [rectangle, rounded corners, minimum width=1cm, minimum height=0.5cm, 
                         inner xsep = 5pt, inner ysep=3pt, text centered, draw=black, fill=red!30]
\tikzstyle{io}        = [rectangle, rounded corners, minimum width=1cm, minimum height=0.5cm, 
inner xsep = 5pt, inner ysep=3pt, text centered, draw=black, fill=blue!30]
\tikzstyle{process}   = [rectangle, rounded corners, minimum width=1cm, minimum height=0.5cm, 
                         inner xsep = 5pt, inner ysep=3pt, text centered, draw=black, fill=orange!30]
\tikzstyle{decision}  = [rectangle, rounded corners, minimum width=1cm, minimum height=0.5cm, 
						 inner xsep = 5pt, inner ysep=3pt, text centered, draw=black, fill=green!30]
\tikzstyle{arrow}     = [thick,->,>=stealth]

\newcommand{\RomanNumeralCaps}[1]
\linenumbers

\journal{Computer Physics Communications}

\begin{document}

\begin{frontmatter}
	
	\title{A high-fidelity and efficient framework for point-particle direct numerical simulation based on multi-block overset grids} %% Article title
	
	\author[a]{Taiyang Wang} %% Author name
	\author[b,c]{Baoqing Meng} %% Author name
	\author[d]{Baolin Tian} %% Author name
	\author[a]{Yaomin Zhao\corref{cor1}} %% Author name
	\ead{yaomin.zhao@pku.edu.cn}
	
	\affiliation[a]{
	organization={HEDPS and Center for Applied Physics and Technology, College of Engineering, Peking University},
	city={Beijing},
	postcode={100871}, 
	country={China}}

	\affiliation[b]{
	organization={Institute of Mechanics, Chinese Academic of Sciences},
	city={Beijing},
	postcode={100190}, 
	country={China}}
	
	\affiliation[c]{
	organization={University of Chinese Academy of Sciences},
	city={Beijing},
	postcode={101408}, 
	country={China}}
	
	\affiliation[d]{
	organization={School of Aeronautic Science and Engineering, Beihang University},
	city={Beijing},
	postcode={101191}, 
	country={China}}
	
	\cortext[cor1]{Corresponding author}
	
	%% Abstract
	\begin{abstract}
		
		In this work, we present a high-fidelity and efficient point-particle direct numerical simulation framework based on a multi-block overset curvilinear grid system \citep{Deuse2020Implementation}, enabling large-scale Lagrangian particle tracking in complex geometries with high-order accuracy and low computational cost.
		To handle the multi-domain topological challenges inherent in such configurations, we develop an efficient particle storage and redistribution framework leveraging overset grid techniques.
		In particular, two optimization strategies have been proposed for particle redistribution: one is an innovative inter-block mapping within overlapping zones, and the other is a fast search-locate algorithm based on particle velocity.
        Together, these approaches significantly reduce the particle tracking overhead, especially for particles passing through interfaces between overlapping grid blocks.
		Moreover, the accuracy and robustness of the present framework are rigorously validated through various cases, including massless particle trajectories, one- and two-way coupled simulations.
		Specifically, we demonstrate the framework’s applicability to the direct numerical simulation of particle-laden flow in a linear compressor cascade at engine-relevant conditions, showcasing its capability to resolve complex particle dynamics in turbomachinery configurations with low computational costs.
	\end{abstract}
	
%	\begin{graphicalabstract}
%		%\includegraphics{grabs}
%	\end{graphicalabstract}
	
%	\begin{highlights}
%		\item Research highlight 1
%		\item Research highlight 2
%	\end{highlights}
	
	%% Keywords
	
\end{frontmatter}

\section{Introduction}\label{sec:Introduction}

Particle-laden turbulent flows have long been a focal point in multiphase fluid dynamics research due to their prevalence in both natural phenomena and industrial applications.
Notable examples include atmospheric pollutant dispersion, colloidal suspension dynamics, and turbine blade fouling by ash deposition \citep{Abhijit2008Transport, Balachandar2010Turbulent, Brandt2022Particle}.
Owing to the limitations of existing experimental methodologies, high-fidelity numerical simulations have become an indispensable tool for studying these complex flows.
Remarkable progress in high-performance computing has further solidified computational fluid dynamics as the primary investigative approach in this field, leading to the development of numerous numerical models and techniques in recent decades.

In early studies, the two-phase system of particle-laden turbulent flow is usually described by the two-fluid model, in which both the continuous and dispersed phases are treated as continua within an Euler–Euler framework \citep{Baer1986two, Druzhinin1998Direct, Slater2003Particle, Crowe2011Multiphase}.
However, formulating a physically consistent closure model for the averaged particle momentum equations poses a formidable challenge.
Additionally, excessive simplification in solid-phase modeling fundamentally compromises the resolution of fine-scale flow structures \citep{Sundaresan2018Toward}.
Consequently, modern computational studies in gas-solid flows predominantly adopt hybrid Euler-Lagrange methods, wherein the fluid phase is resolved through the Eulerian framework while solid particles are typically modeled as discrete spheres and tracked in a Lagrangian sense.
Within this paradigm, multiple distinct numerical strategies are proposed to characterize the particle-laden flow, which are distinguished by resolved scales of both two phases.
For the fluid phase, early studies predominantly relied on large-eddy simulations (LES) \citep{Balachandar2010Turbulent}.
With the advancement of numerical algorithms and high-performance computing techniques, more and more attempts have transitioned to direct numerical simulations (DNS), which provide finer resolution of the smallest eddies \citep{Abhijit2008Transport, Balachandar2010Turbulent}.
Furthermore, according to the particle-to-Kolmogorov scale ratio, two major methodological approaches have been developed.
One is the point-particle direct numerical simulation (PP-DNS), and the other is the particle-resolved direct numerical simulation (PR-DNS) \citep{Maxey2017Simulation, Sundaresan2018Toward, Brandt2022Particle}.
The applicability of PP-DNS remains strictly confined to regimes where the particle diameter is smaller than the Kolmogorov length scales $\eta_\mathrm{f}$.
Besides, the forces and torques acting on the particles are established by modeling the fluid-particle interaction \citep{Crowe2011Multiphase, Norouzi2016Coupled}.
On the contrary, for particles whose size are comparable or larger than $\eta_\mathrm{f}$, the point-particle approximation fundamentally breaks down.
Consequently, it is inevitable to perform fully resolved DNS, thus the simulation is prohibitively expensive and limited to small systems \citep{Balachandar2010Turbulent, Sundaresan2018Toward}.

As demonstrated by Abhijit \citep{Abhijit2008Transport}, the PP-DNS has proven highly effective for investigating the statistical behavior of particles in dilute suspensions, and significant progress has been made in understanding the underlying physical mechanisms on the basis of this approach.
However, the majority of previous studies on PP-DNS remain confined to idealized canonical geometries, including homogeneous isotropic turbulence \citep{Yeung1989Lagrangian, BEC2006Acceleration}, planar channel flow \citep{Lun1997Numerical, Kuerten2011Turbulence, Ruan2024shear}, flat-plate boundary layer flow \citep{Xiao2020Eulerian, Chen2022Two, Yu2024Transport, Liao2024GPU} and circular pipe flow \citep{Marchioli2003Direct, Vreman2007Turbulence}.
As a result, the intricate flow physics inherent to topologically complex configurations which are prevalent in practical engineering applications are critically neglected, such as the turbomachinery flows.

The principal impediments of PP-DNS of geometrically complex flows stem from the key computational challenge, \emph{i.e.} the lack of robust numerical frameworks which are capable of accurately resolving curved boundary layer dynamics while efficiently tracking particle trajectories through geometrically complex domains.
%while maintaining solution fidelity, and the other is difficulties in efficiently tracking particle trajectories through geometrically complex domains.
With unstructured meshes, some efforts have been made and achieved the goal of tracking particles for arbitrary curved elements \citep{Ortwein2019particle, Ching2020two-way, Ching2021Development}.
However, these solvers are far from capable of resolving turbulence and conducting large-scale particle simulations due to the limitation of accuracy or efficiency.
%Besides, no results have been published to confirm their ability to handle the intricate geometric characteristics of turbine blade passages.
In order to remedy some of these shortcomings, Kopper \textit{et al} \citep{Kopper2023framework} recently proposed a particle-tracking framework based on a fluid solver using the discontinuous Galerkin (DG) method, focusing on efficiently achieving particle tracking on massively parallel systems. 
%Although the authors claim high-fidelity of their framework, no DNS data of turbomachinery applications have been published so far.
Nevertheless, to the best of the authors' knowledge, no PP-DNS of turbomachinery flows have been reported so far.
Instead, existing studies mostly employed the large-eddy simulation \citep{Beck2019Towards}, despite its inherent limitations in capturing small-scale turbulent scales.
This is likely because of the inherent challenges for the DG method in maintaining the required high-order accuracy for high Reynolds number turbulence simulations, since increasing the formal order of the scheme will introduce numerical instabilities and incur prohibitive computational costs \citep{Ferrer2023HORSES3D}.

In comparison, the finite difference method based on structured grids exhibits remarkable advantages for implementing high-order schemes while maintaining computational efficiency \citep{Bodony2011Provably}.
Guided by this consideration, different numerical strategies are devised based on multi-block implementations to extend its flexibility when applied to complex geometries.
Traditionally, the computational domain is divided into individual blocks with exactly matched inter-block interfaces \citep{Sorenson1982Grid}.
Nevertheless, it poses a great challenge to the mesh generation of high quality computational grids, since the existence of singularity or unsmoothness of the overall grid will significantly reduce the accuracy of simulations \citep{Gross2008Poisson}.
In addition, the accuracy of interface conditions cannot be effectively guaranteed at inter-block boundaries \citep{Kim2003CIC}.
Consequently, the overset grid approach gradually becomes prevalent in industrial applications owing to its relatively simple mesh generation procedure and interface treatments \citep{Deuse2020Implementation}.
However, to the authors' best knowledge, no corresponding particle-tracking framework based on the overset framework has been proposed in the literature to date.
Therefore, to fill the void in the present study, we aim to develop a PP-DNS framework based on the overset grid methodology.
The objectives of the present framework are two-folds: one is resolving geometrically complex turbulent flows, including curved blade boundary layer dynamics, with high-order finite difference schemes based on structured grids; and the other is accurately and efficiently tracking large-scale particles through geometrically complex domains.
These advantages are expected to enable physically realistic simulations of particle-laden turbomachinery flows at engine-relevant conditions.

The outline of this paper is as follows. 
The theory is proposed in \S~\ref{sec:Theory}, which includes the governing equations for both the continuous and dispersed phase, the coupling strategies and different models for particle-wall interactions.
In \S~\ref{sec:Numerical_Methodology}, the numerical methodologies are discussed in detail, including the overset method for the fluid phase and the newly implemented framework for the particle phase dynamics.
Particularly, the acceleration algorithms for particle redistribution in geometrically complex domains proposed in the present study are discussed in detail.
Subsequently, the numerical validations of the present framework are provided in \S~\ref{sec:Validation}, where the function of Lagrange tracking of massless particles, implementations of one- and two-way coupling strategies are verified.
Besides, the numerical acceleration effects of the proposed algorithms are validated in a large-scale turbomachinery application, and the results are exhibited in \S~\ref{sec:acceleration}.
Conclusions and outlooks are discussed in \ref{sec:Conclusion_and_outlook}.

\section{Theory}\label{sec:Theory}

The computational framework adopts a coupled Eulerian-Lagrangian approach for simulating particle-laden flows \citep{Balachandar2010Turbulent, Brandt2022Particle}, where the continuous and dispersed phases are handled in the Eulerian and Lagrangian viewpoint, respectively.
The continuous phase usually refers to the thermally perfect gas, whose dynamics are described by the three-dimensional compressible Navier-Stokes equations.
Besides, the dispersed phase is treated as discrete Lagrangian particle, whose dynamics are rigorously governed by two governing equations: Newton's second law for the translation and the Euler's equations for rotation \citep{Crowe2011Multiphase, Norouzi2016Coupled}.
Moreover, subscripts $\mathrm{f}$ and $\mathrm{p}$ are used to distinguish variables corresponding to the flow or particle.

\subsection{Continuous phase}

The fluid field is governed by the three-dimensional compressible Navier-Stokes equations.
For a dilute gas-particle flow \citep{Zhang2001Explosive, Tian2020Compressible, Poroshyna2021Numerical}, the complete set of non-dimensional governing equations for the continuous phase consists of
\begin{align}
	\label{NS}
	\frac{\partial \rho_\mathrm{f}}{\partial t}+\nabla\cdot(\rho_\mathrm{f}\bm{u}_\mathrm{f})&=0,\notag\\
	\frac{\partial(\rho_\mathrm{f}\bm{u}_\mathrm{f})}{\partial t}+\nabla\cdot(\rho_\mathrm{f}\bm{u}_\mathrm{f}\bm{u}_\mathrm{f}+p_\mathrm{f}\bm{\delta})&=\nabla\cdot\bm{\tau}_\mathrm{f}+\bm{S}_m, \\
	\frac{\partial(\rho_\mathrm{f} e_\mathrm{f})}{\partial t}+\nabla\cdot[\bm{u}_\mathrm{f}(\rho_\mathrm{f} e_\mathrm{f}+p_\mathrm{f})]&=\nabla\cdot[(\bm{\tau}_\mathrm{f}\cdot\bm{u}_\mathrm{f})-\bm{q}_\mathrm{f}]+S_e,\notag
\end{align}
whose solutions are calculated using the Uni-Melb solver HiPSTAR \citep{Sandberg2015Compressible, Sandberg2022Fluid}.

Here, $\rho_\mathrm{f}, \bm{u}_\mathrm{f}$ and $p_\mathrm{f}$ represent the non-dimensionalized flow density, velocity components and pressure, respectively.
$\bm{\delta}$ is the Kronecker symbol.
Moreover, the pressure $p_\mathrm{f}$ is obtained from the non-dimensional equation of state
\begin{equation}
	p_{\mathrm{f}}=\frac{\rho_\mathrm{f}T_\mathrm{f}}{\gamma\Mach^2_\mathrm{f}},
\end{equation}
where $T_\mathrm{f}$ is the non-dimensionalized flow temperature, $\gamma=1.4$ is the specific heat ratio and $\Mach_\mathrm{f}$ is the flow mach number.
Besides, the total energy $e_\mathrm{f}$ is given by
\begin{equation}
	e_{\mathrm{f}}=\frac{1}{2}\bm{u}_\mathrm{f}\cdot\bm{u}_\mathrm{f}+\frac{T_\mathrm{f}}{\gamma(\gamma-1) \Mach^2_\mathrm{f}}.
\end{equation}
The viscous stress tensor $\bm{\tau}_\mathrm{f}$ and heat flux $\bm{q}_\mathrm{f}$ are written as
\begin{equation}
	\begin{aligned}
		\bm{\tau}_\mathrm{f}&=\frac{\mu_\mathrm{f}}{\Rey_\mathrm{f}}\left[\nabla(\bm{u}_\mathrm{f}+\bm{u}_\mathrm{f}^\mathrm{T})-\frac{2}{3}(\nabla\cdot\bm{u}_\mathrm{f})\bm{\delta}\right],\\
		\bm{q}_\mathrm{f}&=-\frac{\mu_\mathrm{f}}{(\gamma-1)\Pran_\mathrm{f} \Rey_\mathrm{f}\Mach^2_\mathrm{f}}\nabla T_\mathrm{f},
	\end{aligned}
\end{equation}
where $\Rey_\mathrm{f}$ and $\Pran_\mathrm{f}$ are the flow Reynolds number and Prandtl number, respectively.
Based on the temperature $T_\mathrm{f}$, the molecular viscosity $\mu_\mathrm{f}$ is calculated according to the Sutherland's law \citep{White1991viscous}
\begin{equation}
	\mu_\mathrm{f}=T^{3/2}_\mathrm{f}\frac{1+C_\mathrm{Suth}}{T_\mathrm{f}+C_\mathrm{Suth}},
\end{equation}
where $C_\mathrm{Suth}$ is the Sutherland's constant.
Additionally, source terms $\bm{S}_m$ and $S_e$ account for the influence of the dispersed phase on the fluid, which will be discussed detailedly in subsection \ref{subsec:coupling_strategy}.

\subsection{Dispersed phase}

The particles are modeled as spheres with identical diameter $d_\mathrm{p}$ and density $\rho_\mathrm{p}$.
With the assumption that the particle size $d_\mathrm{p}$ is smaller than the flow Kolmogorov scale $\eta_\mathrm{f}$, the point-particle approximation becomes valid, justifying our adoption of the Lagrangian point-particle approach \citep{Balachandar2010Turbulent, Brandt2022Particle}.
This treatment allows particles to be represented as discrete points while retaining their inertial and dynamic effects.

Following the previous studies \citep{Li2016Modulation, Chen2022Two}, the non-dimensional governing equations for the dispersed phase are formulated as:
\begin{equation}
	\label{particle_equations}
	\begin{aligned}
		\frac{\mathrm{d}\bm{r}_\mathrm{p}}{\mathrm{d}t}&=\bm{u}_\mathrm{p},\\
		m_\mathrm{p}\frac{\mathrm{d}\bm{u}_\mathrm{p}}{\mathrm{d}t}&=\bm{f}_\mathrm{p},\\
		I_\mathrm{p}\frac{\mathrm{d}\bm{\omega}_\mathrm{p}}{\mathrm{d}t}&=\bm{M}_\mathrm{p},\\
		c^*_{\!p\!,\mathrm{p}}m_\mathrm{p}\frac{\mathrm{d}T_\mathrm{p}}{\mathrm{d}t}&=c^*_{\!p\!,\mathrm{f}}q_\mathrm{p}.
	\end{aligned}
\end{equation}
Here, $\bm{r}_\mathrm{p}, \bm{u}_\mathrm{p}, \bm{\omega}_\mathrm{p}$ and $T_\mathrm{p}$ denote the position vector, translational velocity vector, angular velocity vector and temperature of the particle, respectively.
The particle mass is given by $m_\mathrm{p}=\rho_\mathrm{p}V_\mathrm{p}$ with the particle volume $V_\mathrm{p}=\pi d_\mathrm{p}^3/6$, and the net hydrodynamic force acting on the particle is denoted by $\bm{f}_\mathrm{p}$.
For rotational dynamics, the moment of inertia $I_\mathrm{p}=m_\mathrm{p}d_\mathrm{p}^2/10$ governs the particle's response to the applied torque $\bm{M}_\mathrm{p}$.
Besides, neglecting internal thermal resistance, an uniform particle temperature is assumed \citep{Norouzi2016Coupled}.
In the thermal equation, $q_\mathrm{p}$ represents the interphase heat transfer rate.
Additionally, $c^*_{\!p\!,\mathrm{p}}$ and $c^*_{\!p\!,\mathrm{f}}$ are respectively the dimensional specific heat capacities of the particle and fluid, where asterisked quantities throughout this work indicate dimensional parameters.

Owing to the complex operation conditions in modern aircraft engines, such as strong pressure gradients, high vorticity regions and elevated thermal environments \citep{Sandberg2022Fluid}, many different factors need to be considered to model the force $\bm{f}_\mathrm{p}$ \citep{Maxey1983Equation, Li2016Modulation}.
Consequently, focusing on the solid particles which are significantly heavier than the fluid $(\rho_\mathrm{g}/\rho_\mathrm{f}\gg1)$, $\bm{f}_\mathrm{p}$ in the present framework is systematically formulated as
\begin{equation}
		\bm{f}_\mathrm{p}=\bm{f}_\mathrm{drag}+\bm{f}_\mathrm{pre}+\bm{f}_\mathrm{body}+\bm{f}_\mathrm{saff}+\bm{f}_\mathrm{mag}+\bm{f}_\mathrm{therm},
\end{equation}
where $\bm{f}_\mathrm{drag}, \bm{f}_\mathrm{pre}, \bm{f}_\mathrm{body}, \bm{f}_\mathrm{saff}, \bm{f}_\mathrm{mag}$ and $\bm{f}_\mathrm{therm}$ represent the drag force, the pressure gradient force, the body force, the Saffman lift force, the Magnus lift force and the thermophoretic force, respectively.
In fact, the modeling for force $\bm{f}_\mathrm{p}$ is still an open question \citep{Norouzi2016Coupled}, thus the specific form of $\bm{f}_\mathrm{p}$ depends on the flow configuration to be faced.
In the following, we will introduce each force in detail, including their physical meaning and mathematical expression.

\subsubsection{Drag force}

The drag force $\bm{f}_\mathrm{drag}$ is proportional to the relative velocity, which can be expressed as
\begin{equation}
	\bm{f}_\mathrm{drag}=\frac{1}{\Rey_\mathrm{f}}\kappa_\mathrm{p}(\bm{u}_\mathrm{f}-\bm{u}_\mathrm{p}),
\end{equation}
where $\kappa_\mathrm{p}$ is the interphase momentum transfer coefficient per unit volume, and $\bm{u}_\mathrm{f}$ denotes the theoretical flow velocity neglecting disturbances caused by the particle, which is obtained by interpolating the flow velocity at the particle center of mass.
For an assembly of particles at finite Reynolds numbers, there is no exact theoretical formulation of $\kappa_\mathrm{p}$.
Nevertheless, many empirical drag models have been established based on the particle Reynolds number $\Rey_\mathrm{p}=\Rey_\mathrm{f}|\bm{u}_\mathrm{f}-\bm{u}_\mathrm{p}|d_\mathrm{p}/\nu_\mathrm{f}$, where $\nu_\mathrm{f}=\mu_\mathrm{f}/\rho_\mathrm{f}$ is the fluid kinematic viscosity.
For sufficiently dilute flows, the following model
\begin{equation}
	\frac{\kappa_\mathrm{p}}{6\pi\mu_\mathrm{f}R_\mathrm{p}}=\frac{C_\mathrm{d}}{24}\Rey_\mathrm{p}
\end{equation}
proposed by Wen and Yu \cite{Wen1966Mechanics} is employed, where $R_\mathrm{p}$ is the particle radius.
Additionally, $C_\mathrm{d}$ is the drag coefficient, which is expressed as
\begin{equation}
	C_\mathrm{d}=\left\{
	\begin{aligned}
		&\dfrac{24}{\Rey_\mathrm{p}}\left(1+0.15\Rey_\mathrm{p}^{0.687}\right)   &\Rey_\mathrm{p}\le1000\\
		&0.44                                                                   &\Rey_\mathrm{p}>1000
	\end{aligned}\right..
\end{equation}
Be primarily concerned with the subsonic gas turbine operation without shocks, we neglect compressibility effects and real gas effects in the drag coefficient formulation \citep{Loth2008Compressibility, Loth2021Supersonic}.
Such a simplification is justified for the majority of operating conditions, though these effects would require reconsideration in specific extreme flow configurations \citep{Capecelatro2022Modeling, Capecelatro2024Gas}.

Furthermore, in most cases, the drag force typically plays a dominant role \citep{Elghobashi1992, Marchioli2008Statistics}.
Under this condition, the particle momentum equation in equation \eqref{particle_equations} simplifies to
\begin{equation}
	\label{drag_equation}
	\frac{\mathrm{d}\bm{u}_\mathrm{p}}{\mathrm{d}t}=\frac{C_\mathrm{d}}{24}\frac{\Rey_\mathrm{p}}{\Rey_\mathrm{f}}\frac{(\bm{u}_\mathrm{f}-\bm{u}_\mathrm{p})}{\tau_\mathrm{p}},
\end{equation}
where $\tau_\mathrm{p}=\rho_\mathrm{p}d_\mathrm{p}^2/(18\mu_\mathrm{f})$ evaluates the relaxation time for a particle to respond to the change of flow field.
With the characteristic time scale of the flow is denoted as $\tau_\mathrm{f}$, an important dimensionless parameter is introduced, namely the Stokes number
\begin{equation}
	\Stokes=\frac{\tau_\mathrm{p}}{\tau_\mathrm{f}},
\end{equation}
which governs the particle dynamics.
The particles with larger Stokes numbers are predominantly driven by the inertia, while those with small Stokes numbers tend to follow the flow streamlines \citep{Kopper2023framework}.

\subsubsection{Pressure gradient force}

In turbomachinery flows, blade boundary layers often experience strong pressure gradients \citep{Zhao2020Bypass, Wang2023Direct}. 
Thus the pressure gradient force might be necessary in the particle momentum equation, and it is formulated as
\begin{equation}
	\bm{f}_\mathrm{pre}=-V_\mathrm{p}\nabla p_\mathrm{f},
\end{equation}
which hinders the movement of particles.

\subsubsection{Body force}

The body force usually refers to the volume force caused by the gravitational acceleration $\bm{g}$, which is computed as
\begin{equation}
	\bm{f}_\mathrm{body}=(m_\mathrm{p}-m_\mathrm{f})\bm{g}.
\end{equation}
Here, $m_\mathrm{f}=\pi \rho_\mathrm{f}d_\mathrm{p}^3/6$ accounts for the influence of buoyancy \citep{Maxey1983Equation, Kopper2023framework}.
Besides, in rotating reference frames relevant to turbomachinery applications, additional fictitious forces must be considered, including the centrifugal and Coriolis forces \citep{Abhijit2008Transport}.

\subsubsection{Saffman lift force}

For a non‐rotating spherical particle in a non‐uniform shear flow, a lateral lift force arises due to pressure asymmetry, known as the the Saffman lift force \citep{Saffman1965lift,Saffman1968lift}.
This force becomes particularly relevant in turbomachinery flows with strong vorticity \citep{Sandberg2022Fluid, Ruan2024shear}, and is typically modeled as
\begin{equation}
	\bm{f}_\mathrm{saff}=1.615C_\mathrm{l,s}\frac{1}{\sqrt{\Rey_\mathrm{f}}}d_\mathrm{p}^2\sqrt{\frac{\mu_\mathrm{f}\rho_\mathrm{f}}{|\bm{\omega}_\mathrm{f}|}}(\bm{u}_\mathrm{f}-\bm{u}_\mathrm{p})\times\bm{\omega}_\mathrm{f},
\end{equation}
where $C_\mathrm{l,s}$ is the Saffman lift coefficient and $\bm{\omega}_\mathrm{f}=\nabla\times\bm{u}_\mathrm{f}$ represents the local flow vorticity interpolated at the particle position.
Moreover, under the classical assumptions that $\Rey_\mathrm{p}\ll1$, shear Reynolds number $\Rey_\mathrm{s}=\Rey_\mathrm{f}|\bm{\omega}_\mathrm{f}|d_\mathrm{p}^2/\nu_\mathrm{f}\ll1$ and $\Rey_\mathrm{s}\ll\sqrt{\Rey_\mathrm{p}}$, it can be derived that $C_\mathrm{l,s}=1$.
However, for the study of particulate motions in turbulent flows, an expression for the shear lift force at larger $\Rey_\mathrm{p}$ is necessary.
Based on the numerical results reported by Dandy and Dwyer \cite{Dandy1990sphere}, the model proposed by Mei \cite{Mei1992approximate} is used:
\begin{equation}
	C_\mathrm{l,s}=\left\{
	\begin{aligned}
		&\exp\left(-\frac{\Rey_\mathrm{p}}{10}\right)+0.3314\sqrt{\xi_\mathrm{p}}\left[1-\exp\left(-\frac{\Rey_\mathrm{p}}{10}\right)\right]&\Rey_\mathrm{p}\le40\\
		&0.0524\sqrt{\xi_\mathrm{p}\Rey_\mathrm{p}}&\Rey_\mathrm{p}>40
	\end{aligned}\right.,
\end{equation}
where the parameter $\xi_\mathrm{p}=\Rey_\mathrm{s}/(2\Rey_\mathrm{p})=|\bm{\omega}_\mathrm{f}|R_\mathrm{p}/|\bm{u}_\mathrm{f}-\bm{u}_\mathrm{p}|$ is the shear rate.

\subsubsection{Magnus lift force}

For a rotating particle in a uniform flow field, the velocity difference between its upper and lower surfaces creates an asymmetric pressure distribution, generating the Magnus lift force \citep{Rubinow1961transverse}, which is expressed as
\begin{equation}
	\bm{f}_\mathrm{mag}
	=\frac{1}{8}\pi C_\mathrm{l,m}\rho_\mathrm{f}d_\mathrm{p}^2|\bm{u}_\mathrm{f}-\bm{u}_\mathrm{p}|
	\frac{(\bm{\omega}_\mathrm{f}-\bm{\omega}_\mathrm{p})\times(\bm{u}_\mathrm{f}-\bm{u}_\mathrm{p})}{|\bm{\omega}_\mathrm{f}-\bm{\omega}_\mathrm{p}|}.
\end{equation}
Here, the Magnus lift coefficient $C_\mathrm{l,m}$ is given by Lun and Liu \citep{Lun1997Numerical} in the form of
\begin{equation}
	C_\mathrm{l,m}=\left\{
	\begin{aligned}
		&\dfrac{\Rey_\mathrm{r}}{\Rey_\mathrm{p}}                                      &&\Rey_\mathrm{p}\le1\\
		&\dfrac{\Rey_\mathrm{r}}{\Rey_\mathrm{p}}\left(0.178+0.822\Rey^{-0.522}_\mathrm{p}\right) &&1<\Rey_\mathrm{p}
	\end{aligned}\right.,
\end{equation}
where $\Rey_\mathrm{r}=\Rey_\mathrm{f}|\bm{\omega}_\mathrm{f}-\bm{\omega}_\mathrm{p}|d_\mathrm{p}^2/\nu_\mathrm{f}$ is the particle rotation Reynolds number.

\subsubsection{Thermophoretic force}

Owing to the presence of substantial temperature gradients in gas turbines, the thermophoretic force maybe important to model the particle motion \citep{Abhijit2008Transport, Talbot1980Thermophoresis}, and its mathematical expression is
\begin{equation}
	\bm{f}_\mathrm{therm} = -\frac{1}{\Rey^2_\mathrm{f}}\frac{6\pi\mu_\mathrm{f}\nu_\mathrm{f} d_\mathrm{p}C_s\left(\varLambda+2C_t\Knud\right) }{\left(1+6 C_m\Knud\right)\left(1+2\varLambda+4C_t\Knud\right)}\dfrac{\nabla T_\mathrm{f}}{T_\mathrm{f}}.
\end{equation}
Here,  $C_m=1.14, C_s=1.17, C_t=2.18$ are the coefficients corresponding to the velocity slip, thermal creep and temperature jump \citep{Chen2022Two}, respectively. 
Besides, $\varLambda=k^*_\mathrm{f}/k^*_\mathrm{p}$ is the ratio of the dimensional thermal conductivity of the fluid and particle, and the Knudsen number $\Knud=\lambda^*/d^*_\mathrm{p}$ characterizes rarefaction effects through the fluid mean free path $\lambda^*$.
Since thermophoresis is a consequence of the Brownian movement, it dominates primarily in systems of very small particles, typically of equivalent diameters less than 10 $\mu$m.

\subsubsection{Torque}

For the angular momentum equation, the torque exerted by the fluid on a rotating sphere is determined by
\begin{equation}
	\bm{M}_\mathrm{p}=\frac{1}{64}C_\mathrm{r}\rho_\mathrm{f}d^5_\mathrm{p}|\bm{\omega}_\mathrm{f}-\bm{\omega}_\mathrm{p}|(\bm{\omega}_\mathrm{f}-\bm{\omega}_\mathrm{p}).
\end{equation}
According to previous results \citep{Rubinow1961transverse, Sawatzki1970Flow, Dennis1980steady, Li2016Modulation}, the rotational drag coefficient $C_\mathrm{r}$ is modeled as
\begin{equation}
	C_\mathrm{r}=\left\{
	\begin{aligned}
		&\dfrac{64\pi}{\Rey_\mathrm{r}}                               &&\Rey_\mathrm{r}\le32\\
		&\dfrac{12.9}{\sqrt{\Rey_\mathrm{r}}}+\dfrac{128.4}{\Rey_\mathrm{r}}                              &&32<\Rey_\mathrm{r}
	\end{aligned}\right..
\end{equation}

\subsubsection{Interphase heat-transfer}

For the energy equation, the convective heat flow between the particle and surrounding fluid is given by \citep{Norouzi2016Coupled}
\begin{equation}
	q_\mathrm{p}=\frac{1}{\Pran_\mathrm{f}\Rey_\mathrm{f}}\Nuss_\mathrm{p}\pi d_\mathrm{p}(T_\mathrm{f}-T_\mathrm{p}).
\end{equation}
Here, $\Nuss_\mathrm{p}$ is the particle Nusselt number, which is a function of particle Reynolds number $\Rey_\mathrm{p}$ and flow Prandtl number $\Pran_\mathrm{f}$ \citep{Li2000computational, Norouzi2016Coupled}:
\begin{equation}
	\Nuss_\mathrm{p}=\left\{
	\begin{aligned}
		&2+0.6\Rey_\mathrm{p}^{1/2}\Pran_\mathrm{f}^{1/3}                               &&\Rey_\mathrm{p}\le200\\
		&2+\left(0.5\Rey_\mathrm{p}^{1/2}+0.02\Rey_\mathrm{p}^{0.8}\right)\Pran_\mathrm{f}^{1/3}   &&200<\Rey_\mathrm{p}\le1500\\
		&2+0.000045\Rey_\mathrm{p}^{1.8}                                     &&\Rey_\mathrm{p}>1500
	\end{aligned}\right..
\end{equation}

\subsection{Coupling strategies}\label{subsec:coupling_strategy}
 
The particle volume fraction $\varPhi_\mathrm{p}$ represents a fundamental parameter in two-phase flow systems, which not only characterizes the concentration degree of the dispersed phase, but also determines the level of interphase interaction \citep{Balachandar2010Turbulent, Brandt2022Particle}.
For dilute mixtures with a small $\varPhi_\mathrm{p}$, the key physical mechanism is the effect of the turbulent carrier flow on the dynamics of the dispersed phase, and the particle-laden flow in this regime is termed as one-way coupled.
With the increase of $\varPhi_\mathrm{p}$, the back-reaction of inertial particles on the continuous phase gradually plays a significant role, thus the resulting modulation effect cannot be ignored anymore.
In such a case, the flow is termed as two-way coupled.
Based on the regime map proposed by Elghobashi \citep{Elghobashi1994On}, the threshold distinguishing the one-way and two-way coupled regimes is set as $\varPhi_\mathrm{p}=10^{-6}$ for the gas-solid systems \citep{Petersen2019Experimental}.
Moreover, for denser suspensions with $\varPhi_\mathrm{p}$ exceeding $10^{-3}$, four-way coupling emerges as the dominant regime, where inter-particle collisions fundamentally alter system dynamics \citep{Balachandar2010Turbulent, Brandt2022Particle}.
However, owing to the significant demand of computational resources resulted by the collision partners detection, contemporary numerical investigations remain predominantly concentrated on dilute flows.
Consequently, the current study only focus on the one- and two- way coupled regimes.

Following the particle-source-in cell methodology proposed by Crowe \citep{Crowe1977Particle}, we account for two-way coupling through source terms in the governing equations \eqref{NS}.
In the present compressible framework, particle feedback effects on both velocities and temperature fields of the surrounding fluid regime $\Omega$ are modeled, and $\bm{S}_m$ represents particle-to-fluid momentum transfer while $S_e$ captures thermal energy exchange \citep{Kuerten2011Turbulence, Chen2022Two}.
Moreover, their mathematical formulations are
\begin{equation}
	\begin{aligned}
		\bm{S}_m&=-\frac{1}{V_\mathrm{cell}}\sum_{k=1}^{N_\mathrm{p}}m_\mathrm{p}\frac{\mathrm{d}}{\mathrm{d}t}\left(\bm{u}^{k}_\mathrm{p}\right)\chi\left(\bm{r}_{\{l,m,n\}}-\bm{r}^{k}_\mathrm{p}\right),\\
		S_e&=-\frac{1}{V_\mathrm{cell}}\sum_{k=1}^{N_\mathrm{p}}m_\mathrm{p}\frac{\mathrm{d}}{\mathrm{d}t}\left(c^*_{\!p\!,\mathrm{p}}T^{k}_\mathrm{p}+\frac{1}{2}\bm{u}^{k}_\mathrm{p}\!\cdot\!\bm{u}^{k}_\mathrm{p}\right)\chi\left(\bm{r}_{\{l,m,n\}}-\bm{r}^{k}_\mathrm{p}\right).
	\end{aligned}
\end{equation}
Here, the subscript $k$ denotes the $k$-th particle, $V_\mathrm{cell}$ is the volume of the grid cell and $N_\mathrm{p}$ represents the total number of particles within the finite volume $\Omega$. 
Furthermore, $\chi$ is the weighing function that distributes the influence of each particle on the carrier flow through a normalized projection scheme \citep{Jacobs2009high, Kozak2020WENO, Ching2021Development}. 

Because of the geometric disparity between overlapping grids, it's hard to define an identical $\chi$ with the smoothing kernel extrapolation approach \citep{Capecelatro2013Euler} based on a uniform characteristic length scale.
Instead, the affected domain $\Omega$ is selected as the region consisting of the stencil-points for interpolation $\bm{r}_{\{l,m,n\}}$, where $l, m$ and $n$ are cell indices in each direction.
Moreover, the projection kernel $\chi$ is constructed using the interpolation weights themselves.
Thus, the mass conservation relationship 
\begin{equation}
	\sum_{l}\sum_{m}\sum_{n}\chi\left(\bm{r}_{\{l,m,n\}}-\bm{r}^{k}_\mathrm{p}\right)=1
\end{equation}
can be satisfied naturally. 

\subsection{Particle-wall interactions}\label{subsec:boundary_conditions}

In order to accommodate different flow configurations, two distinct strategies are integrated in the present code to model the particle-wall interactions, one is the hard-sphere model and the other is the soft-sphere model.
Both of these strategies guarantee that particles could rebound to the flow field after colliding with the solid wall.

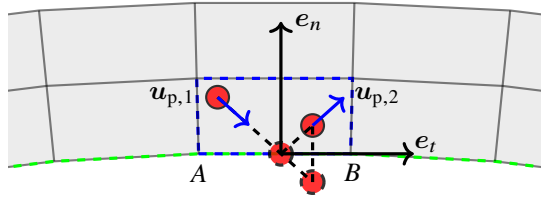
\begin{figure}
	\begin{center}
		\begin{tikzpicture}[scale=0.5]
			\clip (-7,-1.2) rectangle (7,4.5);
			
			\path[fill=gray,opacity=0.15] (-9.8,-0.7) -- (-5.8,-0.2) -- (-2,0) -- (2,0) -- (5.8,-0.2) -- (9.8,-0.7) -- (10,1.3) -- (10.2,3.3) -- (6.2,3.8) -- (2.1,4) -- (-2.1,4) -- (-6.2,3.8) -- (-10.2,3.3) -- (-10,1.3);
			
			\draw [thick, opacity=0.5] (-9.8,-0.7) -- (-5.8,-0.2) -- (-2,0) -- (2,0) -- (5.8,-0.2) -- (9.8,-0.7);
			\draw [thick, opacity=0.5] (-10,1.3) -- (-6,1.8) -- (-2.05,2) -- (2.05,2) -- (6,1.8) -- (10,1.3);
			\draw [thick, opacity=0.5] (-10.2,3.3) -- (-6.2,3.8) -- (-2.1,4) -- (2.1,4) -- (6.2,3.8) -- (10.2,3.3);
			
			\draw [very thick, green, dashed] (-9.8,-0.7) -- (-5.8,-0.2) -- (-2,0) -- (2,0) -- (5.8,-0.2) -- (9.8,-0.7);
			
			\draw [very thick, blue, dashed] (-2,0) -- (2,0) -- (2.05,2) -- (-2.05,2) -- (-2,0);
			
			\draw [thick, opacity=0.5] (-2,0) -- (-2.05,2) -- (-2.1,4);
			\draw [thick, opacity=0.5] (-5.8,-0.2) -- (-6,1.8) -- (-6.2,3.8);
			\draw [thick, opacity=0.5] (2,0) -- (2.05,2) -- (2.1,4);
			\draw [thick, opacity=0.5] (5.8,-0.2) -- (6,1.8) -- (6.2,3.8);
			
			\draw [fill=red, opacity=0.8, thick] (-1.5,1.5) circle (0.3);
			\draw [fill=red, opacity=0.8, thick] (1,0.75) circle (0.3);
			\draw [fill=red, opacity=0.8, very thick, dashed] (0.1667,0) circle (0.3);
			\draw [fill=red, opacity=0.8, very thick, dashed] (1,-0.75) circle (0.3);
			
			\draw [very thick, dashed] (-1.5,1.5) -- (1,-0.75);
			\draw [very thick, dashed] (0.1667,0) -- (1,0.75);
			\draw [very thick, dashed] (1,-0.75) -- (1,0.75);
			
			\draw [very thick, ->] (0.1667,0) -- (3.6667,0);
			\draw [very thick, ->] (0.1667,0) -- (0.1667,3.5);
			
			\draw [very thick, ->, blue] (1,0.75) -- (2-0.1667,1.5);
			\draw [very thick, ->, blue] (-1.5,1.5) -- (-0.6667,0.75);
			
			\node at (4,0.25) {$\bm{e}_t$};
			\node at (0.9,3.5) {$\bm{e}_n$};
			
			\node at (-2.7,1.5) {$\bm{u}_{\mathrm{p},1}$};
			\node at (2.7,1.5) {$\bm{u}_{\mathrm{p},2}$};
			
			\node at (-2,-0.5) {$A$};
			\node at (2.05,-0.5) {$B$};
%			\node at (-5.9,-0.8) {$l_0-1$};
%			\node at (-2,-0.5) {$l_0$};
%			\node at (2.2,-0.5) {$l_0+1$};
%			\node at (5.8,-0.8) {$l_0+2$};
		\end{tikzpicture}
	\end{center}
	\caption{Schematic illustration of particle rebound in a perfectly reflective wall. Here, the particle is denoted by the red circle, and the wall surface is represented by the green dashed line. Moreover, the local cell where the particle is positioned at is marked by blue dashed lines.}
	\label{rebound}
\end{figure}

\subsubsection{Hard-sphere model}\label{subsubsec:hard-model}

The hard-sphere approach is widely adopted in numerical studies \citep{Vreman2007Turbulence, Marchioli2008Statistics, Ruan2024shear, Yu2024Transport}, which assume perfectly elastic particle-wall interactions without the consideration of deformation, and the illustration is presented in figure \ref{rebound}.
Within the discrete framework, the collision with a curved surface is settled based on the local grid cell, whose lower border is modeled as a plane plate.
Consequently, particle locations after the rebound are determined through mirror symmetry, as shown in figure \ref{rebound}.
Moreover, neglecting the impulsive force exerted on particles, the rebound velocities follows from momentum conservation \citep{YAMAMOTO2001Large-eddy}
\begin{equation}m_\mathrm{p}\bm{u}_\mathrm{p,2}=m_\mathrm{p}\bm{u}_\mathrm{p,1}-2m_\mathrm{p}(\bm{u}_\mathrm{p,1}\cdot\bm{e}_n)\bm{e}_n,
\end{equation}
where the subscript 2 denotes the results after the impact while subscript 1 represents the variables before it.
Besides, $\bm{e}_n=\bm{e}_t\times\bm{e}_z$ is the unit normal vector.
Here, $\bm{e}_t=\overrightarrow{AB}/|\overrightarrow{AB}|$ is the unit tangent vector and $\bm{e}_z$ is the unit vector in the spanwise direction.
Additionally, according to the previous results \citep{Crowe2011Multiphase, Kopper2023framework}, the angular velocities during a wall impact are determined by
\begin{equation}
	I_\mathrm{p}\bm{\omega}_\mathrm{p,2}
	=I_\mathrm{p}\bm{\omega}_\mathrm{p,1}-R_\mathrm{p}[\bm{n}\times(m_\mathrm{p}\bm{u}_\mathrm{p,2}-m_\mathrm{p}\bm{u}_\mathrm{p,1})].
\end{equation}

\subsubsection{Soft-sphere model}\label{subsubsec:Soft-model}

Despite the hard-sphere model, the rebound can also be described by the soft-sphere model by modeling the contact force $\bm{f}_\mathrm{cont}$ \citep{Crowe2011Multiphase, Norouzi2016Coupled, Tian2020Compressible}.

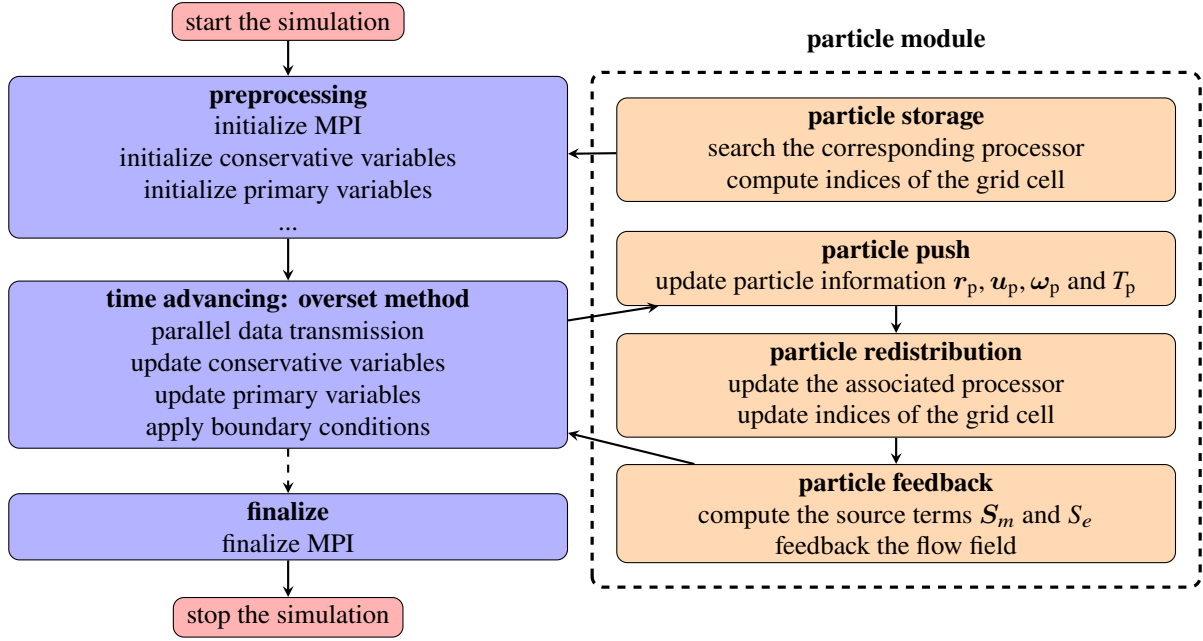
\begin{figure*}[!t]
	\begin{center}
		\begin{tikzpicture}[node distance=1cm]
			% adding nodes
			\node (start) [startstop] {start the simulation};
			\node (io1)   [io, below of=start, yshift=-0.8cm] {\parbox{7cm}{\centering \textbf{preprocessing}\\initialize MPI\\
					initialize conservative variables\\initialize primary variables\\
					...}};
			\node (pro1)  [process, below of=start, yshift=-0.7cm, xshift=8cm] {\parbox{7cm}{\centering \textbf{particle storage}\\search the corresponding processor\\compute indices of the grid cell}};
			\node (io2)  [io, below of=io1, yshift=-1.75cm] {\parbox{7cm}{\centering \textbf{time advancing: overset method}\\parallel data transmission\\update conservative variables\\update primary variables\\apply boundary conditions}};
			\node (io3)  [io, below of=io2, yshift=-1.15cm] {\parbox{7cm}{\centering \textbf{finalize}\\finalize MPI}};
			\node (pro2)  [process, below of=io1, yshift=-0.475cm, xshift=8cm] {\parbox{7cm}{\centering \textbf{particle push}\\update particle information $\bm{r}_\mathrm{p}, \bm{u}_\mathrm{p}, \bm{\omega}_\mathrm{p}$ and $T_\mathrm{p}$}};
			\node (pro3)  [process, below of=pro2, yshift=-0.55cm] {\parbox{7cm}{\centering \textbf{particle redistribution}\\update the associated processor\\update indices of the grid cell}};
			\node (pro4)   [process, below of=pro3, yshift=-0.7cm] {\parbox{7cm}{\centering \textbf{particle feedback}\\compute the source terms $\bm{S}_m$ and $S_e$\\feedback the flow field}};
			\node (stop) [startstop, below of=io3, yshift=-0.2cm] {stop the simulation};
			
			% adding arrows
			\draw [arrow] (start) -- (io1);
			\draw [arrow] (pro1) -- (io1);
			\draw [arrow] (io1) -- (io2);
			\draw [arrow, dashed] (io2) -- (io3);
			\draw [arrow] (io2) -- (pro2);
			\draw [arrow] (pro2) -- (pro3);
			\draw [arrow] (pro3) -- (pro4);
			\draw [arrow] (io3) -- (stop);
			\draw [arrow] (pro4) -- (io2);
			
			\draw [rounded corners, very thick, dashed] (4,-7.5) -- (4,-0.675) -- (12,-0.675) -- (12,-7.5) -- cycle;
			\node at (8,-0.250) {\textbf{particle module}};
			
		\end{tikzpicture}
	\end{center}
	\caption{Flowchart of current framework and computation procedure, where the left part refers to the original fluid solver, and the right part refers to the implanted particle module.}
	\label{framework}
\end{figure*}

In the present framework, the nonlinear viscoelastic contact force model is employed to characterize the collision between particles and the wall surface, incorporating both elastic and dissipative forces.
Assuming an ideally smooth wall, only normal contact forces are considered.
According to the combined theory proposed by Hertz \citep{Hertz1882Ueber} and Tsuji \textit{et al.},  \citep{Tsuji1992Lagrangian}, the mathematical expression of $\bm{f}_\mathrm{cont}$ is
\begin{equation}
	\bm{f}_\mathrm{cont}=[k_{n}\delta _{n}-\eta _{n}(\bm{u}_\mathrm{p}\cdot\bm{e}_n)]\bm{e}_n,
\end{equation}
where $\delta _{n}$ denotes the normal overlap distance.
$k_{n}$ and $\eta _{n}$ respectively represent the spring stiffness and damping coefficient, which are given by
\begin{equation}
	\begin{aligned}
		k_{n}&=\frac{4}{3}E_\mathrm{eff}\sqrt{R_\mathrm{eff}\delta _{n}},\\
		\eta _{n}&=\sqrt{\frac{5}{4}}\beta_{n}\sqrt{m_\mathrm{eff}k_{n}}.
	\end{aligned}
\end{equation}
Here, the parameter $\beta_{n}$ is the function of the restitution coefficient $e_n$, which is expressed as
\begin{equation}
	\beta_{n}=\frac{-2\ln e_n}{\sqrt{\ln ^2e_n+\pi ^2}}.
\end{equation}
Besides, $E_\mathrm{eff}$ denotes the effective Young's modulus
\begin{equation}
	E_\mathrm{eff}=[(1-\nu_\mathrm{p}^2)/E_\mathrm{p}+(1-\nu_\mathrm{b}^2)/E_\mathrm{b}]^{-1}.
\end{equation}
where $\nu_\mathrm{p}, E_\mathrm{p}$ and $\nu_\mathrm{b}, E_\mathrm{b}$ refer to the Poisson ratio and Young's modulus of the particle and wall surface, respectively.
Additionally, treating the wall as an infinitely large plane, the effective radius $R_\mathrm{eff}$ and effective mass $m_\mathrm{eff}$ are approximated as
\begin{equation}
	\begin{aligned}
		R_\mathrm{eff}&=(1/R_\mathrm{p}+1/R_\mathrm{b})^{-1}\approxeq R_\mathrm{p},\\
		m_\mathrm{eff}&=(1/m_\mathrm{p}+1/m_\mathrm{b})^{-1}\approxeq m_\mathrm{p},
	\end{aligned}
\end{equation}
where $R_\mathrm{b}=\infty$ and $m_\mathrm{b}=\infty$ respectively represent the radius and mass of the wall.

\section{Numerical Methodologies}\label{sec:Numerical_Methodology}

The current particle framework is newly developed for the fluid solver based on the overset structured grids, and released on GitHub under the GPLv3 license (\url{https://github.com/Joe0425/Particle}).
The main flow chart of the present fluid solver combined with the particle module is given in figure \ref{framework}.
During the preprocessing step, the Message Passing Interface (MPI) is initialized for parallel computing, as well as the conservative and primary variables.
Additionally, the particle storage strategy is implemented, where the particle is allocated into the corresponding blocks based on the spatial position.
Moreover, indices of the local grid cell are computed.
Then the flow field information is obtained with the overset method \cite{Deuse2020Implementation}, where both the conservative and primary variables are updated.
After that, affected by the carrier flow, all information of the particle is updated, which includes $\bm{r}_\mathrm{p}, \bm{u}_\mathrm{p}, \bm{\omega}_\mathrm{p}$ and $T_\mathrm{p}$.
Subsequently, the particle is redistributed to the corresponding processor, and indices of the grid cell are recalculated.
Specifically, for the two-way coupling regimes, source terms $\bm{S}_m$ and $S_e$ are computed and added to the right hand side of the Navier-Stokes equations to feedback the flow field.
This time advancing process continues until the simulation stops.
In the following, implementation details will be elaborated.

\subsection{Overset grid method}

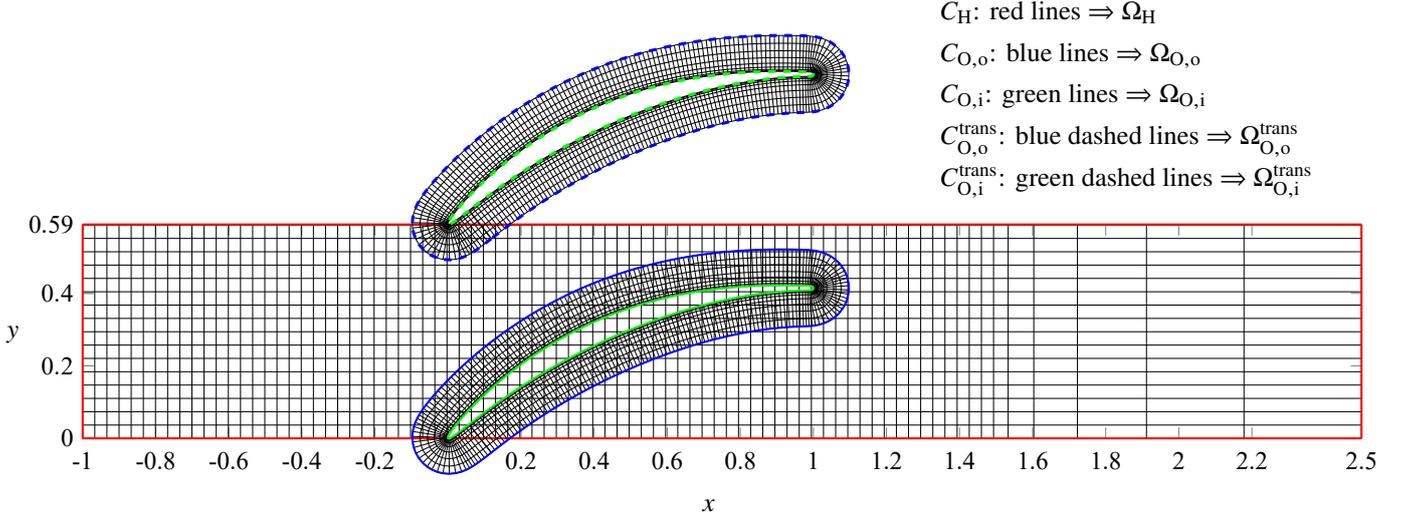
\begin{figure*}[!t]
	\centering
	\begin{overpic}[width=\textwidth]{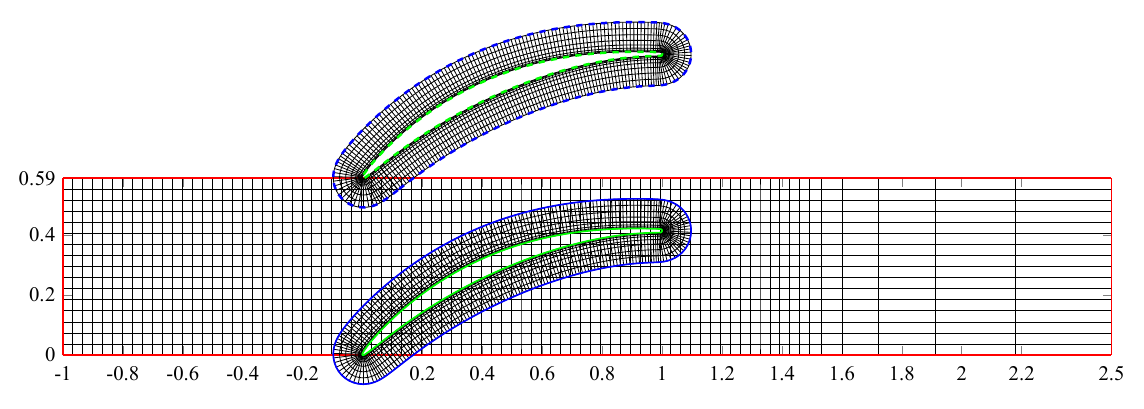}
		\put(50,-0.5){$x$}
		\put(0,12){$y$}
		
		\put(67,35){$C_\mathrm{H}$: red lines $\Rightarrow$ $\Omega_\mathrm{H}$}
		\put(67,32){$C_\mathrm{O,o}$: blue lines $\Rightarrow$ $\Omega_\mathrm{O,o}$}
		\put(67,29){$C_\mathrm{O,i}$: green lines $\Rightarrow$ $\Omega_\mathrm{O,i}$}
		\put(67,26){$C_\mathrm{O,o}^\mathrm{trans}$: blue dashed lines $\Rightarrow$ $\Omega_\mathrm{O,o}^\mathrm{trans}$}
		\put(67,23){$C_\mathrm{O,i}^\mathrm{trans}$: green dashed lines $\Rightarrow$ $\Omega_\mathrm{O,i}^\mathrm{trans}$}
	\end{overpic}
	\caption{Schematic of the axial and pitchwise $(x-y)$ plane for a linear compressor cascade: the computational grid consists of a background H-type grid and an O-type grid. Here, the H-type block is indicated by red lines, while the O-type block is indicated by blue and green lines. Moreover, the transformed O-type grid is also shown.}
	\label{sketch_2D}
\end{figure*}

According to the previous work of Deuse and Sandberg \citep{Deuse2020Implementation}, a stable high-order overset grid method has been successfully implemented in HiPSTAR \citep{Sandberg2015Compressible}, enabling the simulation of turbomachinery flows with complex geometries using multi-block structured grids.
For linear cascade configurations, the computational domain typically consists of a background H-type grid and an O-type grid surrounding the stator blade, which are respectively denoted by red, blue and green lines, as illustrated in figure \ref{sketch_2D}.
The usage of H-type grid maintains the pitchwise periodicity, while the O-type grid ensures accurate resolution of blade boundary layers.

Given the spanwise homogeneity of the flow, only the axial and pitchwise $(x-y)$ plane is shown, and periodic boundary conditions are applied in the spanwise direction $z$.
Owing to the limitation of pitch-to-chord ratio, the pitchwise domain of the background grid may be small, thus the H-type block may not be large enough to contain the whole O-type block.
Therefore, a transformed O-type block is introduced to ensure the periodicity in the pitchwise direction, which is denoted by dashed blue and green lines, as exhibited in figure \ref{sketch_2D}.
Additionally, the left and right boundaries of the H-type block correspond to the inlet and outlet, respectively.

The temporal integration in HiPSTAR employs an ultra-low storage, five-stage, fourth-order accurate Runge–Kutta scheme \citep{Kennedy2000Low}, optimized for large-scale simulations.
Besides, spatial discretizations are achieved through a fourth-order compact finite-difference scheme \citep{Kim2012Efficient}, specifically designed to minimize dispersion and dissipation errors through wavenumber optimization.
Moreover, these grid blocks remain entirely independent during the time advancing process except in their overlapping regions, where data exchange occurs through a fourth-order Lagrange interpolation.
Consequently, this combination of high-order temporal and spatial schemes ensures the fourth-order accuracy in both space and time, which is appropriate for long-time accurate simulations of transition and turbulence \citep{Sandberg2015Compressible, Zhao2020Bypass}.

Building upon this overset grid framework, extensive studies \citep{Deuse2020Different, Zhao2021High, Shubham2022Surface} have demonstrated the code's reliability for single-phase flow simulations, establishing a foundation for particle-laden flow investigations.

\subsection{Particle storage}

In this subsection, the present particle storage strategy will be discussed in detail.
Nevertheless, before the formal discussion, we need to figure out the topological relationships of the common geometric configuration.
Given the spanwise uniformity of the three-dimensional computational domain, we focus exclusively on the $(x-y)$ plane grid topology, and we will take figure \ref{sketch_2D} as an example in the following.

For simplicity of presentation, borders of the H-type block are denoted by $C_\mathrm{H}$, as indicated by red lines in figure \ref{sketch_2D}. 
Besides, the inner boundary of the O-type block is denoted by $C_\mathrm{O,i}$ while the outer boundary is denoted by $C_\mathrm{O,o}$, which are represented by green and blue lines, respectively.
Correspondingly, the inner and outer boundaries of the transformed O-type block are denoted by $C_\mathrm{O,i}^\mathrm{trans}$ and $C_\mathrm{O,o}^\mathrm{trans}$.

Furthermore, we can define the enclosed regions $\Omega_\mathrm{H}$, $\Omega_\mathrm{O,o}$, $\Omega_\mathrm{O,i}$, $\Omega_\mathrm{O,o}^\mathrm{trans}$ and $\Omega_\mathrm{O,i}^\mathrm{trans}$, which respectively represent the interior areas and the borders of curves $C_\mathrm{H}$, $C_\mathrm{O,o}$, $C_\mathrm{O,i}$, $C_\mathrm{O,o}^\mathrm{trans}$ and $C_\mathrm{O,i}^\mathrm{trans}$.
In addition, the computational domain $\Omega_\mathrm{H}$ is further divided into three parts, the valid domain governed by the H-type block $\Omega_\mathrm{H}^\mathrm{val}$, the valid domain governed by the O-type block $\Omega_\mathrm{O}^\mathrm{val}$ and the invalid domain $\Omega^\mathrm{inv}$, which are determined by:
\begin{equation}
	\begin{aligned}
		\Omega_\mathrm{H}^\mathrm{val}&=\Omega_\mathrm{H}-\Omega_\mathrm{H}\cap(\Omega_\mathrm{O,o}\cup\Omega_\mathrm{O,o}^\mathrm{trans}),\\
		\Omega_\mathrm{O}^\mathrm{val}&=\Omega_\mathrm{H}\cap((\Omega_\mathrm{O,o}-\Omega_\mathrm{O,i})\cup(\Omega_\mathrm{O,o}^\mathrm{trans}-\Omega_\mathrm{O,i}^\mathrm{trans})),\\
		\Omega^\mathrm{inv}&=\Omega_\mathrm{H}\cap(\Omega_\mathrm{O,i}\cup\Omega_\mathrm{O,i}^\mathrm{trans}).
	\end{aligned}
\end{equation}
Here, symbols $\cup$, $\cap$ and $-$ are used to represent the union, intersection and difference of two sets \citep{Munkres2000Topology}, respectively.

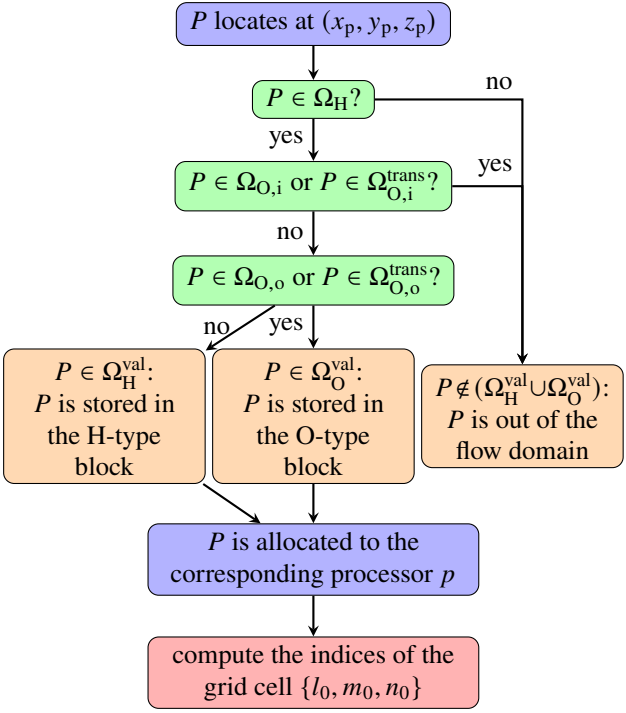
\begin{figure}[!t]
	\begin{center}
		\begin{tikzpicture}[node distance=1cm]
			% adding nodes
			\node (io1)   [io] {$P$ locates at $(x_\mathrm{p},y_\mathrm{p},z_\mathrm{p})$};
			\node (dec1)  [decision, below of=io1] {$P\in\Omega_\mathrm{H}$?};
			\node (dec2)  [decision, below of=dec1, yshift=-0.15cm] {$P\in\Omega_\mathrm{O,i}$ or $P\in\Omega_\mathrm{O,i}^\mathrm{trans}$?};
			\node (dec3)  [decision, below of=dec2, yshift=-0.25cm] {$P\in\Omega_\mathrm{O,o}$ or $P\in\Omega_\mathrm{O,o}^\mathrm{trans}$?};
			
			\node (pro1) [process, below of=dec3, xshift=-2.75cm, yshift=-0.8cm] {\parbox{2.3cm}{\centering $P\in\Omega_\mathrm{H}^\mathrm{val}$:\\ $P$ is stored in the H-type block}};
			\node (pro2) [process, below of=dec3, yshift=-0.8cm] {\parbox{2.3cm}{\centering 
					$P\in\Omega_\mathrm{O}^\mathrm{val}$:\\ $P$ is stored in the O-type block}};
			\node (pro3) [process, below of=dec3, xshift=2.75cm, yshift=-0.8cm] {\parbox{2.3cm}{\centering $P\!\notin\!(\Omega_\mathrm{H}^\mathrm{val}\!\cup\!\Omega_\mathrm{O}^\mathrm{val})$:\\ $P$ is out of the flow domain}};
			
			\node (io2)  [io, below of=pro2, yshift=-0.9cm] {\parbox{4cm}{\centering$P$ is allocated to the corresponding processor $p$}};
			
			\node (stop) [startstop, below of=io2, yshift=-0.5cm] {\parbox{4cm}{\centering compute the indices of the grid cell $\{l_0, m_0, n_0\}$}};
			
			% adding arrows
			\draw [arrow] (io1) -- (dec1);
			\draw [arrow] (dec1) -- node[anchor=east] {yes} (dec2);
			\draw [arrow] (dec2) -- node[anchor=east] {no} (dec3);
			\draw [arrow] (dec3) -- node[anchor=east] {no}(pro1);
			\draw [arrow] (dec3) -- node[anchor=east] {yes}(pro2);
			\draw [arrow] (dec1) -| node[anchor=south east] {no}(pro3);
			\draw [arrow] (dec2) -| node[anchor=south east] {yes}(pro3);
			\draw [arrow] (pro1) -- (io2);
			\draw [arrow] (pro2) -- (io2);
			\draw [arrow] (io2) -- (stop);
			
		\end{tikzpicture}
	\end{center}
	\caption{Flow chart for particle storage during the preprocessing step.}
	\label{particle_storage}
\end{figure}

\begin{figure}[!t]
	\begin{center}
		\tdplotsetmaincoords{60}{120}
		\begin{tikzpicture}[tdplot_main_coords,scale=3]
			\fill[red] (0.75,0.5,0.35) circle (0.75pt);
			
			\fill (0,0,0) circle (0.75pt);
			\fill (0,0.9993,-0.0312) circle (0.75pt);
			\fill (0,0.9993,0.9668) circle (0.75pt);
			\fill (0,0,1) circle (0.75pt);
			\fill (1,0,0) circle (0.75pt);
			\fill (1,0.9993,-0.0312) circle (0.75pt);
			\fill (1,0.9993,0.9668) circle (0.75pt);
			\fill (1,0,1) circle (0.75pt);
			
			\draw[thick] (0,0,0) -- (0,0.9993,-0.0312) -- (0,0.9993,0.9668) -- (0,0,1) -- (0,0,0);
			\draw[thick] (1,0,0) -- (1,0.9993,-0.0312) -- (1,0.9993,0.9668) -- (1,0,1) -- (1,0,0);
			\draw[thick] (0,0,0) -- (1,0,0);
			\draw[thick] (0,0.9993,-0.0312) -- (1,0.9993,-0.0312);
			\draw[thick] (0,0.9993,0.9668) -- (1,0.9993,0.9668);
			\draw[thick] (0,0,1) -- (1,0,1);
			\node at (0,0,0.1) {$\{l_0,m_0,n_0\}$};	
			\node at (0,0,1.1) {$\{l_0,m_0+1,n_0\}$};	
			\node at (0,0.9993,-0.0312+0.1) {$\{l_0+1,m_0,n_0\}$};
			\node at (0,0.9993,0.9668+0.1) {$\{l_0+1,m_0+1,n_0\}$};	
			
			\node at (1,0,0.1) {$\{l_0,m_0,n_0+1\}$};	
			\node at (1,0,1.1) {$\{l_0,m_0+1,n_0+1\}$};	
			\node at (1,0.9993,-0.0312+0.1) {$\{l_0+1,m_0,n_0+1\}$};
			\node at (1,0.9993,0.9668+0.1) {$\{l_0+1,m_0+1,n_0+1\}$};	
			
			\draw[thick,->] (0,1.5,0) -- (0.5,1.5,0) node [left] {$z$};
			\draw[thick,->] (0,1.5,0) -- (0,2,0) node [right] {$x$};  	
			\draw[thick,->] (0,1.5,0) -- (0,1.5,0.5) node [right] {$y$};
		\end{tikzpicture}
	\end{center}
	\caption{Schematic of the particle in the local grid cell, where the particle is denoted by the red sphere and eight grid points are represented by black spheres.}
	\label{Cell}
\end{figure}
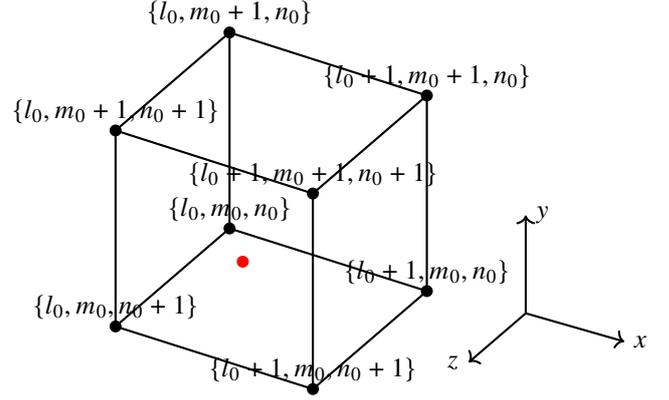

During the preprocessing step, particles that are positioned in the valid domain need to be stored in the corresponding grid cell, while those locate in the invalid domain are removed from the simulation in advance.
In order to illuminate the complete procedure clearly, a flow chart is necessary, and figure \ref{particle_storage} displays the whole process, where the particle is denoted by the point $P$ locates at $(x_\mathrm{p},y_\mathrm{p},z_\mathrm{p})$.
As presented in figure \ref{particle_storage}, firstly, we need to identify whether $P$ is a valid point or not, which can be mathematically expressed as $P\in(\Omega_\mathrm{H}^\mathrm{val}\cup\Omega_\mathrm{O}^\mathrm{val})$ or $P\notin(\Omega_\mathrm{H}^\mathrm{val}\cup\Omega_\mathrm{O}^\mathrm{val})$.
Here, $\in$ and $\notin$ are used to denote whether an element belongs to a set or not \citep{Munkres2000Topology}.
And if it is, then we need to distinguish which block the particle is stored in, the H-type or the O-type block, namely $P\in\Omega_\mathrm{H}^\mathrm{val}$ or $P\in\Omega_\mathrm{O}^\mathrm{val}$.
In the Cartesian coordinate system, it is straightforward to identify whether the particle $P$ is positioned in the rectangular background grid $\Omega_\mathrm{H}$ or not.
However, for the O-type grid, making such a judgment becomes extremely difficult owing to its curved geometry.
Consequently, the ray-casting algorithm \citep{Shimrat1962Algorithm} is adopted, which is effective to identify whether $P$ is inside or outside a closed polygon.
With a preference for the O-type block which has a larger block id in the overset method \citep{Deuse2020Implementation}, the particle is stored in the corresponding grid.
Subsequently, the particle is allocated to the processor $p$ if $P\in\Omega_p$, where $\Omega_p$ represents the subdomain governed by the processor $p$.
Furthermore, indices of the grid cell $\{l_0, m_0, n_0\}$ which contains the particle are calculated, and the storage rule is presented in figure \ref{Cell}.
Specifically, for the particle $P$ which is located at the curved geometry, $\{l_0, m_0\}$ is determined by the angle summation algorithm \citep{Hormann2001point}.

Despite from the information $\bm{r}_\mathrm{p}, \bm{u}_\mathrm{p}, \bm{\omega}_\mathrm{p}$, $T_\mathrm{p}$ and the acceleration of the particle $a_\mathrm{p}=\mathrm{d}\bm{u}_\mathrm{p}/\mathrm{d}t$, two extra Lagrangian variables are included in the current framework.
One is used to denote the processor id $p$ where the particle is stored, which also indicates the block id at the same time.
The other marks the particle that resides in the transformed O-type grid, which is necessary for the subsequent calculations, since the interpolation is performed within the original grid.

\subsection{Particle push}

In this part, the force $\bm{f}_\mathrm{p}$, torque $\bm{M}_\mathrm{p}$ and heat transfer $q_\mathrm{p}$ are computed, then the particle information $\bm{r}_\mathrm{p}, \bm{u}_\mathrm{p}, \bm{\omega}_\mathrm{p}$ and $T_\mathrm{p}$ are updated based on equations \eqref{particle_equations}.

\subsubsection{Interpolation scheme}

During the process of simulation, interpolating variables of the fluid parameters $\varGamma_\mathrm{f}$ at the particle's position $\bm{r}_\mathrm{p}$ is an inevitable part.
In order to satisfy different accuracy requirements for research, three interpolation approaches are integrated in the present framework, which are the trilinear scheme based on surrounding 8 grid points \citep{Hoomans1996Discrete, Xiao2020Eulerian, Yu2024Transport}, third- and fourth-order Lagrange polynomials interpolation based on 64 or 125 grid points \cite{Zhao2013Interphasial, Li2016Modulation, Chen2022Two}.

For ease of notations, the interpolation stencil is denoted by $\mathcal{L}$, and $\mathcal{L}=\mathcal{L}_l\times\mathcal{L}_m\times\mathcal{L}_n$.
Here, $\mathcal{L}_l$, $\mathcal{L}_m$ and $\mathcal{L}_n$ respectively represent the set of stencil-points in $i, j$ and $k$ directions.
Moreover, they are defined by
\begin{equation}
	\begin{aligned}
		\mathcal{L}_l&=\{l|l-l_0+\lfloor(1+N_g)/2\rfloor\in[1,N_g], l\in\mathbb{Z}\},\\
		\mathcal{L}_m&=\{m|m-m_0+\lfloor(1+N_g)/2\rfloor\in[1,N_g], m\in\mathbb{Z}\},\\
		\mathcal{L}_n&=\{n|n-n_0+\lfloor(1+N_g)/2\rfloor\in[1,N_g], n\in\mathbb{Z}\}.
	\end{aligned}
\end{equation}
Here, $N_g$ is the number of stencil-points in each direction, $\lfloor \rfloor$ is the floor function and the symbol $\mathbb{Z}$ denotes the integers.
Furthermore, for the rectilinear H-type grid \citep{Balachandar1989Methods}, the interpolation weights $\chi(\bm{r}_{\{l,m,n\}}-\bm{r}_\mathrm{p})$ can be computed as
\begin{equation}
	\chi(\bm{r}_{\{l,m,n\}}-\bm{r}_\mathrm{p})=
	\mathcal{\zeta}_l(x_\mathrm{p})
	\mathcal{\zeta}_m(y_\mathrm{p})
	\mathcal{\zeta}_n(z_\mathrm{p}),
\end{equation}
where $\bm{r}_{\{l,m,n\}}$ represents the position vector of the grid point $\{l,m,n\}$.
Besides, $\mathcal{\zeta}_l(x_\mathrm{p})$, $\mathcal{\zeta}_m(y_\mathrm{p})$ and $\mathcal{\zeta}_n(z_\mathrm{p})$ are basis functions, which are selected as the Lagrange polynomials, and their mathematical expressions are
\begin{equation}
	\begin{aligned}
		\mathcal{\zeta}_l(x_\mathrm{p})&=\prod_{i\in\mathcal{L}_l,i\neq l}\frac{x_\mathrm{p}-x_i}{x_l-x_i},\\
		\mathcal{\zeta}_m(y_\mathrm{p})&=\prod_{j\in\mathcal{L}_m,j\neq m}\frac{y_\mathrm{p}-y_j}{y_m-y_j},\\
		\mathcal{\zeta}_n(z_\mathrm{p})&=\prod_{k\in\mathcal{L}_n,k\neq n}\frac{z_\mathrm{p}-z_k}{z_n-z_k}.
	\end{aligned}
\end{equation}
Consequently, $\varGamma_\mathrm{f}$ is formulaically expressed as
\begin{equation}
	\varGamma_\mathrm{f}(\bm{r}_\mathrm{p})=\sum_{l}\sum_{m}\sum_{n}\varGamma_\mathrm{f}(\bm{r}_{\{l,m,n\}})\chi(\bm{r}_{\{l,m,n\}}-\bm{r}_\mathrm{p}),
\end{equation}
where $\varGamma_\mathrm{f}(\bm{r}_{\{l,m,n\}})$ represents the fluid parameter at the position $\bm{r}_{\{l,m,n\}}$.
Additionally, under the current framework, the O-type grid is presumed to maintain favorable orthogonality properties, thus the same interpolation scheme is imposed to approximate the local flow information.

\subsubsection{Integration in time}

The trajectory of the particle is obtained by integrating the equation \eqref{particle_equations} with the Velocity-Verlet algorithm \cite{Tartakovsky2005Modeling, Tartakovsky2016Pairwise, Tian2020Compressible}:
\begin{equation}
	\begin{aligned}
		\bm{r}_\mathrm{p}(t+\Delta t)&=\bm{r}_\mathrm{p}(t)+\Delta t\bm{u}_\mathrm{p}+\frac{\Delta t^2}{2m_\mathrm{p}}\bm{f}_\mathrm{p}(t),\\
		\bm{u}_\mathrm{p}(t+\Delta t)&=\bm{u}_\mathrm{p}(t) +\frac{\Delta t}{2m_\mathrm{p}}\left[\bm{f}_\mathrm{p}(t)+\bm{f}_\mathrm{p}(t+\Delta t) \right],\\
	\end{aligned}
\end{equation}
where $\Delta t$ is the time step, and $\bm{f}_\mathrm{p}(t+\Delta t)$ represents the force which is computed using the particle position at time $(t+\Delta t)$.
Moreover, the angular velocity and temperature of the particle are updated with the same strategy:
\begin{equation}
	\begin{aligned}
		\bm{\omega}_\mathrm{p}(t+\Delta t)&=\bm{\omega}_\mathrm{p}(t) +\frac{\Delta t}{2I_\mathrm{p}}\left[\bm{M}_\mathrm{p}(t)+\bm{M}_\mathrm{p}(t+\Delta t)\right],\\
		T_\mathrm{p}(t+\Delta t)&=T_\mathrm{p}(t)+\frac{\Delta t}{2m_\mathrm{p}}\frac{c^*_{\!p\!,\mathrm{f}}}{c^*_{\!p\!,\mathrm{p}}}\left[q_\mathrm{p}(t)+q_\mathrm{p}(t+\Delta t) \right].
	\end{aligned}
\end{equation}
This multi‐step algorithm ensures second-order accuracy of temporal discretizations and third-order accuracy in spatial region \citep{Norouzi2016Coupled}, which satisfies the accuracy requirements for current research \cite{Ruan2024shear}.

\subsection{Optimization strategies for particle redistribution}\label{sec:opti-strategy}

\begin{figure}
	\begin{center}
		\begin{tikzpicture}[node distance=1cm]
			% adding nodes
			\node (start)  [startstop] {\parbox{7.25cm}{\centering $P\in\Omega_\mathrm{H}^\mathrm{val}$: $P$ is originally stored in the H-type block}};
			
			\node (io0)   [io, below of=start, yshift=-0.1cm] {$P$ is originally stored in the processor $p_\mathrm{ori}$};
			
			\node (io1)   [process, below of=io0, yshift=-0.05cm] {update and transform the coordinate $(x_\mathrm{p},y_\mathrm{p},z_\mathrm{p})$};
			\node (dec1)  [decision, below of=io1] {$P\in\Omega_\mathrm{H}$?};
			\node (dec2)  [decision, below of=dec1, yshift=-0.15cm] {$P\in\Omega_\mathrm{O,o}$ or $P\in\Omega_\mathrm{O,o}^\mathrm{trans}$?};
			
			\node (pro1) [process, below of=dec2, xshift=-2.75cm, yshift=-0.8cm] {\parbox{2.3cm}{\centering $P\in\Omega_\mathrm{H}^\mathrm{val}$:\\ $P$ is stored in the H-type block}};
			\node (pro2) [process, below of=dec2, yshift=-0.8cm] {\parbox{2.3cm}{\centering 
					$P\in\Omega_\mathrm{O}^\mathrm{val}$:\\ $P$ is stored in the O-type block}};
			\node (pro3) [process, below of=dec2, xshift=2.75cm, yshift=-0.45cm] {\parbox{2.3cm}{\centering $P\notin\Omega_\mathrm{H}^\mathrm{val}$:\\ $P$ is out of the flow domain}};
			
			\node (dec3)  [decision, below of=pro1, yshift=-0.75cm] {$P\in\Omega_{p_\mathrm{ori}}$?};
			\node (pro4)  [process, below of=pro3, yshift=-0.85cm] {\parbox{2.3cm}{\centering delete $P$ from the processor $p_\mathrm{ori}$}};
			
			\node (io2)  [io, below of=pro2, yshift=-1.9cm] {$P$ is redistributed to the processor $p_\mathrm{new}$};
			
			\node (stop) [startstop, below of=io2, yshift=-0.1cm] {\parbox{4.5cm}{\centering update the indices $\{l_0, m_0, n_0\}$}};
			
			% adding arrows
			\draw [arrow] (start) -- (io0);
			\draw [arrow] (io0) -- (io1);
			\draw [arrow] (io1) -- (dec1);
			\draw [arrow] (dec1) -- node[anchor=east] {yes} (dec2);
			\draw [arrow] (dec2) -- node[anchor=east] {no} (pro1);
			\draw [arrow] (dec2) -- node[anchor=east] {yes} (pro2);
			\draw [arrow] (dec1) -| node[anchor=west] {no} (pro3);
			\draw [arrow] (pro1) -- (dec3);
			\draw [arrow] (pro3) -- (pro4);
			\draw [arrow, blue] (dec3) -- node[anchor=south] {no} (io2);
			\draw [arrow, blue] (dec3) -- ++(-1.425,0) |- node[anchor=north] {yes} (stop);
			\draw [arrow, red] (pro2) -- (io2);
			\draw [arrow, green] (io2) -- (stop);
			
			\node at (-4.175,0.1) {$(a)$};
		\end{tikzpicture}
	\end{center}
	\begin{center}
		\begin{tikzpicture}[node distance=1cm]
			% adding nodes
			\node (start)  [startstop] {\parbox{7.25cm}{\centering $P\in\Omega_\mathrm{O}^\mathrm{val}$: $P$ is originally stored in the O-type block}};
			
			\node (io0)   [io, below of=start, yshift=-0.1cm] {$P$ is originally stored in the processor $p_\mathrm{ori}$};
			
			\node (io1)   [process, below of=io0, yshift=-0.05cm] {update and transform the coordinate $(x_\mathrm{p},y_\mathrm{p},z_\mathrm{p})$};
			\node (dec1)  [decision, below of=io1, yshift=-0.15cm] {$P\in\Omega_\mathrm{O,i}$ or $P\in\Omega_\mathrm{O,i}^\mathrm{trans}$?};
			
			\node (dec2)  [decision, below of=dec1, xshift=-2.25cm, yshift=-0.35cm] {$P\in\Omega_\mathrm{O,o}$ or $P\in\Omega_\mathrm{O,o}^\mathrm{trans}$?};
			\node (pro1)  [process, below of=dec1, xshift=2.25cm, yshift=-0.35cm] {\parbox{2.5cm}{\centering $P\in\Omega^\mathrm{inv}$:\\ particle rebound}};
			
			\node (dec4)  [decision, below of=dec1, yshift=-1.65cm] {$P\in\Omega_{p_\mathrm{ori}}$?};
			
			\node (pro2)  [process, below of=dec1, xshift=-2.75cm, yshift=-3.4cm] {\parbox{2.3cm}{\centering $P\in\Omega_\mathrm{H}^\mathrm{val}$:\\ $P$ is stored in the H-type block}};
			\node (pro3)  [process, below of=dec1, yshift=-3.4cm] {\parbox{2.3cm}{\centering $P\in\Omega_\mathrm{O}^\mathrm{val}$:\\ $P$ is stored in the O-type block}};
			
			\node (io2)  [io, below of=dec1, yshift=-5.1cm] {$P$ is redistributed to the processor $p_\mathrm{new}$};
			
			\node (stop) [startstop, below of=io2, yshift=-0.1cm] {\parbox{4.5cm}{\centering update the indices $\{l_0, m_0, n_0\}$}};
			
			% adding arrows
			\draw [arrow] (start) -- (io0);
			\draw [arrow] (io0) -- (io1);
			\draw [arrow] (io1) -- (dec1);
			\draw [arrow] (dec1) -- node[anchor=east] {no} (dec2);
			\draw [arrow] (dec1) -- node[anchor=west] {yes} (pro1);
			\draw [arrow] (dec2) -- node[anchor=west] {yes} (dec4);
			\draw [arrow] (dec2) -- node[anchor=west] {no} (pro2);
			\draw [arrow] (pro1) -- (dec4);
			\draw [arrow] (dec4) -- node[anchor=west] {no} (pro3);
			\draw [arrow, blue] (dec4) -| ++(4.175,0) |- node[anchor=north] {yes} (stop);
			\draw [arrow, red] (pro2) -- (io2);
			\draw [arrow, blue] (pro3) -- (io2);
			\draw [arrow, green] (io2) -- (stop);
			
			\node at (-4.175,0.1) {$(b)$};
		\end{tikzpicture}
	\end{center}
    \vspace{-0.5cm}
	\caption{Flow chart for particle redistribution during the time advancing process if $P$ is originally stored in the $(a)$ H-type block or $(b)$ O-type block.}
	\label{particle_redistribution}
\end{figure}

During the simulation, each particle's position $(x_\mathrm{p},y_\mathrm{p},z_\mathrm{p})$ is updated at each time step.
Consequently, the processor id $p$ may change and cell indices $\{l_0, m_0, n_0\}$ need recalculation.
Owing to the different geometry and boundary conditions, we need to separately deal with the particle $P$ in the H- or O-type block, and the flow charts are respectively displayed in figure \ref{particle_redistribution}$(a)$ and figure \ref{particle_redistribution}$(b)$.
After the particle location $(x_\mathrm{p},y_\mathrm{p},z_\mathrm{p})$ has been updated with an integration in the time domain as discussed before, the coordinate is then transformed with the boundary condition if $y_\mathrm{p}$ or $z_\mathrm{p}$ is out of the computational domain.
Subsequently, we need to distinguish in which domain the particle is located.
If $P\notin\Omega_\mathrm{H}^\mathrm{val}$, then the particle will be removed from the simulation.
Besides, if $P\in\Omega^\mathrm{inv}$, which means that the particle $P$ collides with the blade surface, we have to carefully handle with the collision to ensure that the particle will not penetrate the wall.
Furthermore, if the particle $P$ is located in the valid domain $\Omega_\mathrm{H}^\mathrm{val}$ or $\Omega_\mathrm{O}^\mathrm{val}$, we need to identify whether it is positioned in the original processor $p_\mathrm{ori}$ or not.
If $P\in\Omega_{p_\mathrm{ori}}$, then the indices $\{l_0, m_0, n_0\}$ are calculated directly, which is shown by dashed flow lines in figure \ref{particle_redistribution}.
On the contrary, if $P\notin\Omega_{p_\mathrm{ori}}$, this particle needs to be redistributed to a new processor $p_\mathrm{new}$ first.

Due to the sophisticated conditional statements, the procedure of the particle redistribution will impose a heavy computational overhead on the hardware, leading to an unaffordable computation cost.
Therefore, it is necessary to speed up this process, and two optimization strategies are introduced in the current study.
One is constructing a mapping between the H- and O-type blocks in the nearby regions of curves $C_\mathrm{O,o}$ and $C_\mathrm{O,o}^\mathrm{trans}$, and the other is substituting the common global search method with the search-locate algorithm.
In figure \ref{particle_redistribution}, the numerical implementations accelerated by these two algorithms are highlighted by red and blue arrows, respectively, with green arrows indicating the process where both approaches can contribute.
These two strategies are discussed in detail below.

\subsubsection{Mapping between blocks}

\begin{figure}[!t]
	\centering
	\begin{overpic}[width=0.5\textwidth]{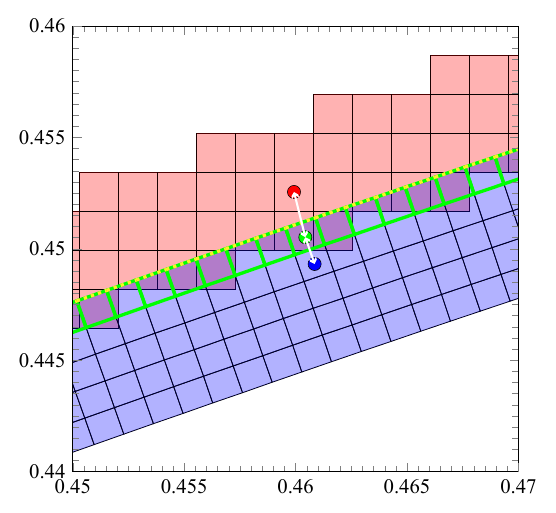}
		\put(52.5,-1){$x$}
		\put(0,46.5){$y$}
		
		\put(20,75){\parbox{3cm}{\centering mapping cell\\in the H-type grid}}
		\put(35,59){
			\begin{tikzpicture}[>=stealth]
				\draw[->,very thick,dashed] (0,0) to [bend left]  (1.5,-1);
			\end{tikzpicture}
		}
		
		\put(30,13){\parbox{3cm}{\centering mapping cell\\in the O-type grid}}
		\put(42,19){
			\begin{tikzpicture}[>=stealth]
				\draw[->,very thick,dashed] (0,0) to [bend left]  (1,2.25);
			\end{tikzpicture}
		}
		
		\put(62,24){\parbox{3cm}{\centering target cell}}
		\put(55.5,27.5){
			\begin{tikzpicture}[>=stealth]
				\draw[->,very thick,dashed] (0,0) to [bend right]  (-2,2);
			\end{tikzpicture}
		}
		
	\end{overpic}
	\caption{Schematic of the mapping between the H- and O-type blocks, where only target and mapping cells are presented, and the interface is marked by the yellow dotted line. Here, target cells are represented by green quadrilaterals. Besides, mapping cells correspond to the H- and O-type grid are colored by red and blue, respectively. Moreover, red, green and blue circles are used to mark the centroid of current grid cell, and the distance between them are denoted by white arrows.}
	\label{mapping}
\end{figure}

Because particles are able to penetrate the outer boundary of the O-type block freely, namely the curves $C_\mathrm{O,o}$ and $C_\mathrm{O,o}^\mathrm{trans}$, processors $p_\mathrm{ori}$ and $p_\mathrm{new}$ may correspond to the same or different block during the redistribution process, as presented in figure \ref{particle_redistribution}.
Generally speaking, if they belong to the same block, $p_\mathrm{new}$ will be preferentially selected from neighboring processes of $p_\mathrm{ori}$.
This procedure will not cost much time, and the search-locate algorithm based on the particle velocity to be mentioned later will accelerate this process.
Nevertheless, on the contrary, if they originate from different blocks, $p_\mathrm{new}$ will be searched from the first process of the corresponding block in the order of default array element, which is an inefficient search strategy.
Besides, the recalculation of indices $\{l_0, m_0\}$ suffers from the same fundamental issue owing to complex composite grid topology.
Notably, in the current framework, the update of the index $\{n_0\}$ is handled separately because of the uniform grid in the spanwise direction, thus will not be discussed in the following.
In other words, if the particle $P$ moves from the H-type block to O-type block, or the converse, the redistribution becomes time-consuming.

In the present study, inspired by overset grid interpolation strategies \citep{Deuse2020Implementation}, that require a inter-block mapping in overlapping regions to determine the finite difference stencils, we establish a bidirectional mapping between the H- and O-type blocks in the vicinity of interfaces $C_\mathrm{O,o}$ and $C_\mathrm{O,o}^\mathrm{trans}$.

In order to elaborate the mapping strategy clearly, each grid cell is given a status:
\begin{itemize}
	\item target cell: 
	the outermost grid layer of the O-type block, as presented by green quadrilaterals in figure \ref{mapping}.
	\item mapping cell: 
	particle $P$ locates in this cell will be redistributed based on the mapping when $P$ passes through the interface. 
	Those cells are presented by red and blue quadrilaterals as shown in figure \ref{mapping}, the red ones correspond to the H-type grid, while the blue ones correspond to the O-type block. 
	Specially, it should be noticed that target cell is the mapping cell belong to the O-type block.
	\item non-mapping cell: 
	nothing special needs to be done at this cell. 
	Particle $P$ which locates in this cell is considered to hardly pass through the interface, thus the mapping will not be conducted for those cells. 
\end{itemize}

Moreover, for the convenience of expression, each target cell is denoted as the area $\Omega^\mathrm{tar}$ with its centroid positioned at $\bm{r}^\mathrm{tar}$.
Additionally, $\Omega^\mathrm{cell}$ represents an arbitrary grid cell with its centroid lying at $\bm{r}^\mathrm{cell}$.
Consequently, mapping cells from the H-type block $\Omega_\mathrm{H}^\mathrm{map}$ is determined by
\begin{equation}
	\Omega_\mathrm{H}^\mathrm{map}=\left\{\Omega^\mathrm{cell}\left|
	\begin{aligned}
		&~\Omega^\mathrm{cell}~~\text{belongs to the H-type block},\\
		&~\Omega^\mathrm{cell}\cap\Omega_\mathrm{H}^\mathrm{val}\neq\varnothing,\\
		&~\exists\bm{r}=\bm{r}^\mathrm{tar}~~\text{such that}~~|\bm{r}^\mathrm{cell}-\bm{r}|\le3\Delta x
	\end{aligned}
	\right.
	\right\},
\end{equation} 
where $\Delta x$ is the mesh size of the background grid.
Besides, $\varnothing$ is the empty set and $\exists$ means there exists \citep{Munkres2000Topology}. 
Similarly, mapping cells from the O-type block $\Omega_\mathrm{O}^\mathrm{map}$ can be identified by 
\begin{equation}
	\Omega_\mathrm{O}^\mathrm{map}=\left\{\Omega^\mathrm{cell}\left|
	\begin{aligned}
		&~\Omega^\mathrm{cell}~~\text{belongs to the O-type block},\\
		&~\Omega^\mathrm{cell}\cap\Omega_\mathrm{O}^\mathrm{val}\neq\varnothing,\\
		&~\exists\bm{r}=\bm{r}^\mathrm{tar}~~\text{such that}~~|\bm{r}^\mathrm{cell}-\bm{r}|\le3\Delta x
	\end{aligned}
	\right.
	\right\}\!.
\end{equation}  
Correspondingly, the centroid of each mapping cell in $\Omega_\mathrm{H}^\mathrm{map}$ and $\Omega_\mathrm{O}^\mathrm{map}$ are denoted as $\bm{r}_\mathrm{H}^\mathrm{map}$ and $\bm{r}_\mathrm{O}^\mathrm{map}$.
Notably, the distance threshold is artificially selected as $3\Delta x$ in the present study, which is sufficiently large to cover almost all particles that are  potentially passing through interfaces $C_\mathrm{O,o}$ or $C_\mathrm{O,o}^\mathrm{trans}$, as a result of the constraint of the Courant-Friedrichs-Lewy (CFL) condition \citep{Anderson1995Computational}.

Furthermore, the mapping $\mathcal{M}$ between blocks is mathematically described by
\begin{equation}
	\mathcal{M}:(p,\{l,m\})\rightarrow(p^\prime,\{l^\prime,m^\prime\}).
\end{equation}
Here, $p$, $l$ and $m$ respectively represents the current processor id and indices of the local grid cell in the first two directions of $\Omega_\mathrm{H}^\mathrm{map}$ (or $\Omega_\mathrm{O}^\mathrm{map}$), while $p^\prime$, $l^\prime$ and $m^\prime$ denote the corresponding mapped variables from $\Omega_\mathrm{O}^\mathrm{map}$ (or $\Omega_\mathrm{H}^\mathrm{map}$).
For a specific $(p,\{l,m\})$, $(p^\prime,\{l^\prime,m^\prime\})$ is obtained by searching the cell with the shortest distance between $\bm{r}_\mathrm{H}^\mathrm{map}$ and $\bm{r}_\mathrm{O}^\mathrm{map}$.
Therefore, if $(p,\{l,m\})\in\Omega_\mathrm{H}^\mathrm{map}$, $(p^\prime,\{l^\prime,m^\prime\})$ is determined by:
for all $(q,\{r,s\})\in\Omega_\mathrm{O}^\mathrm{map}$, there exists $(p^\prime,\{l^\prime,m^\prime\})\in\Omega_\mathrm{O}^\mathrm{map}$, such that 
\begin{equation}
	\left|\bm{r}_{\mathrm{H},\{l,m,1\}}^\mathrm{map}-\bm{r}_{\mathrm{O},\{l^\prime,m^\prime,1\}}^\mathrm{map}\right|\le\left|\bm{r}_{\mathrm{H},\{l,m,1\}}^\mathrm{map}-\bm{r}_{\mathrm{O},\{r,s,1\}}^\mathrm{map}\right|.
\end{equation}
Similarly, if $(p,\{l,m\})\in\Omega_\mathrm{O}^\mathrm{map}$, $(p^\prime,\{l^\prime,m^\prime\})$ is determined by:  for all $(q,\{r,s\})\in\Omega_\mathrm{H}^\mathrm{map}$, there exists $(p^\prime,\{l^\prime,m^\prime\})\in\Omega_\mathrm{H}^\mathrm{map}$, such that
\begin{equation}
	\left|\bm{r}_{\mathrm{O},\{l,m,1\}}^\mathrm{map}-\bm{r}_{\mathrm{H},\{l^\prime,m^\prime,1\}}^\mathrm{map}\right|\le\left|\bm{r}_{\mathrm{O},\{l,m,1\}}^\mathrm{map}-\bm{r}_{\mathrm{H},\{r,s,1\}}^\mathrm{map}\right|.
\end{equation}

Based on the mapping $\mathcal{M}:(p,\{l,m\})\rightarrow(p^\prime,\{l^\prime,m^\prime\})$, when the particle which originally locates at the cell with indices $\{l,m\}$ in the processor $p$ penetrates the interface $C_\mathrm{O,o}$ or $C_\mathrm{O,o}^\mathrm{trans}$, it will be redistributed to the cell with indices $\{l^\prime,m^\prime\}$ in the processor $p^\prime$.
Nevertheless, it is necessary to verify whether the particle lies within the current grid cell, since the present strategy only provides the preference of particle redistribution, rather than a correct result.
Moreover, if the mapping is invalid, a search-locate algorithm is conducted to reseek the updated processor and cell.  
 
\subsubsection{Search-locate algorithm}

\begin{figure}[!t]
	\begin{center}
		\begin{tikzpicture}[scale=1.1]
			
			\path[fill=gray,opacity=0.15] (-2,20) -- (3,20) -- (3,25) -- (-2,25) -- (-2,20);
			
			\foreach \x in {0,1,...,4,5}
			{
				\draw [black,thick,opacity=0.5] (\x-2,20) -- (\x-2,25);
			}
			\foreach \y in {0,1,...,4,5}
			{
				\draw [black,thick,opacity=0.5] (-2,\y+20) -- (3,\y+20);
			}
			
			\foreach \x in {0,1,...,4}
			{
				\draw [fill=black] (\x-2,20) circle (0.05);
				\draw [fill=black] (\x-2,21) circle (0.05);
				\draw [fill=black] (\x-2,22) circle (0.05);
				\draw [fill=black] (\x-2,23) circle (0.05);
				\draw [fill=black] (\x-2,24) circle (0.05);
			}
			
			\node at (1.5-2,21.5) {$\{l_0,m_0\}$};	
			\node at (-2,19.7) {$l_0\!-\!2$};	
			\node at (-1,19.7) {$l_0\!-\!1$};	
			\node at (0,19.7) {$l_0$};	
			\node at (1,19.7) {$l_0\!+\!1$};	
			\node at (2,19.7) {$l_0\!+\!2$};
			\node at (3,19.7) {$l_0\!+\!3$};	
			\node at (-2.5,20) {$m_0\!-\!2$};
			\node at (-2.5,21) {$m_0\!-\!1$};
			\node at (-2.5,22) {$m_0$};
			\node at (-2.5,23) {$m_0\!+\!1$};
			\node at (-2.5,24) {$m_0\!+\!2$};
			\node at (-2.5,25) {$m_0\!+\!3$};
			
			\draw[very thick,->] (0,22) -- (0.75,22) node [below left] {$\bm{e}_x$};
			\draw[very thick,->] (0,22) -- (0,22.75) node [below left] {$\bm{e}_y$};
			
			\draw [blue,very thick,->] (0.25,22.25) -- (1.5,22.75);
			\node at (1.7,22.95) {$\bm{u}_\mathrm{p}$};	
			\draw [green,very thick,->] (0.25,22.25) -- (0.25,22.75);
			\node at (0.4,22.95) {$u_{\mathrm{p},y}$};	
			\draw [green,very thick,->] (0.25,22.25) -- (1.5,22.25);
			\node at (1.9,22.25) {$u_{\mathrm{p},x}$};
			
			\draw [fill=red, thick] (0.25,22.25) circle (0.1);

			\path[fill=gray,opacity=0.15] ({14*cos(1.3833 r)},{14*sin(1.3833 r)}) -- ({19*cos(1.3833 r)},{19*sin(1.3833 r)}) -- ({19*cos(1.4458 r)},{19*sin(1.4458 r)}) -- ({19*cos(1.5083 r)},{19*sin(1.5083 r)}) -- ({19*cos(1.5708 r)},{19*sin(1.5708 r)}) -- ({19*cos(1.6333 r)},{19*sin(1.6333 r)}) -- ({19*cos(1.6958 r)},{19*sin(1.6958 r)}) -- ({14*cos(1.6958 r)},{14*sin(1.6958 r)}) -- ({14*cos(1.6333 r)},{14*sin(1.6333 r)}) -- ({14*cos(1.5708 r)},{14*sin(1.5708 r)}) -- ({14*cos(1.5083 r)},{14*sin(1.5083 r)}) -- ({14*cos(1.4458 r)},{14*sin(1.4458 r)});
			
			\foreach \r in {0,1,...,4,5}
			{
				\draw[black,thick,opacity=0.5] plot[domain=1.3833:1.6958,smooth,variable=\t] ({(\r+14)*cos(\t r)},{(\r+14)*sin(\t r)});
			}
			\foreach \x in {1.3833,1.4458,1.5083,...,1.6958}
			{
				\draw [black,thick,opacity=0.5] ({14*cos(\x r)},{14*sin(\x r)}) -- ({19*cos(\x r)},{19*sin(\x r)});
			}
			
			\foreach \x in {1.4458,1.5083,...,1.6958}
			{
				\draw [fill=black] ({14*cos(\x r)},{14*sin(\x r)}) circle (0.05);
				\draw [fill=black] ({15*cos(\x r)},{15*sin(\x r)}) circle (0.05);
				\draw [fill=black] ({16*cos(\x r)},{16*sin(\x r)}) circle (0.05);
				\draw [fill=black] ({17*cos(\x r)},{17*sin(\x r)}) circle (0.05);
				\draw [fill=black] ({18*cos(\x r)},{18*sin(\x r)}) circle (0.05);
			}
			
			\node at ({16*cos(1.5708 r)-0.5},{16*sin(1.5708 r)-0.5}) {$\{l_0,m_0\}$};	
			\node at ({14*cos(1.6958 r)},{14*sin(1.6958 r)-0.3}) {$l_0\!-\!2$};	
			\node at ({14*cos(1.6333 r)},{14*sin(1.6333 r)-0.3}) {$l_0\!-\!1$};	
			\node at ({14*cos(1.5708 r)},{14*sin(1.5708 r)-0.3}) {$l_0$};	
			\node at ({14*cos(1.5083 r)},{14*sin(1.5083 r)-0.3}) {$l_0\!+\!1$};	
			\node at ({14*cos(1.4458 r)},{14*sin(1.4458 r)-0.3}) {$l_0\!+\!2$};	
			\node at ({14*cos(1.3833 r)},{14*sin(1.3833 r)-0.3}) {$l_0\!+\!3$};	
			\node at ({14*cos(1.6958 r)-0.5},{14*sin(1.6958 r)}) {$m_0\!-\!2$};	
			\node at ({15*cos(1.6958 r)-0.5},{15*sin(1.6958 r)}) {$m_0\!-\!1$};	
			\node at ({16*cos(1.6958 r)-0.5},{16*sin(1.6958 r)}) {$m_0$};	
			\node at ({17*cos(1.6958 r)-0.5},{17*sin(1.6958 r)}) {$m_0\!+\!1$};	
			\node at ({18*cos(1.6958 r)-0.5},{18*sin(1.6958 r)}) {$m_0\!+\!2$};	
			\node at ({19*cos(1.6958 r)-0.5},{19*sin(1.6958 r)}) {$m_0\!+\!3$};	
			
			\draw[very thick,->] ({16*cos(1.5708 r)},{16*sin(1.5708 r)}) -- ({16.75*cos(1.5708 r)},{16.75*sin(1.5708 r)}) node [below left] {$\bm{e}_n$};
			\draw[very thick,->] ({16*cos(1.5708 r)},{16*sin(1.5708 r)}) -- ({16*cos(1.5239 r)},{16*sin(1.5239 r)}) node [below left] {$\bm{e}_t$};
			
			\draw [blue,very thick,->] ({16*cos(1.5708 r)+0.25},{16*sin(1.5708 r)+0.25}) -- ({16*cos(1.5708 r)+0.75},{16*sin(1.5708 r)+1.5});
			\node at ({16*cos(1.5708 r)+0.95},{16*sin(1.5708 r)+1.7}) {$\bm{u}_\mathrm{p}$};	
			\draw [green,very thick,->] ({16*cos(1.5708 r)+0.25},{16*sin(1.5708 r)+0.25}) -- ({16*cos(1.5708 r)+0.25},{16*sin(1.5708 r)+1.5});
			\node at ({16*cos(1.5708 r)+0.25},{16*sin(1.5708 r)+1.7}) {$u_{\mathrm{p},n}$};
			\draw [green,very thick,->] ({16*cos(1.5708 r)+0.25},{16*sin(1.5708 r)+0.25}) -- ({16*cos(1.5708 r)+0.75},{16*sin(1.5708 r)+0.25});
			\node at ({16*cos(1.5708 r)+1.15},{16*sin(1.5708 r)+0.25}) {$u_{\mathrm{p},t}$};
			
			\draw [fill=red, thick] ({16*cos(1.5708 r)+0.25},{16*sin(1.5708 r)+0.25}) circle (0.1);
			
			\node at (-3.5,25) {$(a)$};
			\node at (-3.5,19) {$(b)$};
			
			\node at (4.5,19) {~};
		\end{tikzpicture}
	\end{center}
	\caption{Schematic of the search-locate algorithm for the $(a)$ rectangular H-type block and $(b)$ curved O-type block, where the particle is denoted by the red circle and the search region is colored by gray.}
	\label{Search-locate}
\end{figure}
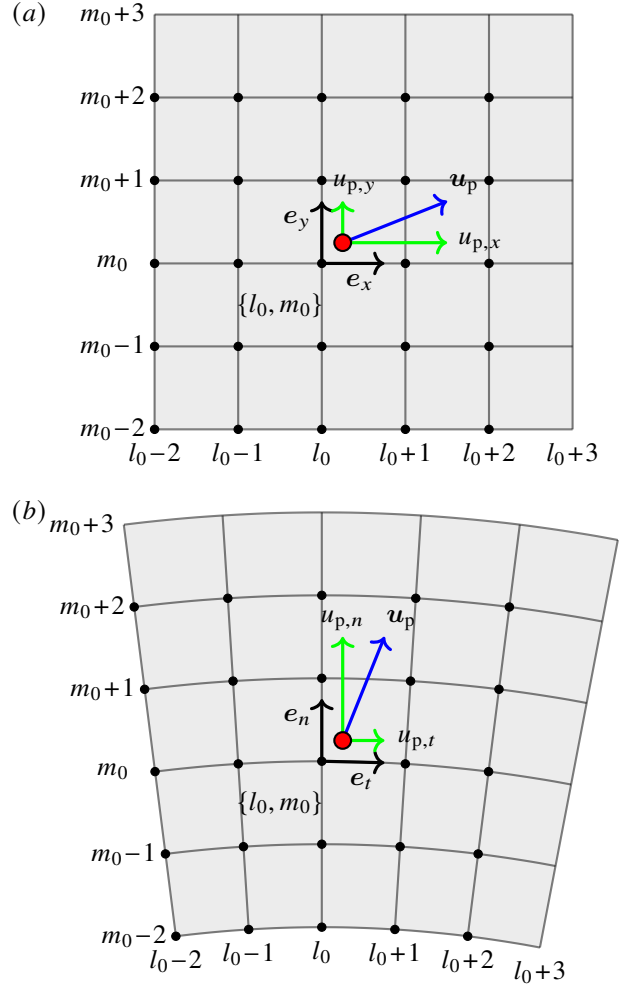

During the redistribution process, in order to identify which cell the particle is located in, one robust way is to perform a global search across the entire computational domain.
Undoubtedly, the presence of nested loops in this algorithm inevitably leads to significant computational inefficiency.
As a result, a search-locate algorithm is implemented to update the processor id $p$ and indices $\{l_0, m_0, n_0\}$, wherein the search is based on the particle velocity $\bm{u}_\mathrm{p}$ rather than the default array element ordering.
Focus on the general idea how the algorithm is conducted, it is assumed that the particle has been redistributed to the correct processor.
Besides, for the convenience of discussion, we will restrict ourselves to two dimensions.
Consequently, only indices $\{l_0, m_0\}$ are taken into consideration, and the schematic is presented in the figure \ref{Search-locate}.

As shown in figure \ref{Search-locate}$(a)$, if the particle is positioned in the H-type block, the velocity $\bm{u}_\mathrm{p}$ is decomposed into
\begin{equation}
\bm{u}_\mathrm{p}=u_{\mathrm{p},x}\bm{e}_x+u_{\mathrm{p},y}\bm{e}_y,
\end{equation}
where $\bm{e}_x$ and $\bm{e}_y$ are unit vector in the $x$ and $y$ direction, respectively.
Owing to the rectangular geometry of the computational domain, the indices $l_0$ and $m_0$ are updated separately, while with the same approach.
Numerically speaking, the index is searched from the current cell by evaluating the condition statement in sequence, terminating until a correct index is obtained.
Moreover, the searching direction is determined by the velocity, which is summarized as
\begin{equation}
	\begin{aligned}
		u_{\mathrm{p},x}\ge0:~&\{l_0,l_0+1,\cdots\}\\
		u_{\mathrm{p},x}<0:~&\{l_0,l_0-1,\cdots\}\\
		u_{\mathrm{p},y}\ge0:~&\{m_0,m_0+1,\cdots\}\\
		u_{\mathrm{p},y}<0:~&\{m_0,m_0-1,\cdots\}
	\end{aligned}
\end{equation}

Differently, due to the curved geometry of the O-type block, the primary search area is restricted to the domain consisting of $5\times5$ cells, as presented by the gray region in figure \ref{Search-locate}$(b)$.
Besides, the decomposition is applied in the local orthogonal coordinate system $(\bm{e}_t,\bm{e}_n)$,
\begin{equation}	\bm{u}_\mathrm{p}=u_{\mathrm{p},t}\bm{e}_t+u_{\mathrm{p},n}\bm{e}_n,
\end{equation}
rather than the Cartesian coordinate system $(\bm{e}_x,\bm{e}_y)$.
Accordingly, this domain is partitioned into four parts consisting of $3\times3$ cells with proper overlap, and the target area is further localized to a specific region based on the sign of $\bm{e}_t$ and $\bm{e}_n$.
Starting from the current grid cell, the search scope is expanded from inner to outer layers, and the searching order for indices is summarized as
\begin{equation}
	\begin{aligned}
		u_{\mathrm{p},t}\ge0,u_{\mathrm{p},n}\ge0:~&\{l_0,l_0+1,l_0+2\}\times\{m_0,m_0+1,m_0+2\}\\
		u_{\mathrm{p},t}\ge0,u_{\mathrm{p},n}<0:~&\{l_0,l_0+1,l_0+2\}\times\{m_0+1,m_0,m_0-1\}\\
		u_{\mathrm{p},t}<0,u_{\mathrm{p},n}\ge0:~&\{l_0+1,l_0,l_0-1\}\times\{m_0,m_0+1,m_0+2\}\\
		u_{\mathrm{p},t}<0,u_{\mathrm{p},n}<0:~&\{l_0+1,l_0,l_0-1\}\times\{m_0+1,m_0,m_0-1\}
	\end{aligned}
\end{equation}
Additionally, in order to ensure the robustness of the present framework, if the correct value cannot be obtained with the search-locate algorithm, then a global search method is applied.

\vspace{1em}
It is noted that the two strategies presented in \S~\ref{sec:opti-strategy} are specifically introduced to speed up the current particle framework, especially to reduce the computational cost for particles redistribution through the interfaces between the overset multi-grids.
The acceleration effects will thus be tested in \S~\ref{sec:acceleration}, in which the codes with and without these optimization strategies are denoted as optimized and unoptimized, respectively.
%These two optimization strategies are specially designed for the current particle framework to speed up the code, thus it is necessary to distinguish the code revision for the subsequent discussion on numerical acceleration effects in \S~\ref{sec:acceleration}.
%Moreover, optimized and unoptimized refer to the code with and without the proposed two algorithms, respectively.

\subsection{Particle feedback}

The feedback effect of the dispersed phase only needs to be considered for two-way coupled particle-laden flows.
In such a case, source terms $\bm{S}_m$ and $S_e$ are computed at each time step.
They summarize the influence of all particles by using the particle-in-cell method \citep{Crowe1977Particle}, and the local flow domain affected by each particle consists of the corresponding stencil-points for interpolation, which has been discussed in the subsection \ref{subsec:coupling_strategy}.
Subsequently, source terms $\bm{S}_m$ and $S_e$ are added to the right-hand side of the original Navier-Stokes equations. %and participate the calculation of Runge-Kutta time integration scheme.}

\section{Numerical Validation}\label{sec:Validation}

In this section, many simulations are performed to validate the function of the present framework for particle-laden flow.
The first test focus on verifying the function of Lagrangian tracking of massless particles within multi-block overset grids, as well as the interpolation scheme.
Besides, three one-way coupled two-phase flow cases are conducted to validate the time integration scheme, the multi-block framework and the ability to perform a direct numerical simulation.
Moreover, a two-way coupled particle-laden turbulent channel flow are conducted to verify the implemented coupling strategy.

\subsection{Lagrangian tracking of massless particles}\label{subsec:Lagrangian_tracking}

At first, the Lagrangian tracking function is tested, and a two-dimensional incompressible Taylor-Green vortex \citep{Taylor1937Mechanism} is selected as the carrier flow.

\begin{figure}
	\centering
	\begin{overpic}[width=0.425\textwidth]{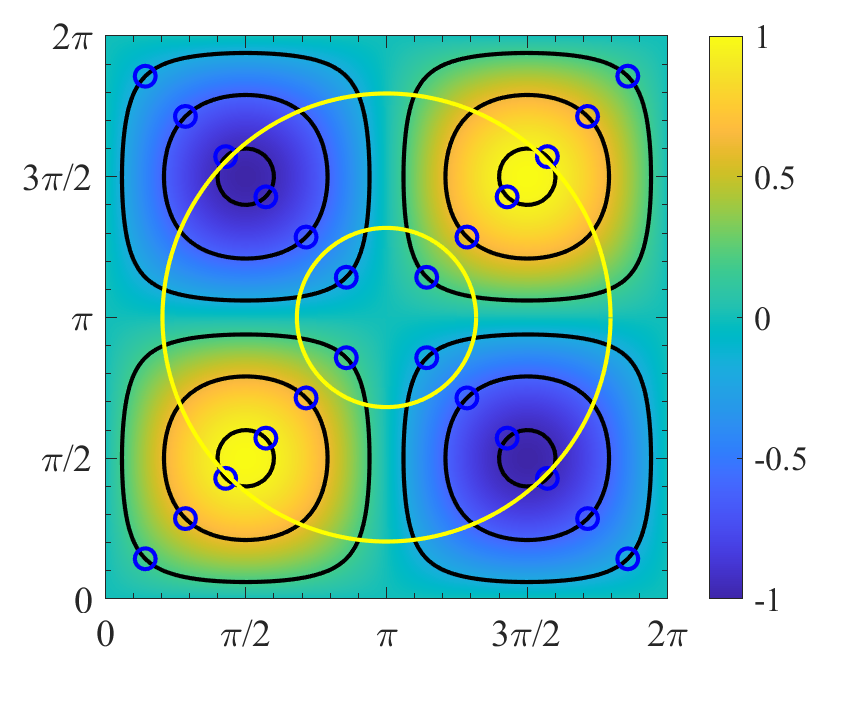}
		{
			\put(2.5,44){$y$}
			\put(44.25,2){$x$}
			\put(95,44){$\varphi$}
		}
	\end{overpic}
	\vspace{-0.2cm}
	\caption{Schematic of the numerical configuration for two-dimensional Taylor-Green vortex: the computational grid consists of a background H-type grid and an O-type grid whose boundaries are shown by yellow lines. Besides, the initial locations of 24 massless particles are indicated by blue circles. Moreover, the diagram is colored by the function $\varphi=\sin(x)\sin(y)$, whose contours are shown by the black lines.}
	\label{TG_grid}
\end{figure}

In order to validate the feasibility of aforementioned strategies based on multi-block overset grids, two blocks are used here, which are shown in the figure \ref{TG_grid}.
One is the background H-type grid, and the other is an O-type grid denoted by yellow lines.
The background grid determines the square computational domain $(x,y)=[0,2\pi L_0]\times[0,2\pi L_0]$, where $L_0=1$ and periodic conditions are applied in each direction.
Besides, both the inner and outer boundaries of the O-type block are set as penetrable, allowing particles to pass through unhindered.

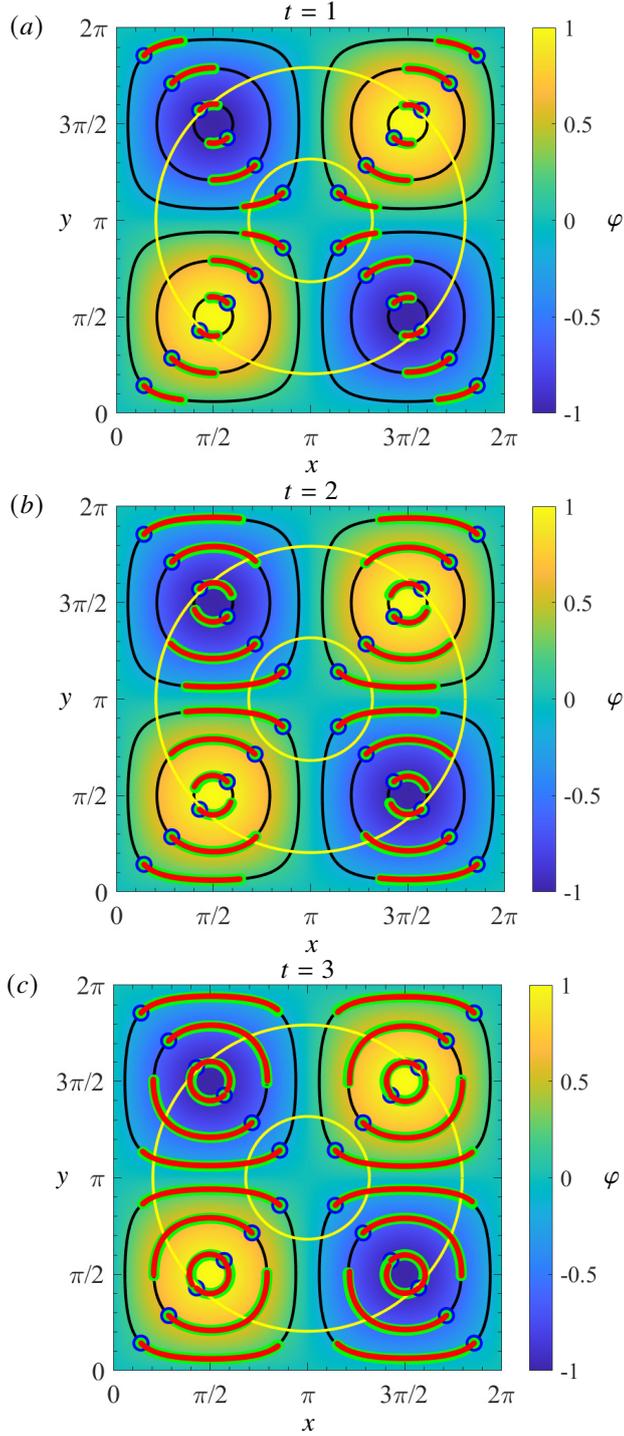
\begin{figure}[!t]
	\centering
	\begin{overpic}[width=0.425\textwidth]{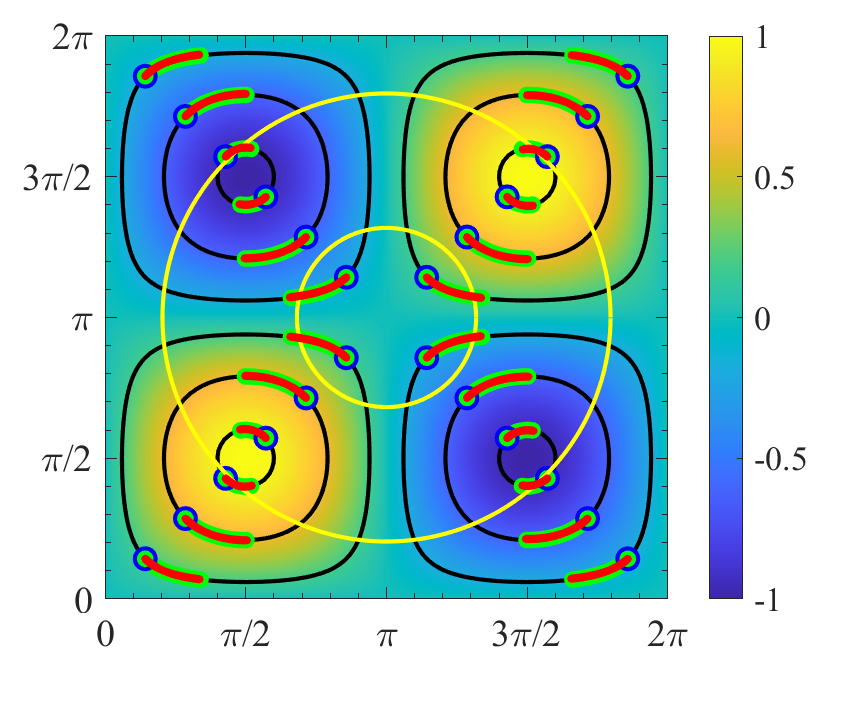}
		{
			\put(-6,76.5){$(a)$}
			\put(2.5,44){$y$}
			\put(44.25,2){$x$}
			\put(40.5,79){$t=1$}
			\put(95,44){$\varphi$}
		}
	\end{overpic}
	\begin{overpic}[width=0.425\textwidth]{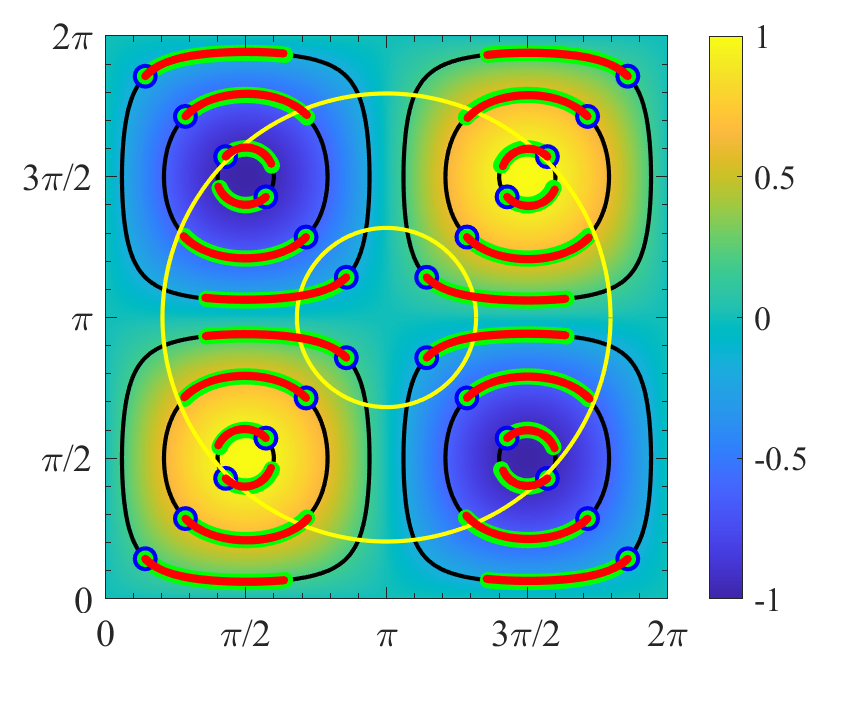}
		{
			\put(-6,76.5){$(b)$}
			\put(2.5,44){$y$}
			\put(44.25,2){$x$}
			\put(40.5,79){$t=2$}
			\put(95,44){$\varphi$}
		}
	\end{overpic}
	\begin{overpic}[width=0.425\textwidth]{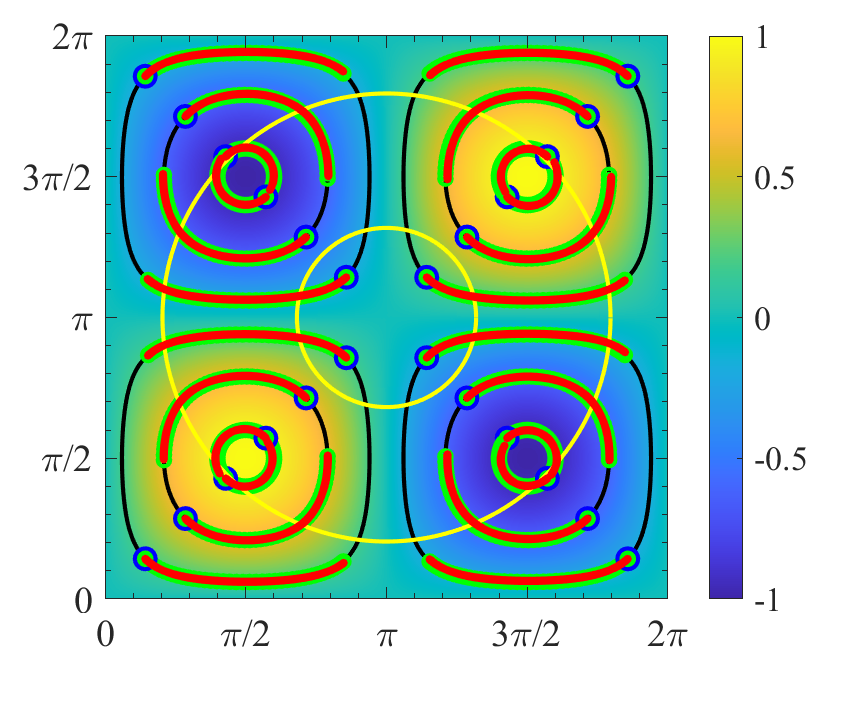}
		{
			\put(-6,76.5){$(c)$}
			\put(2.5,44){$y$}
			\put(44.25,2){$x$}
			\put(40.5,79){$t=3$}
			\put(95,44){$\varphi$}
		}
	\end{overpic}
	\vspace{-0.2cm}
	\caption{Pathlines of 24 tracers obtained by different methods at a timestep $(a)$ $t=1$, $(b)$ $t=2$ and $(c)$ $t=3$. Here, the O-type grid is shown by yellow lines, and the initial locations of 24 tracers are indicated by blue circles. Besides, red circles denote the results obtained by HiPSTAR while green circles denote the results by solving the equations \eqref{ODEs} with Matlab \citep{MATLAB}.	Moreover, the diagram is colored by the function $\varphi=\sin(x)\sin(y)$, whose contours are shown by the black lines.}
	\label{TG}
\end{figure}

Because the governing equations for the continuous phase are compressible, flow parameters adopted here are in accordance with the numerical setup reported in \cite{Deuse2020Implementation, Ge2020Development} to approximate an incompressible flow, where the Reynolds and mach number are respectively set as $\Rey_\mathrm{f}=1600$ and $\Mach_\mathrm{f}=0.1$.
Additionally, the initial conditions are prescribed as follows:
\begin{align}
	u_\mathrm{f}(x,y,t=0)&=U_0\sin\left(x/L_0\right)\cos\left(y/L_0\right),\notag\\
	v_\mathrm{f}(x,y,t=0)&=-U_0\cos\left(x/L_0\right)\sin\left(y/L_0\right),\notag\\
	\rho_\mathrm{f}(x,y,t=0)&=\rho_0,\\
	T_\mathrm{f}(x,y,t=0)&=T_0,\notag\\
	p_\mathrm{f}(x,y,t=0)&=\frac{\rho_0T_0}{\gamma\Mach^2_\mathrm{f}}+\frac{\rho_0U_0^2}{4}\left[\cos\left(2x/L_0\right)+\cos\left(2y/L_0\right)\right],\notag
\end{align}
%\begin{align}
%	u_\mathrm{f}(x,y,t=0)&=U_0\sin\left(\frac{x}{L_0}\right)\cos\left(\frac{y}{L_0}\right),\notag\\
%	v_\mathrm{f}(x,y,t=0)&=-U_0\cos\left(\frac{x}{L_0}\right)\sin\left(\frac{y}{L_0}\right),\notag\\
%	\rho_\mathrm{f}(x,y,t=0)&=\rho_0,\\
%	T_\mathrm{f}(x,y,t=0)&=T_0,\notag\\
%	p_\mathrm{f}(x,y,t=0)&=\frac{\rho_0T_0}{\gamma\Mach^2_\mathrm{f}}+\frac{\rho_0U_0^2}{4}\left[\cos\left(\frac{2x}{L_0}\right)+\cos\left(\frac{2y}{L_0}\right)\right],\notag
%\end{align}
where the parameters $U_0=\rho_0=T_0=1$ are treated as constant.
Manifested as counter-rotating vortices decaying temporally due to viscosity, the incompressible analytical solutions of the velocity components exist \citep{Abdelsamie2021Taylor}, which are given by
\begin{equation}
	\begin{aligned}
		u_\mathrm{f}(x,y,t)&=u_\mathrm{f}(x,y,0)\exp\left(-2t/(\Rey_\mathrm{f}L_0^2)\right),\\
		v_\mathrm{f}(x,y,t)&=v_\mathrm{f}(x,y,0)\exp\left(-2t/(\Rey_\mathrm{f}L_0^2)\right).
	\end{aligned}
\end{equation}

In the present simulation, the trajectories of 24 massless particles are tracked.
Indicated by blue circles in figure \ref{TG_grid}, their initial $x$-coordinates are set as $x_\mathrm{p}(t=0)=i\pi/7$, where $i$ equals to $1,\cdots,6,8,\cdots,13$, and the corresponding $y$-coordinates are $y_\mathrm{p}(t=0)=x_0$ or $2\pi-x_0$.
Under the approximation of incompressible flow, the trajectories can also be obtained by directly solving the ordinary differential equations:
\begin{equation}
	\label{ODEs}
	\left\{
	\begin{aligned}
		\frac{\mathrm{d}}{\mathrm{d}t}x_\mathrm{p}(t)&=u_\mathrm{f}(x,y,t),\\
		\frac{\mathrm{d}}{\mathrm{d}t}y_\mathrm{p}(t)&=v_\mathrm{f}(x,y,t),
	\end{aligned}
	\right.
\end{equation}
whose analytical solutions is expressed as
\begin{equation}
	\sin(x)\sin(y)=\sin(x_\mathrm{p}(t=0))\sin(y_\mathrm{p}(t=0)).
\end{equation}

Figure \ref{TG} displays the numerical results at different time $t$.
It can be seen that as time goes on, Lagrangian particles in both two blocks gradually move from their initial locations with the carrier flow.
During the process, some of them pass through the boundary of the O-type block smoothly.
Importantly, both the results calculated with the present solver and Matlab \citep{MATLAB} agree well with the analytical solutions, and no obvious differences are identified.
The comparison shown in figure \ref{TG} implies that the interpolation approach adopted is available not only in the Cartesian grid but also in the curved grid. 
Furthermore, it is suggested that the current framework has the potential to simulate a particle-laden flow with complex geometry using multi-block overset grids.

\subsection{One-way coupling}

\subsubsection{Taylor–Green vortices particle-laden flow}

\begin{figure}[!t]
	\centering
	\begin{overpic}[width=0.375\textwidth]{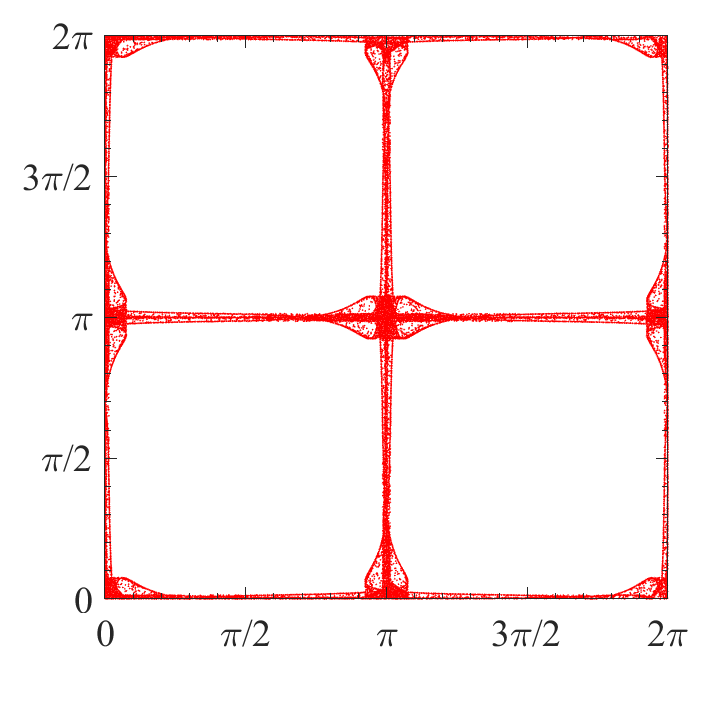}
		{
			\put(-3,93){$(a)$}
			\put(3,53.5){$y$}
			\put(54,3){$x$}
			\put(46,97){$\Stokes=5\Stokes_c$}
		}
	\end{overpic}
	\begin{overpic}[width=0.375\textwidth]{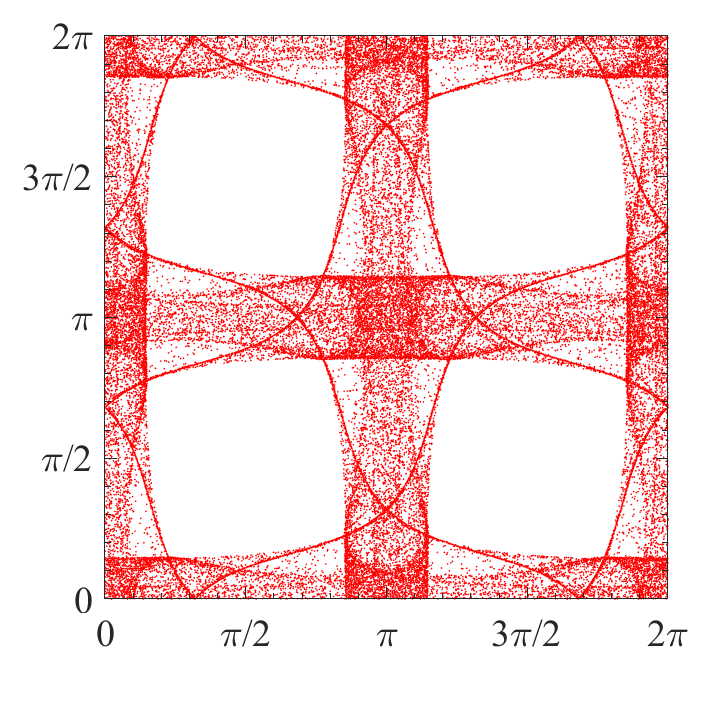}
		{
			\put(-3,93){$(b)$}
			\put(3,53.5){$y$}
			\put(54,3){$x$}
			\put(45,97){$\Stokes=10\Stokes_c$}
		}
	\end{overpic}
	\begin{overpic}[width=0.375\textwidth]{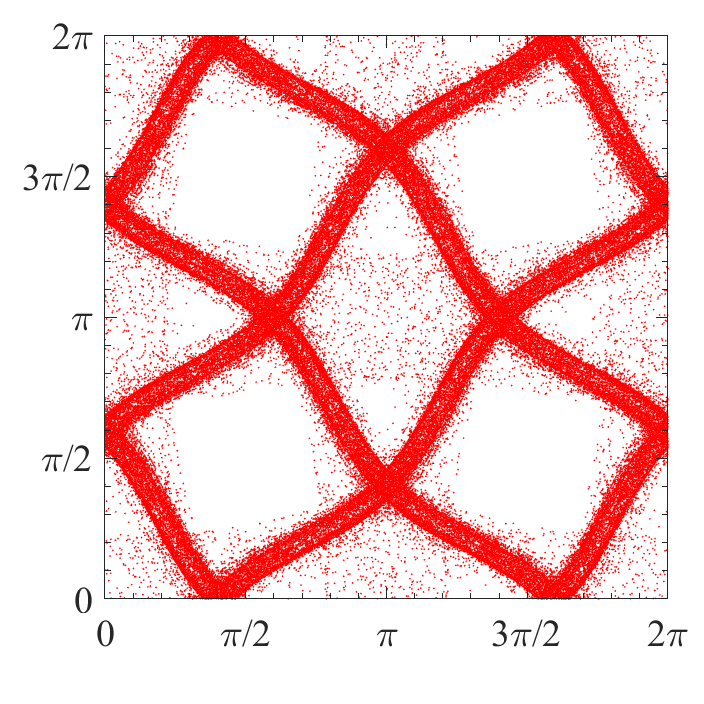}
		{
			\put(-3,93){$(c)$}
			\put(3,53.5){$y$}
			\put(54,3){$x$}
			\put(45,97){$\Stokes=20\Stokes_c$}
		}
	\end{overpic}
	\vspace{-0.2cm}
	\caption{Particle distributions in two-dimensional Taylor-Green vortex with particles at a timestep $t=72\tau_\mathrm{f}$ for three Stokes numbers: $(a)$ $\Stokes=5\Stokes_c$, $(a)$ $\Stokes=10\Stokes_c$ and $(a)$ $\Stokes=20\Stokes_c$. Here, particles are denoted by red point clouds.}
	\label{TG_St}
\end{figure}
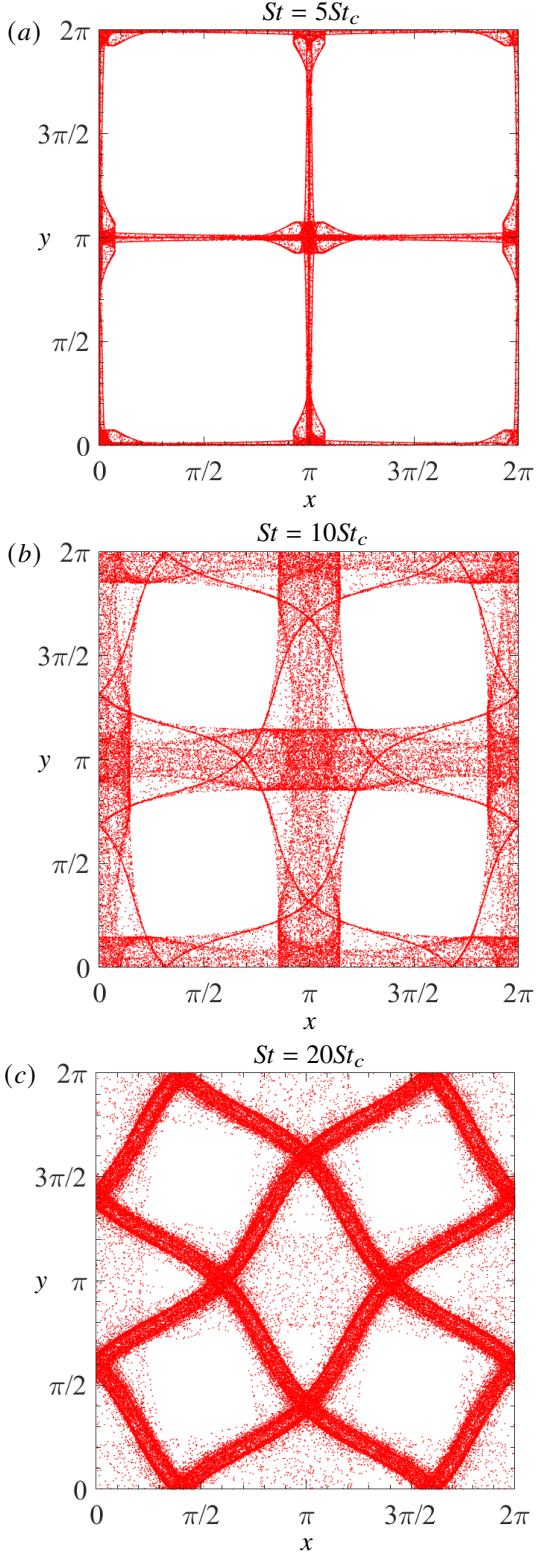

Except for the Lagrangian tracking function, the particle push part was further checked.
According to the previous studies \cite{Chaisemartin2009Mod, Ge2020Development}, forced by a two-dimensional Taylor-Green vortex, particles with different Stokes numbers will advance in time and finally form distinct complex patterns.
Therefore, on the basis of the subsection \ref{subsec:Lagrangian_tracking}, a Taylor–Green vortices particle-laden flow  was conducted to verify the implementation of particle push, where the O-type block was removed for simplicity.

For the carrier flow, the characteristic time scale is defined as $\tau_\mathrm{f}=L_0/U_0=1$.
Thus, for the dispersed phase, the response time is determined by $\tau_\mathrm{p}=\Stokes\tau_\mathrm{f}$.
Moreover, the ratio of the Stokes number $\Stokes$ and the critical Stokes number $\Stokes_c=1/4$ governs the distribution pattern of particles, which are uniformly seeded in the initial time.
In the present validation, only the Stokes drag \citep{Stokes1850effect} are considered, and $\kappa_\mathrm{p}$ equals to the Stokes drag coefficient $6\pi\mu_\mathrm{f}R_\mathrm{p}$.
Additionally, three cases are tested, which are distinguished by the Stokes number, $\Stokes=5\Stokes_c$, $\Stokes=10\Stokes_c$ and $\Stokes=20\Stokes_c$.
Besides, in each simulation, 160,000 particles are added.

Figure \ref{TG_St} presents the results at the time $t=72\tau_\mathrm{f}$, when the patterns are approximately steady.
For the case $\Stokes=5\Stokes_c$, it can be seen that particles predominantly concentrate in the marginal area of the vortices core, which is shown by the figure \ref{TG_St}$(a)$.
As the Stokes number increases to $\Stokes=10\Stokes_c$, particles which originally accumulate in the zero-vorticity region become dispersed, as indicated in the figure \ref{TG_St}$(b)$.
Additionally, crossing trajectories are generated owing to the fact that some particles are ejected from the vortex \citep{Chaisemartin2009Mod}.
Moreover, for the case $\Stokes=20\Stokes_c$, the ejection events become more violent.
As a result, a large number of particles aggregate and crossing trajectories become wider, as presented in figure \ref{TG_St}$(c)$.
The patterns displayed in figure \ref{TG_St} are qualitatively the same as the results published by Ge \textit{et al.} \citep{Ge2020Development}, which implies the reliability of the time stepping scheme used.

\subsubsection{Particle impaction on a cylinder in a crossflow}

After checking the function of the particle push in a one-block flow configuration, a more complex test case is considered: simulating the particle impaction on a cylinder in a crossflow.

\begin{figure}
	\centering
	\begin{overpic}[width=0.466667\textwidth]{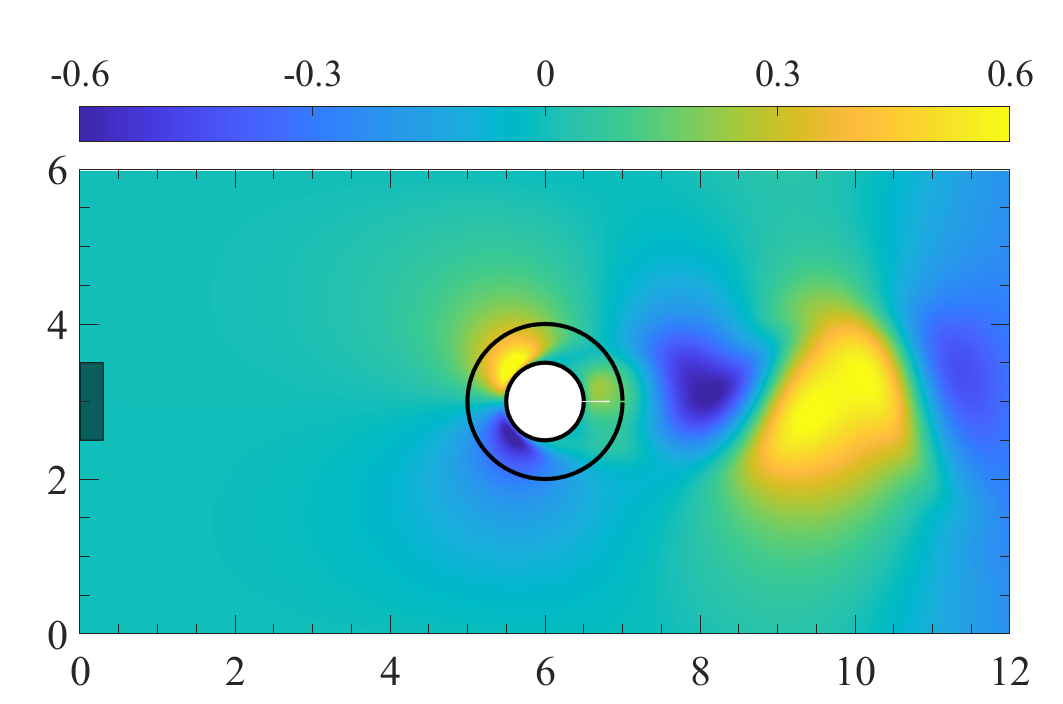}
		{
			\put(1,28){$y$}
			\put(50.5,-2){$x$}
			\put(50,61.5){$v_\mathrm{f}$}
		}
	\end{overpic}
	\caption{Schematic of the numerical configuration for particle impaction on a cylinder in a crossflow: the computational grid consists of a background H-type grid and an O-type grid whose boundaries are shown by black lines. Here, the gray box shown in the left part of the plot corresponds to the subdomain where the particles have been inserted with a random distribution. Moreover, the diagram is colored by the lateral velocity $v_\mathrm{f}$.}
	\label{Cylinder_flow}
\end{figure}

Figure \ref{Cylinder_flow} shows the numerical configuration based on multi-block overset grids.
The background grid determines the computational domain $(x,y)=[0,12D]\times[0,6D]$, which is the same as the setup of \citep{Haugen2010Particle}.
Here, $D=1$ is the cylinder diameter.
The left side of the domain $x=0$ corresponds to the inlet of the flow, where a fixed Dirichlet condition with a uniform velocity $(u_\mathrm{f},v_\mathrm{f})=(1,0)$ is applied, and the right side $x=12D$ corresponds to the outlet.
Besides, in the lateral direction $y$, a periodic boundary condition is applied. 
In order to describe the curved geometry of the cylinder, the usage of an O-typed grid is necessary, whose boundaries are shown by black lines in figure \ref{Cylinder_flow}.
The outer boundary is the penetrable interface, which allows particle to pass through freely.
Differently, the inner boundary corresponds to the surface of the cylinder, which is modeled as an isothermal impenetrable wall.
Moreover, it is assumed that all particles collide with the cylinder will deposit, which is numerically performed by removing them from the simulation.

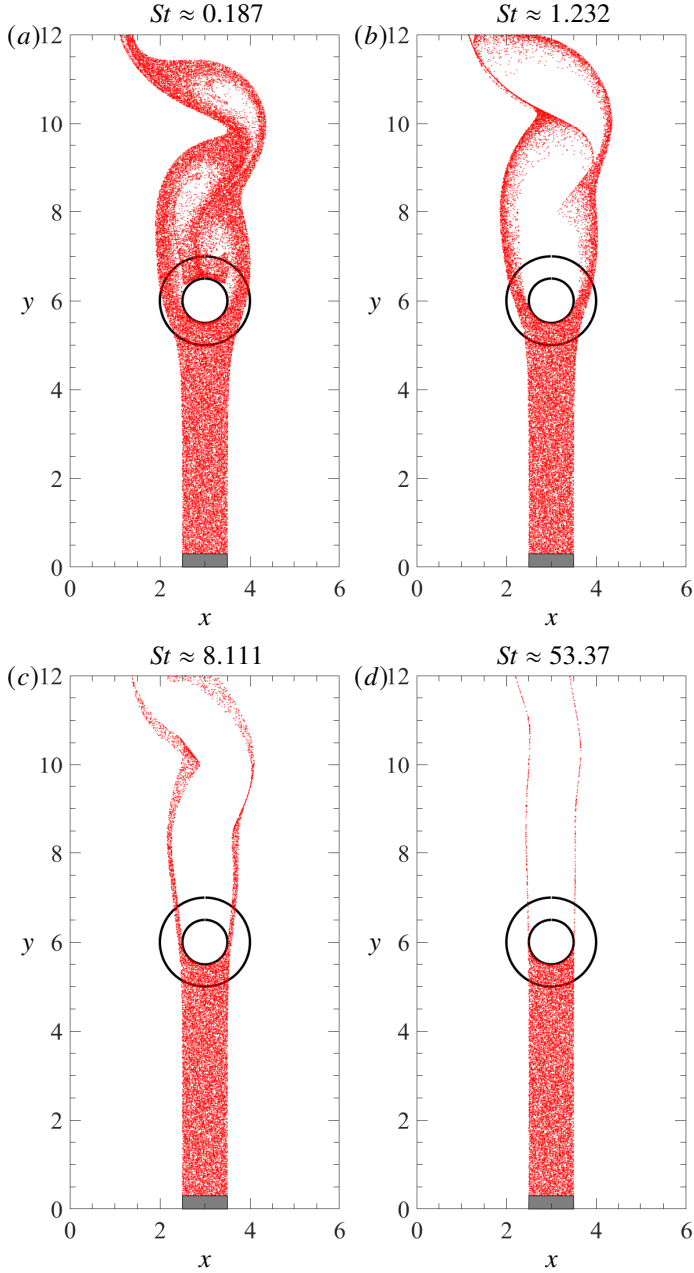
\begin{figure}[!t]
	\centering
	\begin{overpic}[width=0.22\textwidth]{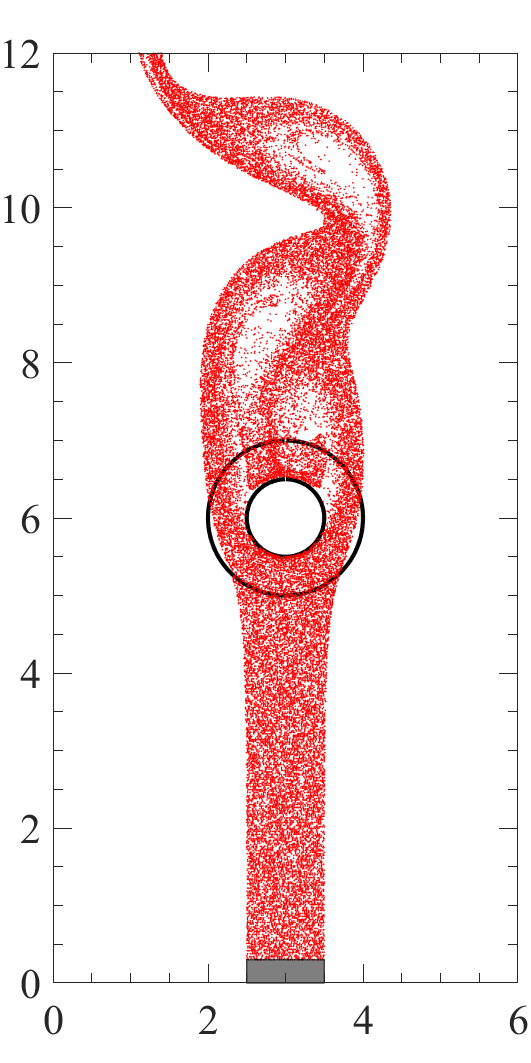}
		{
			\put(-5.5,93.5){$(a)$}
			\put(-3,50){$y$}
			\put(26,-2){$x$}
			\put(18,97){$\Stokes\approx0.187$}
		}
	\end{overpic}
	\vspace{0.4cm}
	\quad
	\begin{overpic}[width=0.22\textwidth]{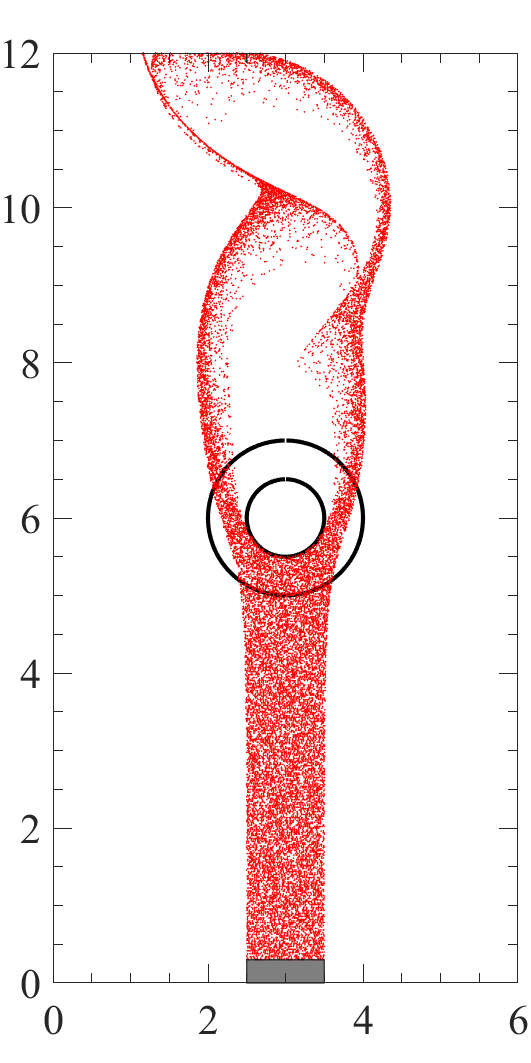}
		{
			\put(-5.5,93.5){$(b)$}
			\put(-3,50){$y$}
			\put(26,-2){$x$}
			\put(18,97){$\Stokes\approx1.232$}
		}
	\end{overpic}
	\vspace{0.4cm}
	\begin{overpic}[width=0.22\textwidth]{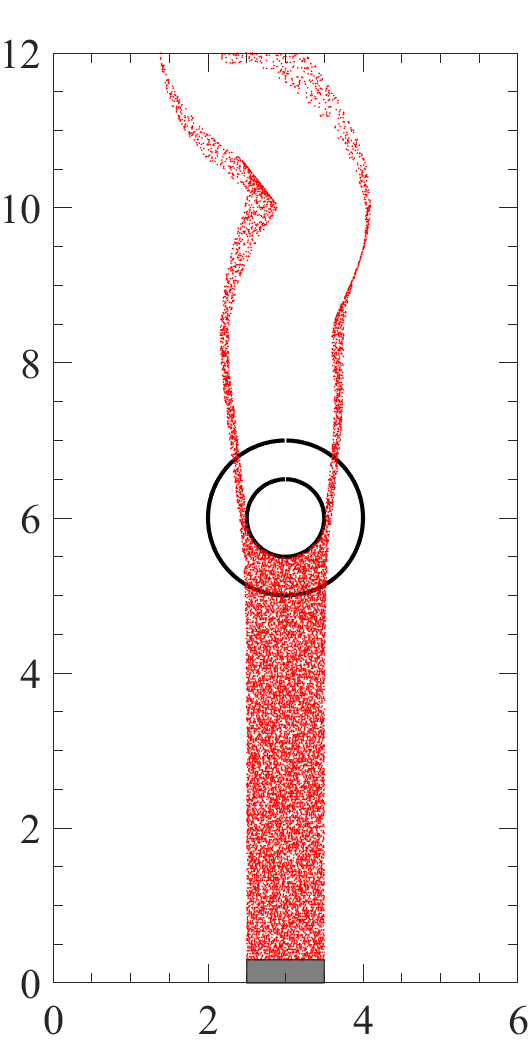}
		{
			\put(-5.5,93.5){$(c)$}
			\put(-3,50){$y$}
			\put(26,-2){$x$}
			\put(18,97){$\Stokes\approx8.111$}
		}
	\end{overpic}
	\quad
	\begin{overpic}[width=0.22\textwidth]{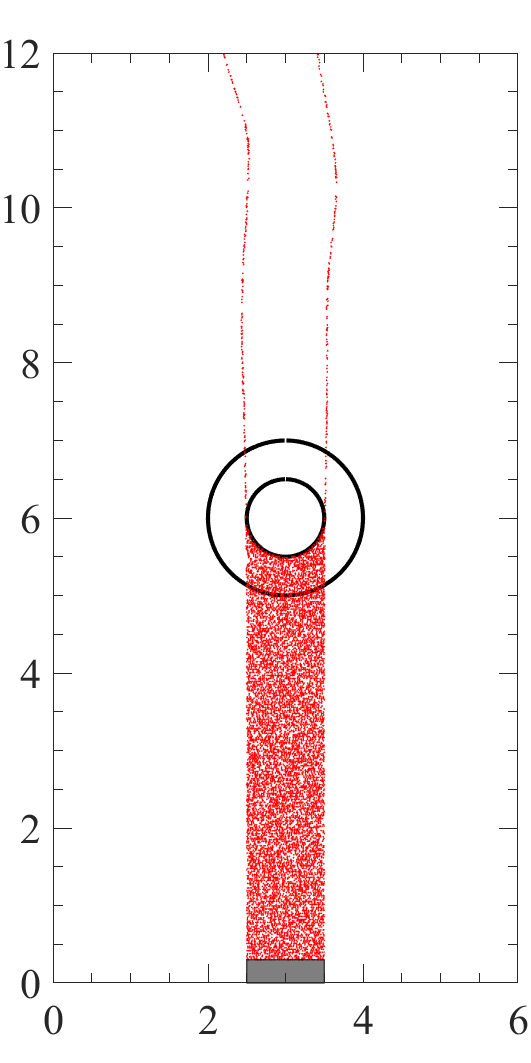}
		{
			\put(-5.5,93.5){$(d)$}
			\put(-3,50){$y$}
			\put(26,-2){$x$}
			\put(18,97){$\Stokes\approx53.37$}
		}
	\end{overpic}
	\vspace{-0.4cm}
	\caption{Time snapshots for different Stokes number: $(a)$ $\Stokes\approx0.187$, $(b)$ $\Stokes\approx1.232$, $(c)$ $\Stokes\approx8.111$ and $(d)$ $\Stokes\approx53.37$. Here, particles are denoted by red point clouds, and the gray box corresponds to the subdomain where the particles have been inserted with a random distribution.}
	\label{Cylinder_St}
\end{figure}

Performed at the Reynolds number of $\Rey_\mathrm{f}=100$, the present simulation consists of two periods.
In the first period, particles are continuously generated with a random distribution in $y\in[2.5D,3.5D]$ and a constant velocity $(u_\mathrm{p},v_\mathrm{p})=(1,0)$, then they are uniformly released into the flow field from $x=0.3D$.
The injection process has been lasted for 36 non-dimensional time units, and $N_\mathrm{init}=90,000$ particles were added in total.
Subsequently, in the second period, no particles are inserted into the flow field, while the simulation continues to run for about 108 non-dimensional time units in order for nearly all particles to leave the domain.
During the whole process, only the drag force is considered, and the particle motion is governed by equation \eqref{drag_equation} \citep{Haugen2010Particle}.

Figure \ref{Cylinder_St} displays the time snapshots for different Stokes numbers in the first period.
It can be observed that the added particles are transported by the carrier flow, part of them impact the cylinder and deposit while others can bypass the cylinder as they move downstream.
For the case with the Stokes number of $\Stokes\approx0.187$, only a small fraction of the particles will collide with wall owing to the small inertia, which is shown in figure \ref{Cylinder_St}$(a)$.
Besides, as a result of the lateral velocity indicated in figure \ref{Cylinder_flow}, some particles may get trapped in the wake behind the cylinder.
Nevertheless, this phenomenon disappears as the Stokes number increases to $\Stokes\approx1.232$, which is shown in the figure \ref{Cylinder_St}$(b)$.
With the further increase of $\Stokes$, it can be seen that the trajectory of particles becomes straighter.
Besides, as presented in figures \ref{Cylinder_St}$(c)$ and \ref{Cylinder_St}$(d)$, more particles will impact the surface of the cylinder and be captured, thus fewer particles appear downstream.

\begin{figure}[!t]
	\centering
	\begin{overpic}[width=0.5\textwidth]{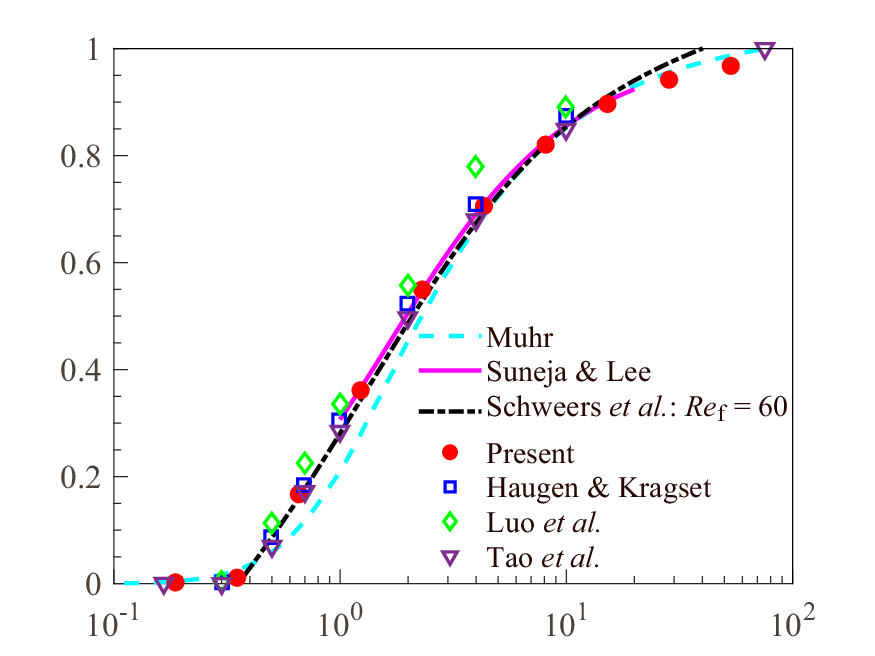}
		{
			\put(2,38){$\eta$}
			\put(51,0){$St$}
		}
	\end{overpic}
	\caption{Impaction efficiency on the front side of the cylinder as function of Stokes number for different Reynolds numbers, where some results from other groups \citep{Muhr1976Theoretical,Suneja1974Aerosol,Schweers1994Experimental,Haugen2010Particle,Luo2015Effects,Tao2018Numerical} are also displayed for comparation.}
	\label{Cylinder_impact}
\end{figure}

In order to describe the probability that particles are captured, the parameter impaction efficiency $\eta$ is introduced, which is defined as
\begin{equation}
	\eta=\frac{N_\mathrm{impact}}{N_\mathrm{init}}.
\end{equation}
Here, $N_\mathrm{impact}$ denotes the number of particles that impact on the front side of the cylinder.
Focus on the change of impaction efficiency $\eta$ as a function of Stoke number $\Stokes$, figure \ref{Cylinder_impact} presents the $(\Stokes,\eta)$ plot.
Additionally, many results obtained by different theoretical, experimental and numerical investigation of particle deposition on this problem \citep{Muhr1976Theoretical,Suneja1974Aerosol,Schweers1994Experimental,Haugen2010Particle,Luo2015Effects,Tao2018Numerical} are also displayed for comparison, and the present result shows excellent agreement with the previous literature.
Therefore, it can be concluded that the implemented one-way coupling strategy is not only applicable for the flow configuration with one block, but also applies to the cases based on multi-block overset grids.

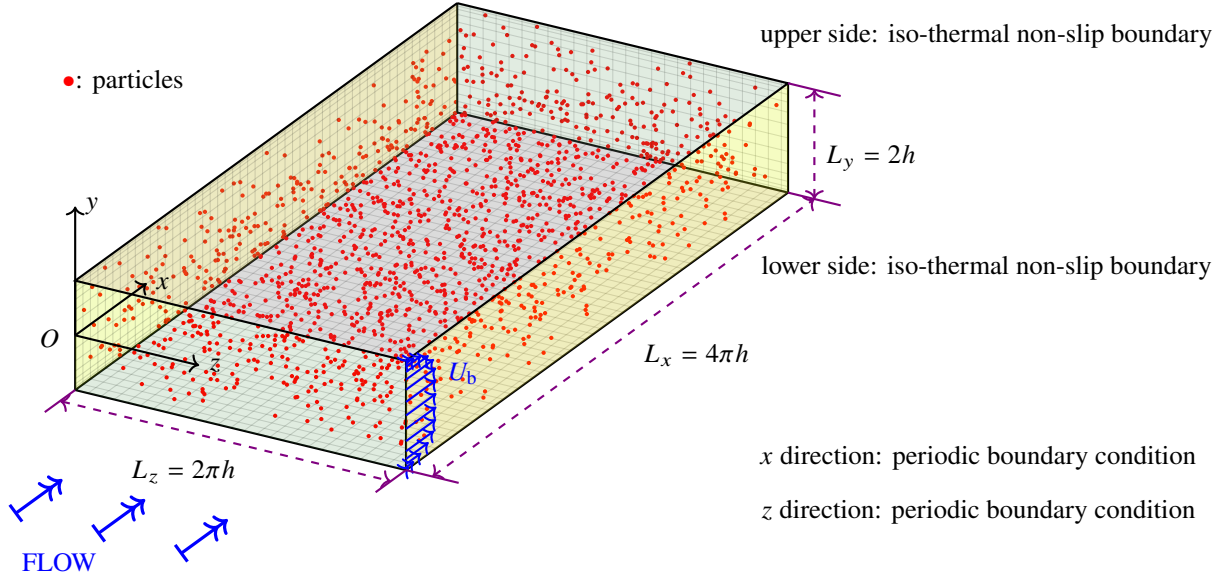
\begin{figure*}[!t]
	\begin{center}
		\tdplotsetmaincoords{65}{120}
		\begin{tikzpicture}[tdplot_main_coords,scale=0.8]
			
			\draw[thick] (0,0,0) -- (0,2*pi,0) -- (4*pi,2*pi,0) -- (4*pi,0,0) -- (0,0,0);
			\path[fill=gray,opacity=0.15] (0,0,0) -- (0,2*pi,0) -- (4*pi,2*pi,0) -- (4*pi,0,0) -- (0,0,0);
			\foreach \x in {0,0.24623359220599516650,...,12.566370614359172}
			{
				\draw [black,opacity=0.15]   (\x,0,0) -- (\x,2*pi,0);
			}
			\foreach \y in {0,0.24623359220599516650,...,6.283185307179586}
			{
				\draw [black,opacity=0.15]   (0,\y,0) -- (4*pi,\y,0);
			}
			
			\foreach \x in {1,2,...,200}
			{
				\fill[red] (12.566370614359172*rnd,6.283185307179586*rnd,2*rnd) circle (1pt);
				\fill[red] (12.566370614359172-12.566370614359172*rnd,6.283185307179586*rnd,2*rnd) circle (1pt);
				\fill[red] (12.566370614359172*rnd,6.283185307179586-6.283185307179586*rnd,2*rnd) circle (1pt);
				\fill[red] (12.566370614359172*rnd,6.283185307179586*rnd,2-2*rnd) circle (1pt);
				\fill[red] (12.566370614359172-12.566370614359172*rnd,6.283185307179586-6.283185307179586*rnd,2*rnd) circle (1pt);
				\fill[red] (12.566370614359172-12.566370614359172*rnd,6.283185307179586*rnd,2-2*rnd) circle (1pt);
				\fill[red] (12.566370614359172*rnd,6.283185307179586-6.283185307179586*rnd,2-2*rnd) circle (1pt);
				\fill[red] (12.566370614359172-12.566370614359172*rnd,6.283185307179586-6.283185307179586*rnd,2-2*rnd) circle (1pt);
			}
			\node at (3.5*pi,0,5) {\textcolor{red}{$\bullet$}: particles};
			
			\draw[thick] (4*pi,0,0) -- (4*pi,0,2);
			\draw[thick] (4*pi,2*pi,0) -- (4*pi,2*pi,2);
			\draw[thick] (0,0,0) -- (0,0,2);
			\draw[thick] (0,2*pi,0) -- (0,2*pi,2);
			
			\path[fill=yellow,opacity=0.2] (4*pi,0,0) -- (0,0,0) -- (0,0,2) -- (4*pi,0,2) -- (4*pi,0,0);
			\foreach \z in {-0.98275011173847781620,
				-0.94922009218539371567,
				-0.88994910948117089688,
				-0.79466663929959191481,
				-0.65536766370082910615,
				-0.47016464471582170015,
				-0.24623359220599516650,
				0,
				0.24623359220599516650,
				0.47016464471582170015,
				0.65536766370082910615,
				0.79466663929959191481,
				0.88994910948117089688,
				0.98275011173847781620,
				0.94922009218539371567}
			{
				\draw [black,opacity=0.15]   (4*pi,0,1+\z) -- (0,0,1+\z);
			}
			\foreach \x in {0,0.24623359220599516650,...,12.566370614359172}
			{
				\draw [black,opacity=0.15]   (\x,0,0) -- (\x,0,2);
			}
			
			\path[fill=green,opacity=0.05] (4*pi,0,0) -- (4*pi,2*pi,0) -- (4*pi,2*pi,2) -- (4*pi,0,2) -- (4*pi,0,0);
			\foreach \z in {-0.98275011173847781620,
				-0.94922009218539371567,
				-0.88994910948117089688,
				-0.79466663929959191481,
				-0.65536766370082910615,
				-0.47016464471582170015,
				-0.24623359220599516650,
				0,
				0.24623359220599516650,
				0.47016464471582170015,
				0.65536766370082910615,
				0.79466663929959191481,
				0.88994910948117089688,
				0.98275011173847781620,
				0.94922009218539371567}
			{
				\draw [black,opacity=0.15]   (0,0,1+\z) -- (0,2*pi,1+\z);
			}
			\foreach \y in {0,0.24623359220599516650,...,6.283185307179586}
			{
				\draw [black,opacity=0.15]   (0,\y,0) -- (0,\y,2);
			}
			
			\path[fill=green,opacity=0.05] (0,0,0) -- (0,2*pi,0) -- (0,2*pi,2) -- (0,0,2) -- (0,0,0);
			
			\path[fill=yellow,opacity=0.2] (4*pi,2*pi,0) -- (0,2*pi,0) -- (0,2*pi,2) -- (4*pi,2*pi,2) -- (4*pi,2*pi,0);
			
			\path[fill=gray,opacity=0.15] (0,0,2) -- (0,2*pi,2) -- (4*pi,2*pi,2) -- (4*pi,0,2) -- (0,0,2);
			\draw[thick] (0,0,2) -- (0,2*pi,2) -- (4*pi,2*pi,2) -- (4*pi,0,2) -- (0,0,2);
			
			\foreach \z in {-0.98275011173847781620,
				-0.94922009218539371567,
				-0.88994910948117089688,
				-0.79466663929959191481,
				-0.65536766370082910615,
				-0.47016464471582170015,
				-0.24623359220599516650,
				0,
				0.24623359220599516650,
				0.47016464471582170015,
				0.65536766370082910615,
				0.79466663929959191481,
				0.88994910948117089688,
				0.98275011173847781620,
				0.94922009218539371567}
			{
				\draw [blue,thick,<-]   (4*pi-1+\z*\z*\z*\z*\z*\z,2*pi,1+\z) -- (4*pi,2*pi,1+\z);
			}
			\node at (3.4*pi,2*pi,1) {\color{blue}{$U_\mathrm{b}$}};	
			
			\draw[violet,thick] (4*pi+1,0,0) -- (4*pi,0,0);
			\draw[violet,thick,dashed,<->] (4*pi+0.5,0,0) -- (4*pi+0.5,2*pi,0);
			\draw[violet,thick] (4*pi+1,2*pi,0) -- (4*pi,2*pi,0);
			\node at (4.6*pi,pi,0) {$L_z=2\pi h$};	
			
			\draw[violet,thick] (4*pi,2*pi+1,0) -- (4*pi,2*pi,0);
			\draw[violet,thick,dashed,<->] (4*pi,2*pi+0.5,0) -- (0,2*pi+0.5,0);
			\draw[violet,thick] (0,2*pi+1,0) -- (0,2*pi,0);	
			\node at (2*pi,2.6*pi,0) {$L_x=4\pi h$};	
			
			\draw[violet,thick] (0,2*pi+1,2) -- (0,2*pi,2);
			\draw[violet,thick,dashed,<->] (0,2*pi+0.5,0) -- (0,2*pi+0.5,2);		
			\node at (0,2.5*pi,1) {$L_y=2h$};	
			
			\node at (4*pi+0.3,-0.3,1) {$O$};	
			\draw[thick,->] (4*pi,0,1) -- (3.25*pi,0,1) node [right] {$x$};
			\draw[thick,->] (4*pi,0,1) -- (4*pi,0.75*pi,1) node [right] {$z$};  	
			\draw[thick,->] (4*pi,0,1) -- (4*pi,0,1+0.75*pi) node [right] {$y$};
			
			\draw[very thick,|->>,blue] (5.5*pi,0.5*pi,0) -- (5*pi,0.5*pi,0);
			\draw[very thick,|->>,blue] (5.5*pi,pi,0) -- (5*pi,pi,0);
			\draw[very thick,|->>,blue] (5.5*pi,1.5*pi,0) -- (5*pi,1.5*pi,0);
			\node at (5.9*pi,pi,0) {\color{blue}{FLOW}};	
			
			\node at (0,3.15*pi,-4) {$x$ direction: periodic boundary condition};
			\node at (0,3.15*pi,-5) {$z$ direction: periodic boundary condition};
			\node at (0,3.2*pi,3.75) {upper side: iso-thermal non-slip boundary};
			\node at (0,3.2*pi,-0.5) {lower side: iso-thermal non-slip boundary};
			
		\end{tikzpicture}
	\end{center}
	\vspace{-0.7cm}
	\caption{Schematic of the numerical configuration for particle-laden channel flow: the computational doamin and the boundary condition. Here, particles are denoted by red point clouds.}
	\label{Channel}
\end{figure*}
\begin{figure*}[!t]
	\centering
	\begin{overpic}[width=0.495\textwidth]{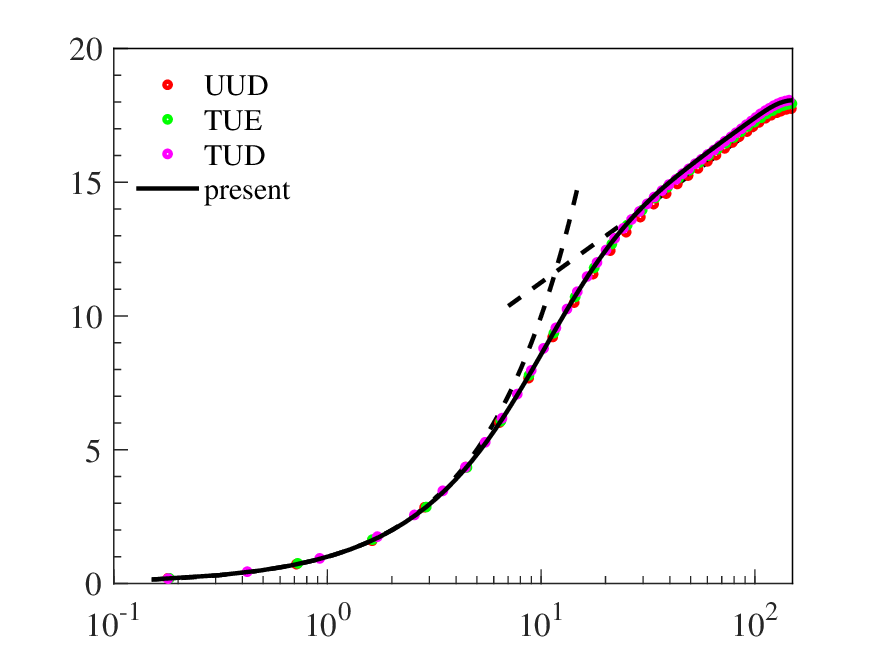}
		{
			\put(0,68.5){$(a)$}
			\put(2,38){$u^+_\mathrm{f}$}
			\put(50,0){$n^+$}
			
			\put(40,30){$u^+_\mathrm{f}=n^+$}
			\put(50,60){$u^+_\mathrm{f}=2.5\ln(n^+)+5.5$}
		}
	\end{overpic}
	\begin{overpic}[width=0.495\textwidth]{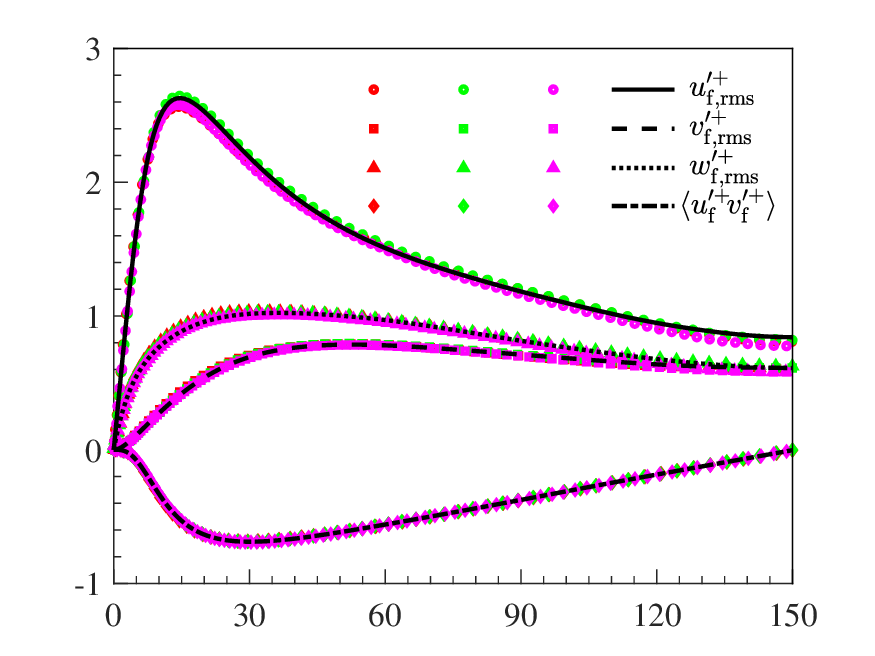}
		{
			\put(0,68.5){$(b)$}
			\put(2,16){\rotatebox{90}{$u^{\prime+}_\mathrm{f,rms},v^{\prime+}_\mathrm{f,rms},w^{\prime+}_\mathrm{f,rms},\langle u^{\prime+}_\mathrm{f}\!v^{\prime+}_\mathrm{f}\rangle$}}
			\put(50,0){$n^+$}
		}
	\end{overpic}
	\caption{Wall normal profile of the flow statistics: $(a)$ streamwise velocity and $(b)$ velocity fluctuations.}
	\label{Channel_statistics}
\end{figure*}
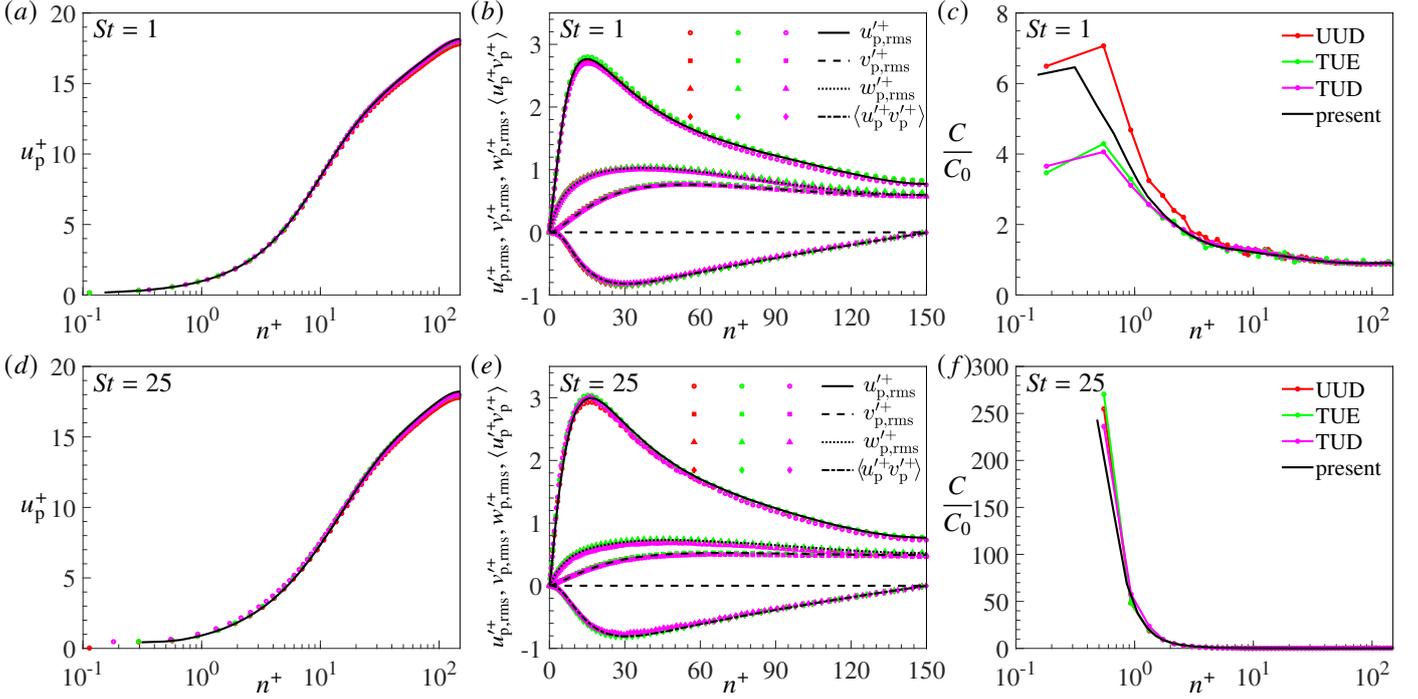
\begin{figure*}[!t]
	\centering
	\begin{overpic}[width=0.325\textwidth]{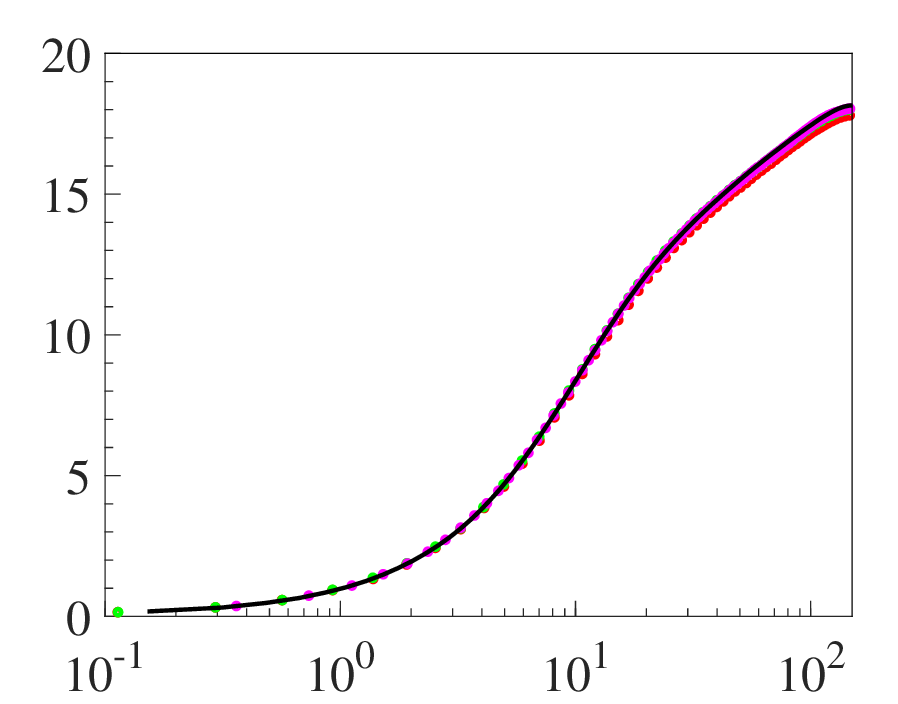}
		{
			\put(-6,71){$(a)$}
			\put(-2,40){$u^+_\mathrm{p}$}
			\put(50,0){$n^+$}
			\put(14,67){$\Stokes=1$}
		}
	\end{overpic}
	\begin{overpic}[width=0.325\textwidth]{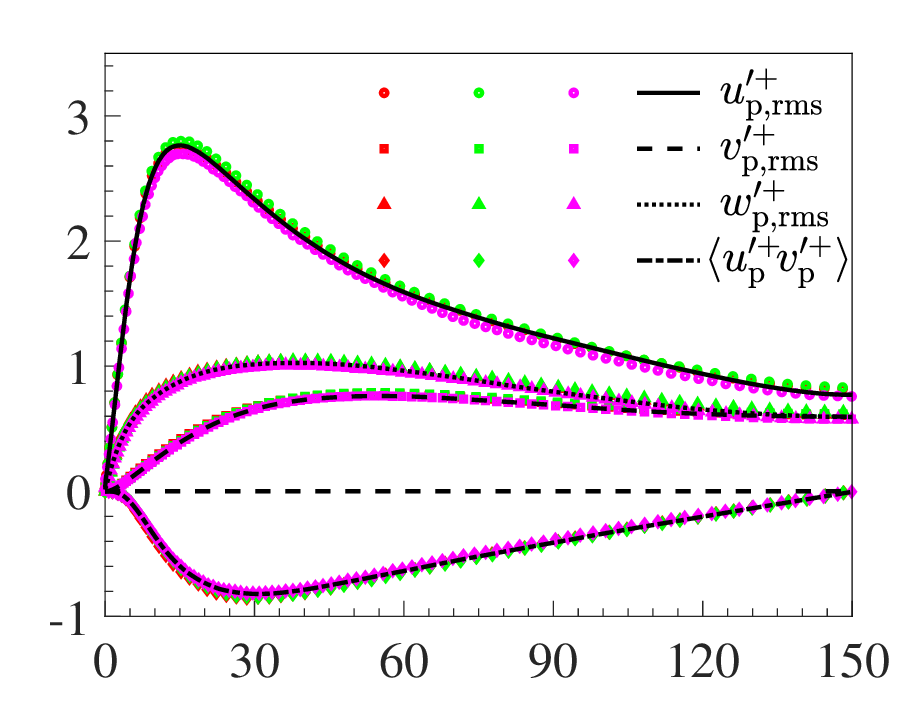}
		{
			\put(-6,71){$(b)$}
			\put(-2,10){\rotatebox{90}{\footnotesize$u^{\prime+}_\mathrm{p,rms},v^{\prime+}_\mathrm{p,rms},w^{\prime+}_\mathrm{p,rms},\langle u^{\prime+}_\mathrm{p}\!v^{\prime+}_\mathrm{p}\rangle$}}
			\put(50,0){$n^+$}
			\put(14,67){$\Stokes=1$}
		}
	\end{overpic}
	\begin{overpic}[width=0.325\textwidth]{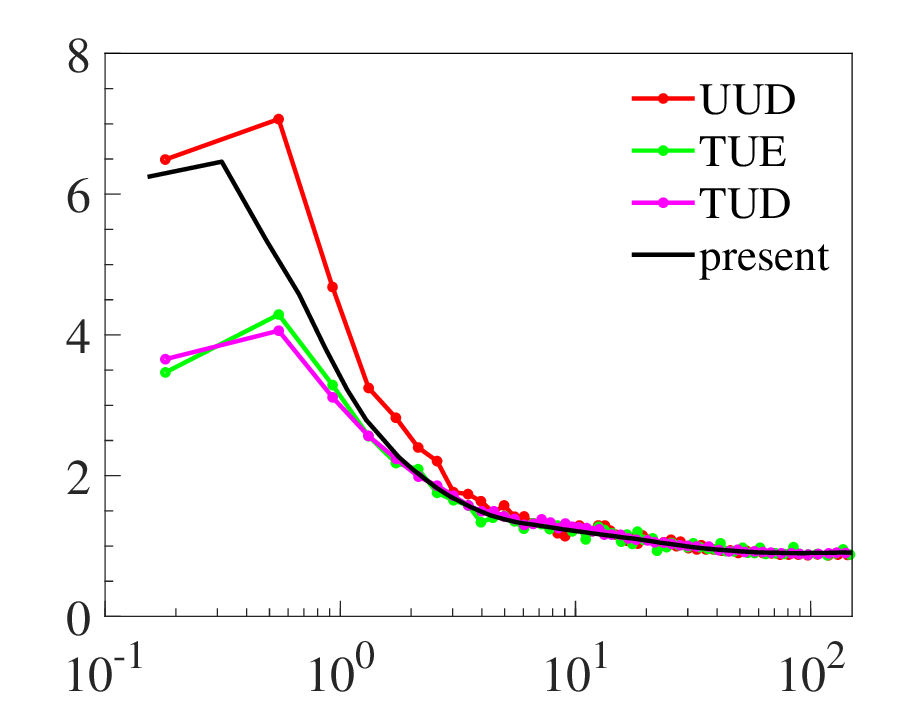}
		{
			\put(-6,71){$(c)$}
			\put(-5,40){$\dfrac{C}{C_0}$}
			\put(50,0){$n^+$}
			\put(14,67){$\Stokes=1$}
		}
	\end{overpic}
	\begin{overpic}[width=0.325\textwidth]{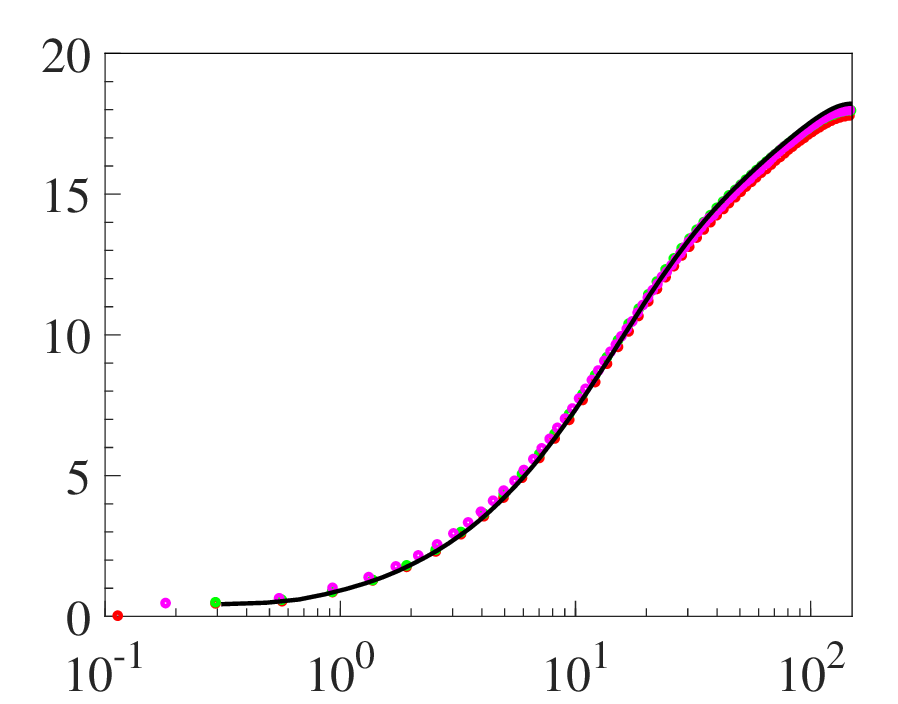}
		{
			\put(-6,71){$(d)$}
			\put(-2,40){$u^+_\mathrm{p}$}
			\put(50,0){$n^+$}
			\put(14,67){$\Stokes=25$}
		}
	\end{overpic}
	\begin{overpic}[width=0.325\textwidth]{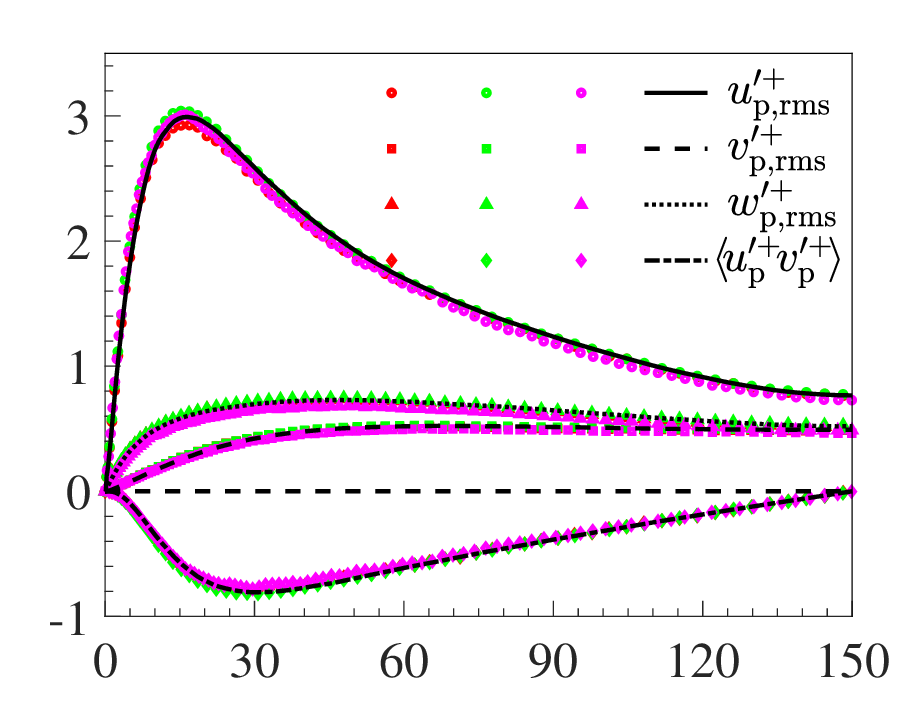}
		{
			\put(-6,71){$(e)$}
			\put(-2,10){\rotatebox{90}{\footnotesize$u^{\prime+}_\mathrm{p,rms},v^{\prime+}_\mathrm{p,rms},w^{\prime+}_\mathrm{p,rms},\langle u^{\prime+}_\mathrm{p}\!v^{\prime+}_\mathrm{p}\rangle$}}
			\put(50,0){$n^+$}
			\put(14,67){$\Stokes=25$}
		}
	\end{overpic}
	\begin{overpic}[width=0.325\textwidth]{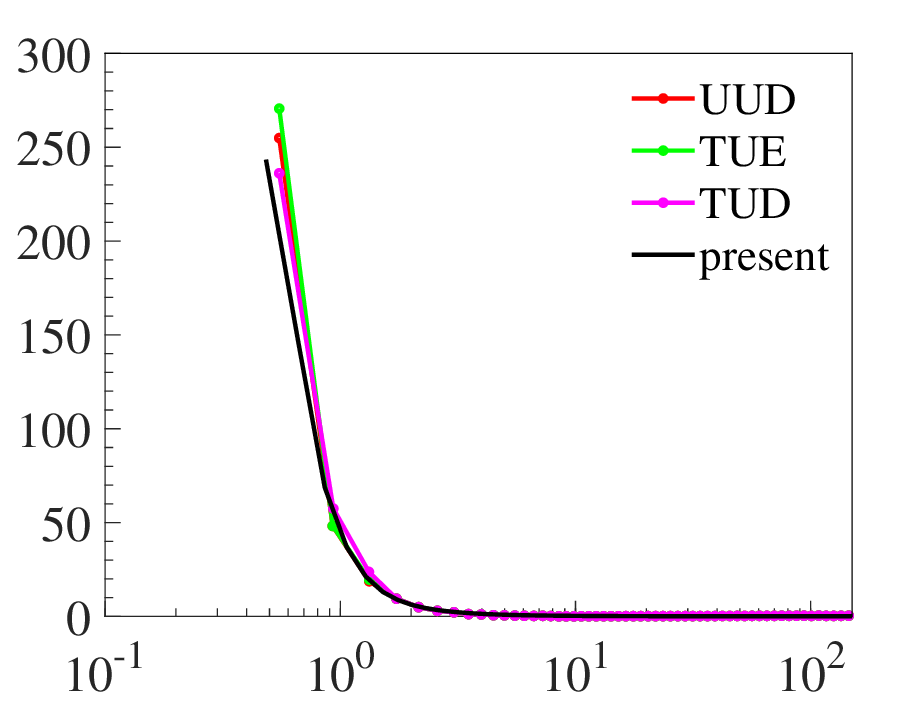}
		{
			\put(-6,71){$(f)$}
			\put(-5,40){$\dfrac{C}{C_0}$}
			\put(50,0){$n^+$}
			\put(14,67){$\Stokes=25$}
		}
	\end{overpic}
	\caption{Wall normal profiles of $(a,d)$ mean streamwise particle velocity, $(b,e)$ particle velocity fluctuations and $(c,f)$ instantaneous particle number density. Here, diagrams in the first row correspond to the results of $\Stokes=1$, and diagrams in the second row correspond to the results of $\Stokes=25$.}
	\label{Channel_par}
\end{figure*}

\subsubsection{One-way coupled particle-laden channel flow}\label{one-way_channel}

The aforementioned test cases are two-dimensional, and the carrier flows are laminar.
Thereafter, a simulation for the three-dimensional particle-laden turbulent flow is further presented to confirm the ability of the current solver to perform a large-scale simulation.

In the following, a simulation of the one-way coupled particle-laden channel flow is performed, and the numerical setup is presented in the figure \ref{Channel}.
As indicated by figure \ref{Channel}, the channel flow is conducted in a rectangular computational domain, thus only one H-type block is used here. 
Moreover, the domain size is $(x,y,z)=[0,4\pi h]\times[0,2h]\times[0,2\pi h]$, where $h=1$ is the half channel height, and $x,y,z$ represent the streamwise, wall-normal and spanwsie directions, respectively.
The upper and lower side of the domain are set as the iso-thermal non-slip wall, while periodic boundary conditions are imposed on both the $x$ and $z$ directions.

In this benchmark calculation, based on the bulk velocity $U_\mathrm{b}$, the bulk Reynolds number is $\Rey_\mathrm{b}=U_\mathrm{b}h/\nu_\mathrm{f}=2320$, which results in a friction Reynolds number of $\Rey_\tau=u_\tau h/\nu_\mathrm{f}\approx150$.
Here, $u_\tau=\sqrt{\tau_\mathrm{w}/\rho_\mathrm{f}}$ is the friction velocity, and $\tau_\mathrm{w}$ is the mean shear stress at the wall.
Besides, the bulk Mach number is set as $\Mach_\mathrm{f}=0.2$ according to the previous results \citep{Liao2024GPU}, so as to approximate the incompressible flow condition \citep{Marchioli2008Statistics}.
In order to ensure the constant mass-flow-rate during the simulation, an external volume force $\bm{f}\cdot\bm{e}_x$ is added to the right hand side of the compressible Navier-Stokes equations \eqref{NS} in the $x$ direction \citep{Bernardini2021STREAmS, Liao2024GPU}.
Correspondingly, the resulting power spent $\bm{f}\cdot\bm{u}_{\mathrm{f},x}$ is added to the total energy equation.
Additionally, the source terms $\bm{S}_m$ and $S_e$ have been neglected owing to the one-way coupling strategy.
Moreover, the computational domain is discretized by a grid with $192\times193\times192$ points.
Normalized by the wall unit $\delta_\nu=\nu_\mathrm{f}/u_\tau$, the grid resolution is (9.85,0.15,4.93) in each direction, thus the boundary layer is well resolved.

\begin{figure*}[!t]
	\centering
	\begin{overpic}[width=0.495\textwidth]{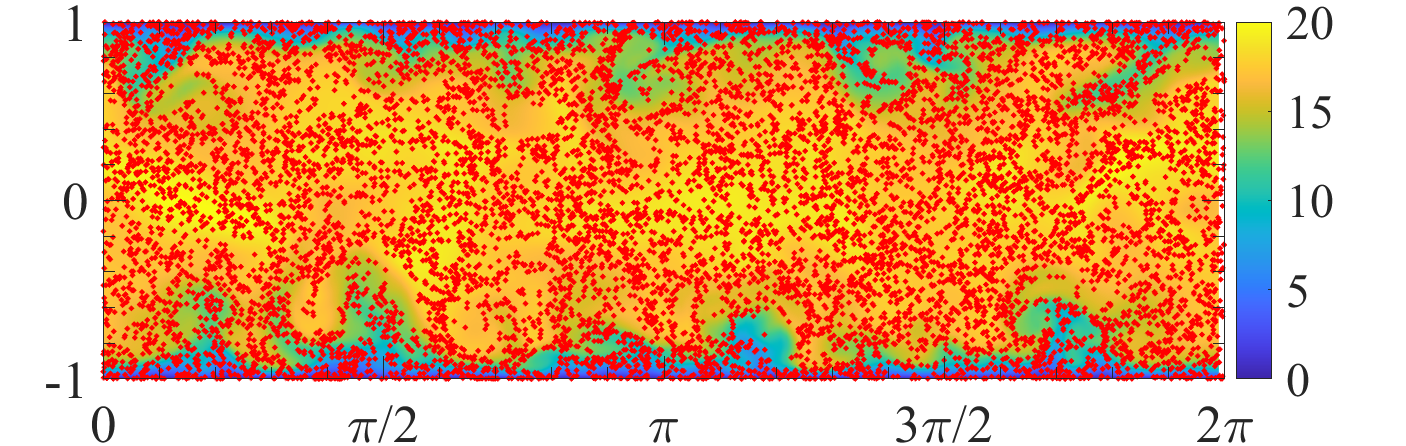}
		{
			\put(-1,29){$(a)$}
			\put(1,16.5){$y$}
			\put(96,16.5){$u^+_\mathrm{f}$}
			\put(47,-3){$z$}
			\put(42.5,31.5){$\Stokes=1$}
		}
	\end{overpic}
	\begin{overpic}[width=0.495\textwidth]{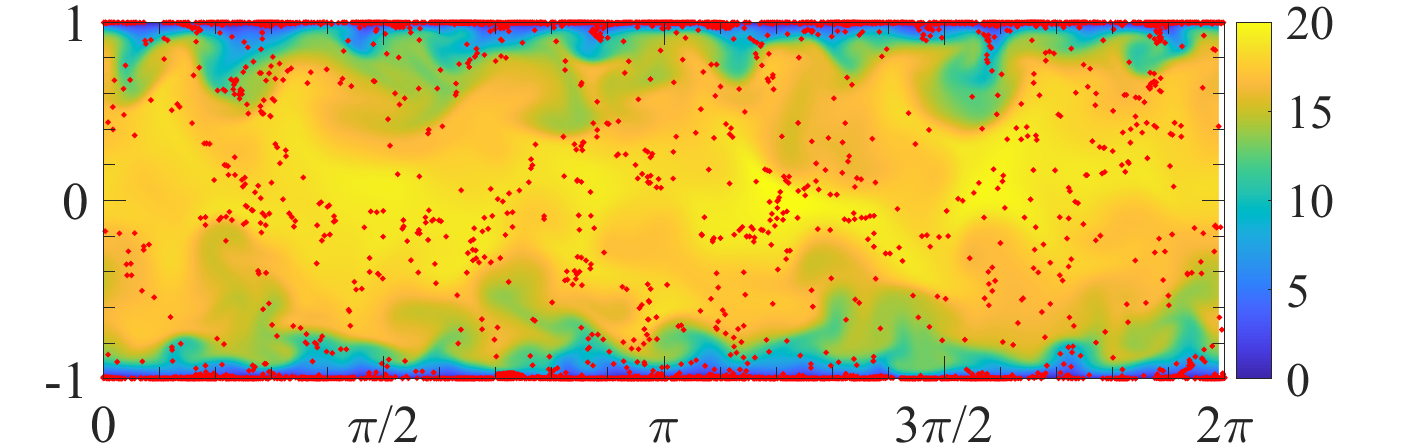}
		{
			\put(-1,29){$(b)$}
			\put(1,16.5){$y$}
			\put(96,16.5){$u^+_\mathrm{f}$}
			\put(47,-3){$z$}
			\put(41.75,31.5){$\Stokes=25$}
		}
	\end{overpic}
	\caption{A snapshot in the cross-section $x=0$ for $(a)$ $\Stokes=1$ and $(b)$ $\Stokes=25$. Here, the diagram is colored by the streamwise velocity $u_\mathrm{f}$ normalized by $u_\tau$, and the particles are denoted by red point clouds.}
	\label{Channel_yz}
\end{figure*}
\begin{figure*}[!t]
	\centering
	\begin{overpic}[width=0.495\textwidth]{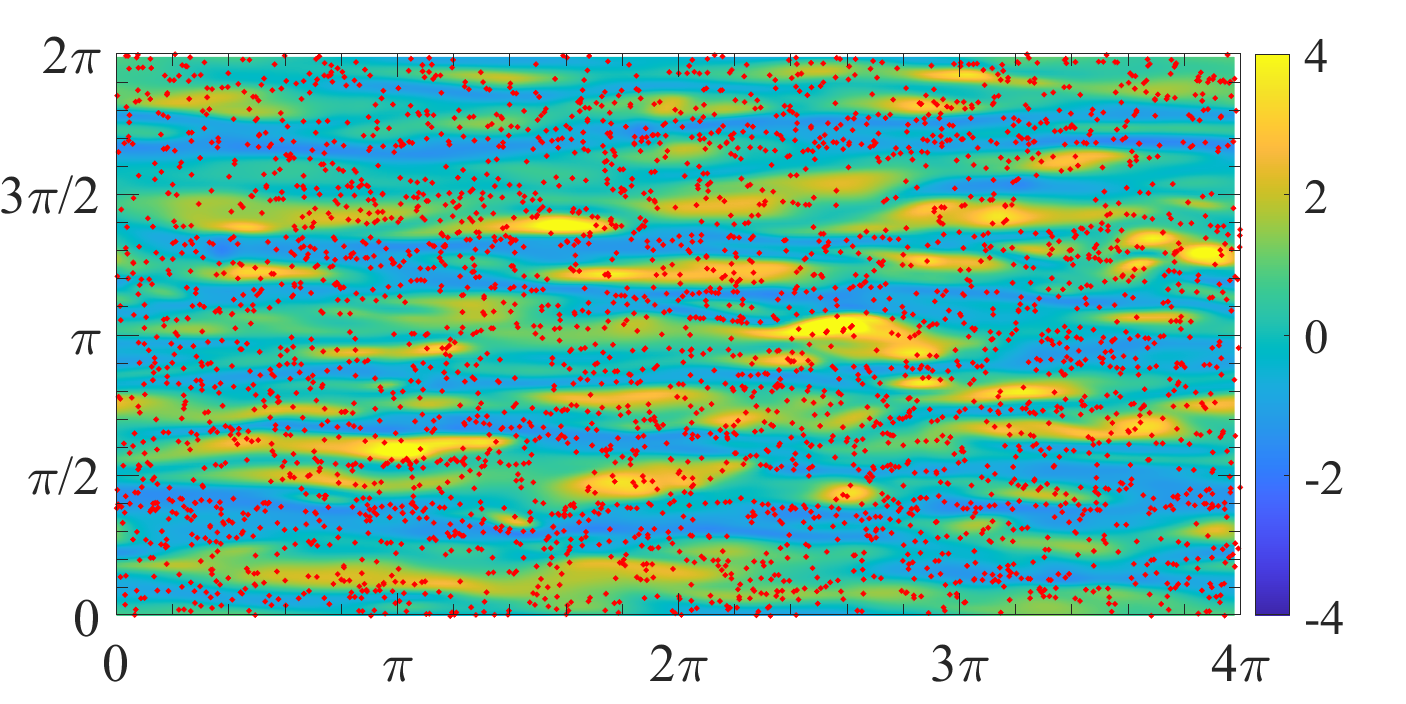}
		{
			\put(-2,45){$(a)$}
			\put(2,25){$z$}
			\put(95.5,25){$u^{\prime+}_\mathrm{f}$}
			\put(48,-2){$x$}
			\put(43.75,47.5){$\Stokes=1$}
		}
	\end{overpic}
	\begin{overpic}[width=0.495\textwidth]{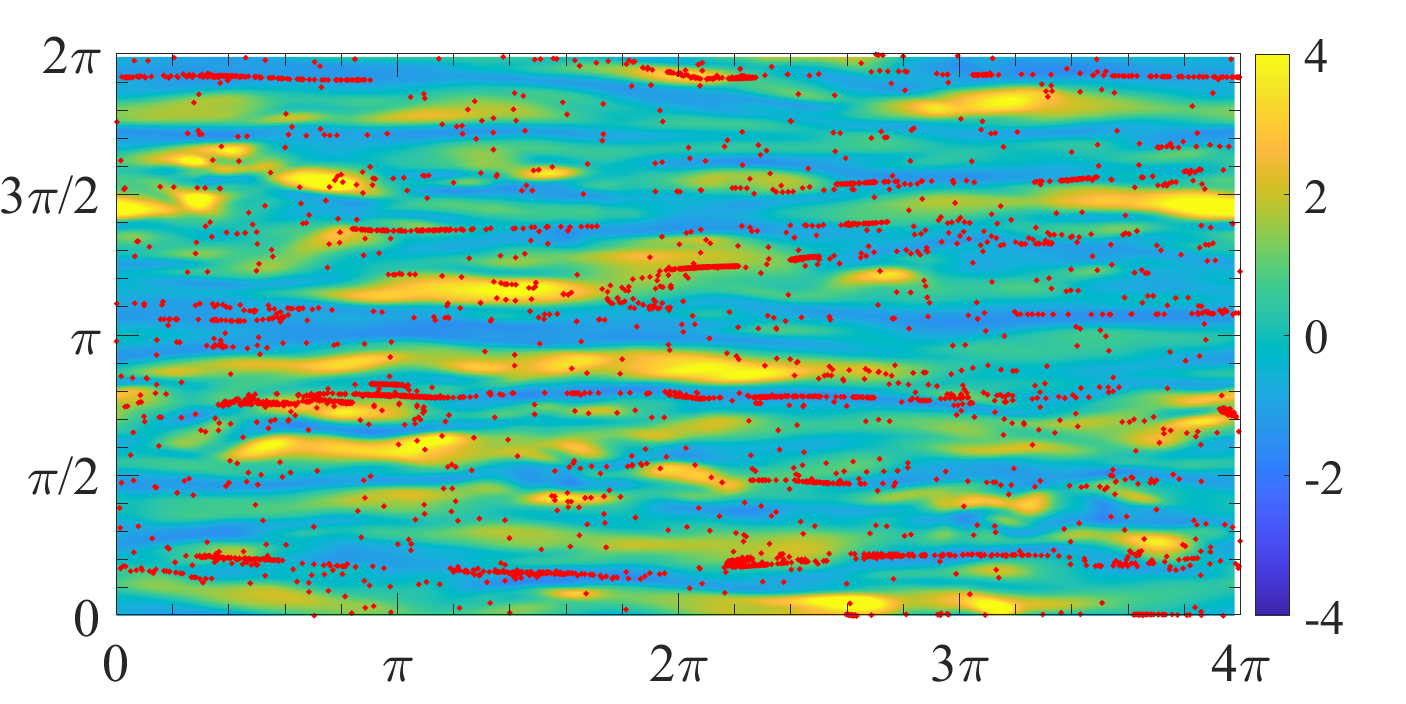}
		{
			\put(-2,45){$(b)$}
			\put(2,25){$z$}
			\put(95.5,25){$u^+_\mathrm{f}$}
			\put(48,-2){$x$}
			\put(43,47.5){$\Stokes=25$}
		}
	\end{overpic}
	\caption{A snapshot in the cross-section $n^+\approx3$ for $(a)$ $\Stokes=1$ and $(b)$ $\Stokes=25$. Here, the diagram is colored by the streamwise velocity $u_\mathrm{f}$ normalized by $u_\tau$, and the particles are denoted by red point clouds.}
	\label{Channel_xz}
\end{figure*}

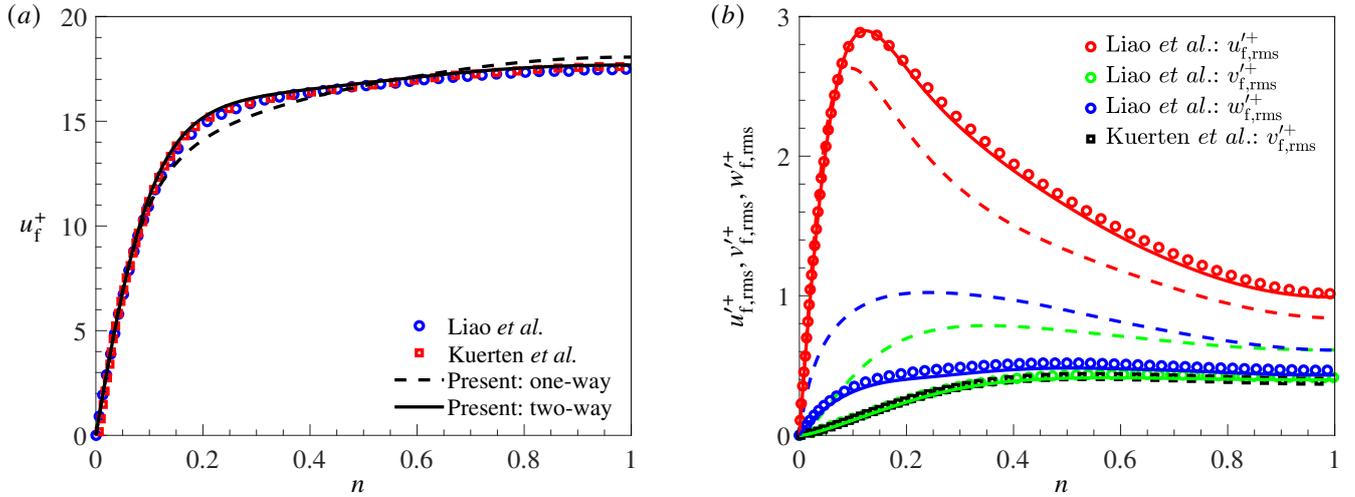
\begin{figure*}[!t]
	\centering
	\begin{overpic}[width=0.495\textwidth]{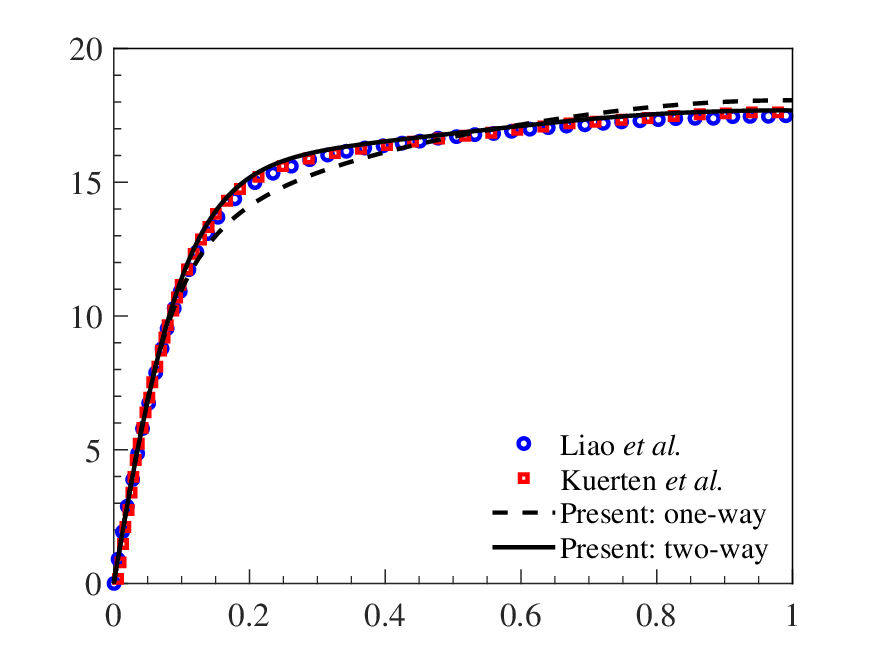}
		{
			\put(0,68.5){$(a)$}
			\put(2,38){$u^+_\mathrm{f}$}
			\put(50,0){$n$}
		}
	\end{overpic}
	\begin{overpic}[width=0.495\textwidth]{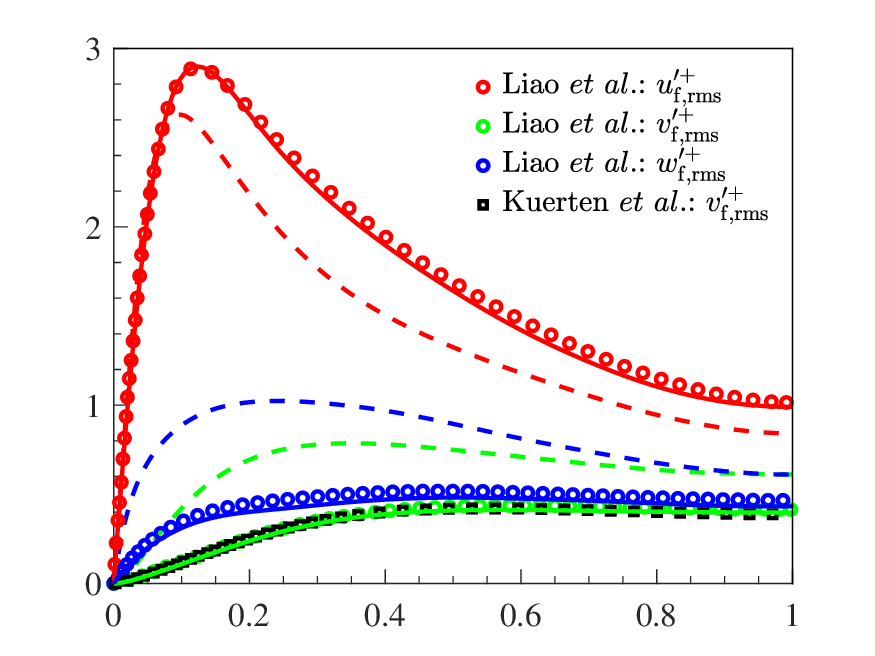}
		{
			\put(0,68.5){$(b)$}
			\put(2,24){\rotatebox{90}{$u^{\prime+}_\mathrm{f,rms},v^{\prime+}_\mathrm{f,rms},w^{\prime+}_\mathrm{f,rms}$}}
			\put(50,0){$n$}
		}
	\end{overpic}
	\caption{Wall normal profile of the flow statistics: $(a)$ streamwise velocity and $(b)$ velocity fluctuations. Here, the results of the one-way coupled flow are presented with dashed lines, while results of the two-way coupled flow are presented with solid lines.}
	\label{Channel_statistics_twoway}
\end{figure*}

Starting from a laminar velocity profile with random perturbations, the simulation is conducted for a long time to guarantee that the flow reaches a turbulent state.
Moreover, figure \ref{Channel_statistics} presents the flow statistics obtained by the present solver, where the superscript $+$ represents that variables are normalized by the viscous scale $u_\tau$ or $\delta_\nu$.
It can be observed that they are remarkably consistent with the results in the literature \citep{Marchioli2008Statistics}, which implies the reliability of the present code.
As a result, we can claim that the flow dynamics are captured, and the turbulent channel flow could be seen as the carrier flow for the subsequent verification of the implemented particle module.

Based on the previous study, two cases are selected as the verification cases, which are distinguished by the Stokes numbers of $\Stokes=1$ and $\Stokes=25$.
For the fixed density ratio $\rho_\mathrm{p}/\rho_\mathrm{f}$, the Stokes number $\Stokes=\tau_\mathrm{p}/\tau_\mathrm{f}$ is changed by varying the particle diameter $d_\mathrm{p}$, where $\tau_\mathrm{f}$ is defined as the viscous time scale $\tau_\mathrm{f}=\nu_\mathrm{f}/u_\tau^2$.
Besides, the particle motion is governed by the equation \eqref{drag_equation}, where only the modified drag force is considered.

Figure \ref{Channel_par} provides a comparison of the present results with the literature, including the mean streamwise particle velocity, particle velocity fluctuations, and instantaneous particle number density.
As displayed in figure \ref{Channel_par}, results from different groups show no obvious differences for both two Stokes numbers.
Furthermore, figure \ref{Channel_yz} and figure \ref{Channel_xz} provide the instantaneous snapshot of the flow and particle distribution in the $x-z$ and $z-y$ cross-sections, respectively. 
By comparing the figure \ref{Channel_yz}$(a)$ and \ref{Channel_yz}$(b)$, it can be observed that particles tend to aggregate in the near-wall region, especially for the large Stokes number.
Besides, as shown in figure \ref{Channel_xz}, the flow within the boundary layer manifests as streak-like structures.
For the small Stokes number $\Stokes=1$, particles behave as an almost uniform distribution, which is indicated by figure \ref{Channel_xz}$(a)$.
Nevertheless, particles with large Stokes number $\Stokes=25$ are manifested as a straight pattern correlated with the low-speed streak, which is presented in figure \ref{Channel_xz}$(b)$.
The observations are also qualitatively consistent with the physical phenomena reported in \citep{Gong2023CP3d}.
Therefore, we can conclude that the one-way coupling strategy has been successfully implemented in the present framework.

\subsection{Two-way coupling}

After the one-way coupling strategy has been adequately validated, a two-way coupled particle-laden channel flow is tested on the basis of the simulation discussed before, thus the source terms $\bm{S}_m$ and $S_e$ are added to consider the feedback effects of particles.

Following the same setups as in case 2W in \cite{Kuerten2011Turbulence}, the flow statistics are focused on, which are illustrated in figure \ref{Channel_statistics_twoway}.
By comparing the results of the solid and dashed lines, it is observed that profiles of mean velocity and fluctuations are altered due to the influence of particles, indicating that the feedback terms have indeed been added in the governing equations. 
Moreover, there is no obvious difference between the present results and the previous literature \citep{Kuerten2011Turbulence, Liao2024GPU}, confirming that the two-way coupling strategy has been successfully implemented in the current framework.

\begin{figure*}[!t]
	\centering
	\begin{overpic}[width=\textwidth]{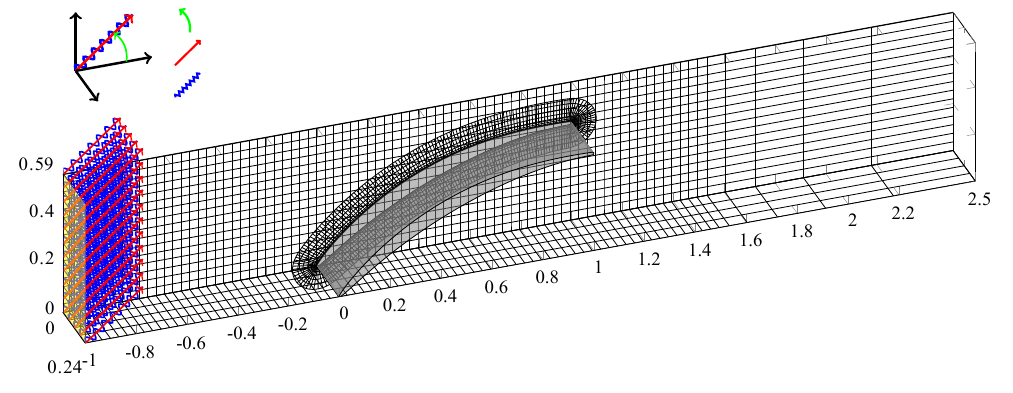}
		\put(52,8){$x$}
		\put(1,14){$y$}
		\put(3,4){$z$}
		
		\put(32.7,9){\rotatebox{-90}{\hdashrule[2pt][x]{1.25cm}{1pt}{0.85mm 0.85mm}}}
		\put(33.25,3){\rotatebox{10.5}{$\Larrow{\rule{1.38cm}{0pt}}c_{\mathrm{ax}}=1\Rarrow{\rule{1.38cm}{0pt}}$}}
		\put(58,23.5){\rotatebox{-90}{\hdashrule[2pt][x]{3.1cm}{1pt}{0.85mm 0.85mm}}}
		
		\put(90.75,36.25){\rotatebox{10.5}{\hdashrule[2pt][x]{0.8cm}{1pt}{0.85mm 0.85mm}}}
		\put(96,35.9){$L_z$}
		\put(92.5,33.125){\rotatebox{10.5}{\hdashrule[2pt][x]{0.8cm}{1pt}{0.85mm 0.85mm}}}
		\put(97,27){$h$}
		\put(92.5,19.5){\rotatebox{10.5}{\hdashrule[2pt][x]{0.8cm}{1pt}{0.85mm 0.85mm}}}
		
		\put(0.5,26.5){inlet}
		\put(92,15){outlet}
		
		\put(9.25,27){$z$}
		\put(8.25,37.5){$y$}
		\put(15,30.75){$x$}
		
		\put(21,36){angle of the inlet velocities $\theta=41^\circ$}
		\put(21,32.75){fixed inlet velocities $(U_\mathrm{in}\cos\theta,U_\mathrm{in}\sin\theta,0)$}
		\put(21,29.5){artifical inflow turbulence}
		
		\put(71,9){outlet: non-reflecting}
		\put(71,5.75){pitchwise direction: periodic}
		\put(71,2.5){spanwise direction: periodic}
		
		\put(4,21.5){
			\begin{tikzpicture}[>=stealth]
				\draw[->,thick,gray,dashed] (0,0) to [bend left]  (0.15,1.9);
			\end{tikzpicture}
		}
	\end{overpic}
	\caption{Schematic of the numerical configuration for the particle-laden flow in a linear compressor cascade: the computational grid consists of a background H-type grid and an O-type grid, which is shown about every sixteenth line in each direction. Besides, the inlet is indicated by the light yellow plane, and the inflow are presented by red arrows with blue wavy lines.}
	\label{sketch}
\end{figure*}
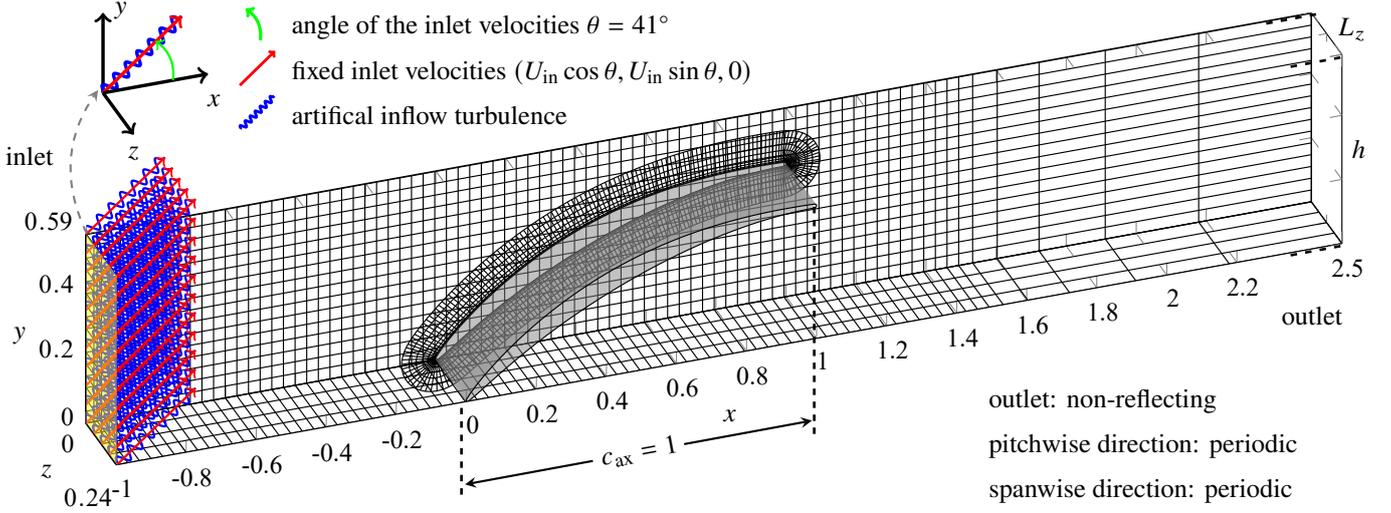

\section{Validation of numerical efficiency in large-scale turbomachinery applications}\label{sec:acceleration}

The aforementioned test cases have clearly shown that the current framework is effective and reliable. 
Furthermore, in order to validate the numerical efficiency of the proposed algorithms, a direct numerical simulation of particle-laden flow in a linear compressor cascade at engine-relevant conditions is proceeded.

Figure \ref{sketch} shows the configuration of the present linear compressor stage, which operates at the axial Reynolds number of $\Rey_\mathrm{c}=138,500$ in room temperature.
Here, two blocks are used, one is the background H-type grid and the other is the O-type grid.
For the background grid, at the inlet indicated by the light yellow plane, a Dirichlet boundary condition is applied.
Moreover, an artificial turbulence generated by the digital filter method \citep{Klein2003filter} is added, so as to simulate the coming free-stream turbulence from upstream in real engines.
Differently, at the outlet, a characteristic boundary condition with a sponge layer is applied to prevent spurious reflection \citep{Sandberg2006Nonreflecting, Zhao2020Bypass}.
Besides, periodic boundary conditions are imposed at both the pitchwise and spanwise direction.
Additionally, the O-type grid well describes the curved geometry of the blade surface, which is modeled as an non-slip adiabatic wall. 

Owing to the high Reynolds number, the total grid points is up to $1.3\times10^8$, and the grid resolution of the blade boundary layer falls below (9, 0.7, 10) in the streamwise, wall-normal and spanwise directions, suggesting that the complex high-Reynolds number boundary layer flows \citep{Zaki2010Direct} are well resolved.
After the flow reaches the turbulent state, sand particles are constantly added into the flow field from the inlet.
The particle density is set as 2690 $\mathrm{kg/m^3}$ \cite{Bansal2015Properties}, and its diameter $d_\mathrm{p}=4\mu\mathrm{m}$ is set to be less than the Kolmogorov scale \citep{Balachandar2010Turbulent}, thus the point-particle approach is applicable.
Due to the very low values of volume fraction $\varPhi_\mathrm{p}\lessapprox6\times10^{-7}<10^{-6}$ \citep{Elghobashi1994On}, the one-way coupling strategy is adopted, which means that the effect of particles on turbulence is neglected.

\begin{figure}[!t]
	\centering
	\begin{overpic}[width=0.495\textwidth]{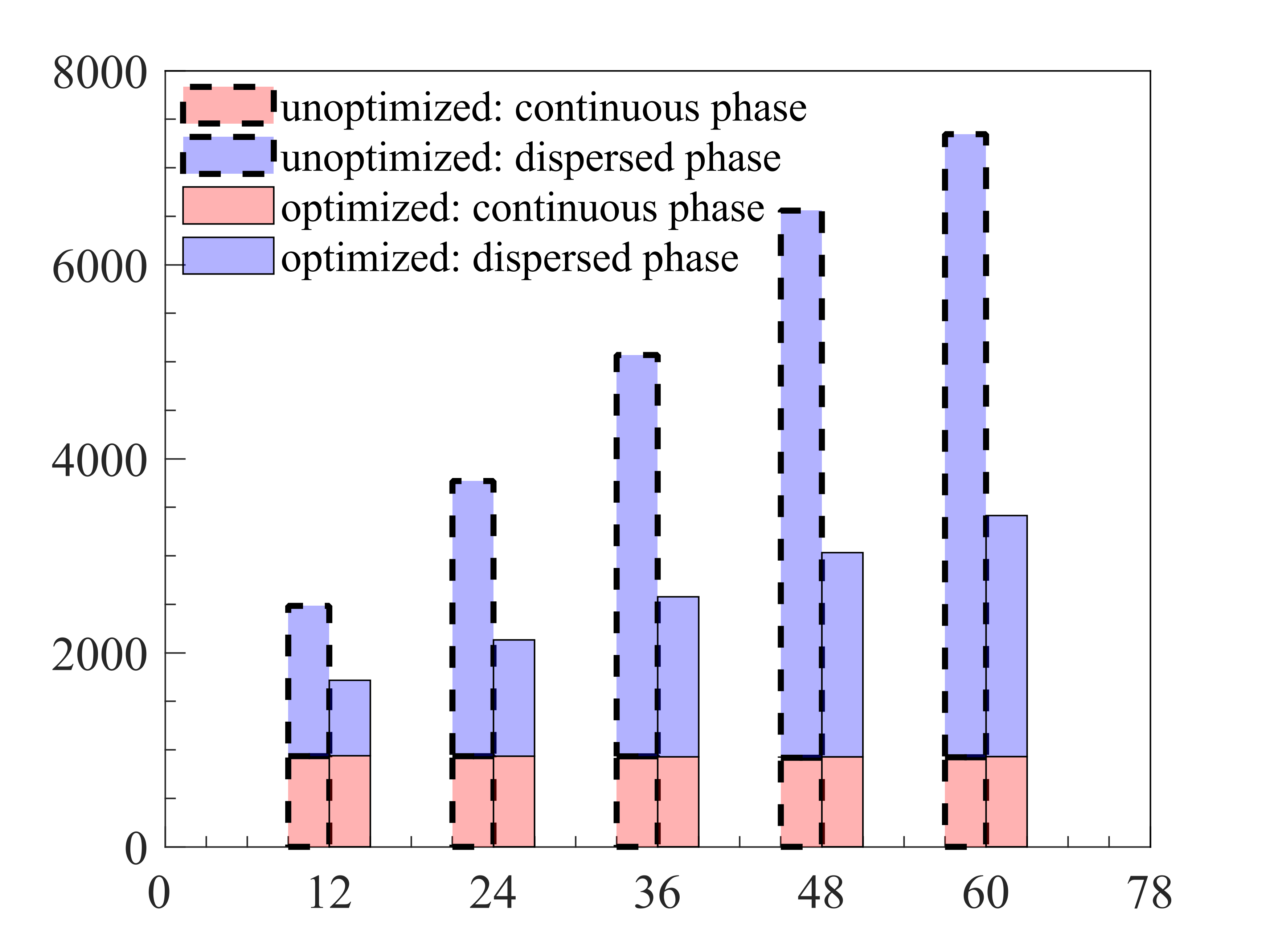}
		{
			\put(13,71){/(s)}
			\put(2,38){$t$}
			\put(39.5,-1){particle number}
			\put(91,9){$\times10^4$}
		}
	\end{overpic}
	\caption{The computational time in seconds spent on the continuous and dispersed phases, where the results obtained with the global search strategy are marked by dashed lines.}
	\label{acceleration}
\end{figure}
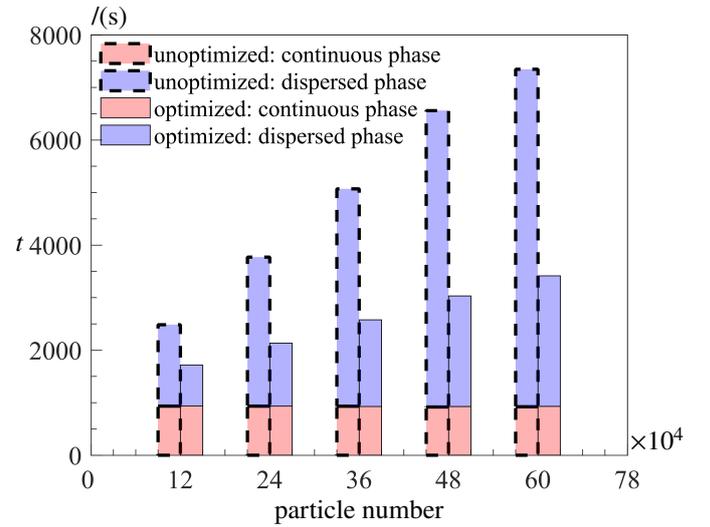

In order to highlight the accelerating effects of the particle redistribution algorithms proposed in the present study, a set of cases were performed with varying particle numbers, and each case was run twice with the same initial conditions: one
using the unoptimized code, and the other using
the optimized code.
Simulated with 3 NVIDIA A800 80 GB graphics processing units (GPUs) and 1 Intel XEON Platinum 8358 central processing unit (CPU), the run-time for one thousand non-dimensional timesteps was collected and further presented in figure \ref{acceleration}.

As shown in figure \ref{acceleration}, for different cases, the time consumed by the continuous phase is nearly the same, which is indicated by the pink bars.
Nevertheless, the time consumed by the dispersed phase increases almost linearly with the particle number, which is indicated by the violet bars.
Remarkably, the computational time for particles is decreased by 50$\sim$63 percent with the proposed algorithms, greatly reducing the computational cost.
In addition, the results imply that such a large-scale simulation can be realized with the aforementioned limited computational resources, which lays the foundation for subsequent simulations.

\begin{figure*}[!t]
	\centering
	\begin{overpic}[width=\textwidth]{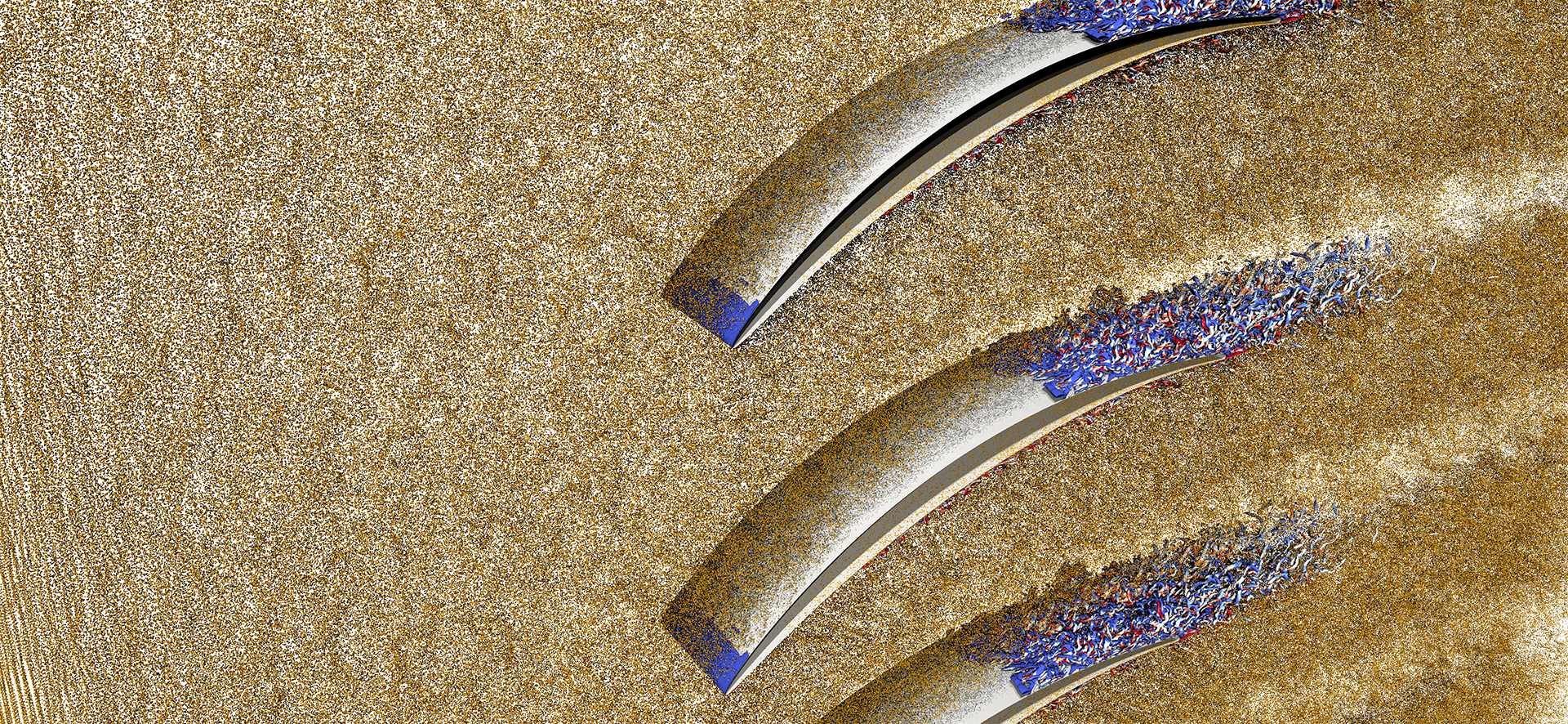}
		{
		}
	\end{overpic}
	\caption{A snapshot of the particle-laden flow in a linear compressor cascade, where particles are denoted by brown spheres. Besides, vortical structures are indicated by the $Q$-criterion \citep{Hunt1988Eddies}, and colored by the spanwise vorticity.}
	\label{Compressor}
\end{figure*}

Moreover, figure \ref{Compressor} presents a snapshot of the particle-laden flow in a linear compressor cascade, where the instantaneous iso-surface of the $Q$-criterion \citep{Hunt1988Eddies} is colored by the spanwise vorticity.
Besides, there are about 6$\times10^5$ sand particles locate in the flow field, which are represented by brown spheres.
The complex flow dynamics and the rich patterns of particles indicate the capability of the present numerical framework.

\section{Conclusion and outlook}\label{sec:Conclusion_and_outlook}

Based on the overset method, a framework for point-particle direct numerical simulation has been developed, which is appropriate for flow configurations consisting of overlapping multi-block grids.
Consequently, the particle dynamics within complex geometries can be successfully captured, facilitating the research of particle-laden turbomachinery flows.

The present particle tracking framework consists of modules with different functionalities, including efficient particle storage, high-fidelity particle push, particle redistribution in complex geometrical configurations, and particle feedback on the fluid phase.
In particular, the particle redistribution in multi-block overset grids poses significant challenges, and optimization strategies have been proposed accordingly.  
Specifically, particles passing through inter-block interfaces are redistributed according to the proposed mapping strategy, greatly reducing the computational cost.
%The current framework allows particles tracking in different blocks separately, and particles passing through inter-block interfaces are redistributed according to the proposed mapping strategy to avoid the significant computational consumptions.
Besides, a fast search-locate algorithm based on the particle velocity is implemented to accommodate the rectangular or curved grid configuration.
The robustness and reliability of the developed code are confirmed by various verification cases, including the Lagrange tracking of massless particles, and one- and two-way coupled simulations.
Moreover, on the basis of the proposed framework, the direct numerical simulation of the particle-laden flow in a linear compressor cascade was successfully conducted, and the numerical accelerating effects of the proposed algorithms are validated.

In the present solver, the computation of the continue phase is conducted on GPUs, while the calculation of dispersed phase is performed on CPUs.
Although this configuration releases the computational overhead to some extent, it also results in unnecessary time expenditure for data transmission.
Consequently, a direction for future studies is the GPU implementation for the particle module.
%In addition, the current framework only applies to cases with two blocks, and the extension to more blocks will be developed.

\section*{Funding}

This work has been supported by the National Natural Science Foundation of China (Grant Nos. 92152202, 12432010, 12588201 and 12472262).

\section*{CRediT authorship contribution statement}

\textbf{Taiyang Wang:} Conceptualization, Data curation, Formal analysis, Investigation, Methodology, Software, Validation, Visualization, Writing original draft, Writing – review \& editing. 
\textbf{Baoqing Meng:} Methodology, Writing – review \& editing.
\textbf{Baolin Tian:} Writing – review \& editing.
\textbf{Yaomin Zhao:} Formal analysis, Funding acquisition, Project administration, Resources, Supervision, Writing – review \& editing.

\section*{Declaration of competing interest}

The authors declare that they have no known competing financial interests or personal relationships that could have appeared to influence the work reported in this paper.

\section*{Data availability}

Data will be made available on request.

\section*{Code availability}
The particle tracking module, which contains the implementation details of the present framework, can be found in the developer’s GitHub repository \url{https://github.com/Joe0425/Particle}.

\section*{Acknowledgments}\label{sec:Acknowledgments}
The authors wish to thank Prof. Richard D. Sandberg for the license of using the code HiPSTAR, and Zi-Mo Liao for helpful instructions on the numerical validation for particle-laden channel flow.

\newpage
\bibliographystyle{elsarticle-num}
\bibliography{tyw}

\end{document}